%% file: main.tex
\documentclass{aa}  

\usepackage{graphicx}
\usepackage{lscape}
\usepackage{subfigure}
\usepackage{natbib}
\usepackage[varg]{txfonts}
\usepackage{longtable}
\usepackage{booktabs}
\usepackage[normalem]{ulem} 
\usepackage{epstopdf}
\usepackage{xcolor}
\usepackage{soul}
\usepackage{multirow}
\usepackage{amsmath}

\usepackage{xspace}
\usepackage{amssymb}
\usepackage{adjustbox}

\usepackage{scalerel}
\usepackage{tikz}
\usetikzlibrary{svg.path}

\definecolor{orcidlogocol}{HTML}{A6CE39}
\tikzset{
  orcidlogo/.pic={
    \fill[orcidlogocol] svg{M256,128c0,70.7-57.3,128-128,128C57.3,256,0,198.7,0,128C0,57.3,57.3,0,128,0C198.7,0,256,57.3,256,128z};
    \fill[white] svg{M86.3,186.2H70.9V79.1h15.4v48.4V186.2z}
                 svg{M108.9,79.1h41.6c39.6,0,57,28.3,57,53.6c0,27.5-21.5,53.6-56.8,53.6h-41.8V79.1z M124.3,172.4h24.5c34.9,0,42.9-26.5,42.9-39.7c0-21.5-13.7-39.7-43.7-39.7h-23.7V172.4z}
                 svg{M88.7,56.8c0,5.5-4.5,10.1-10.1,10.1c-5.6,0-10.1-4.6-10.1-10.1c0-5.6,4.5-10.1,10.1-10.1C84.2,46.7,88.7,51.3,88.7,56.8z};
  }
}

\newcommand\orcidicon[1]{\href{https://orcid.org/#1}{\mbox{\scalerel*{
\begin{tikzpicture}[yscale=-1,transform shape]
\pic{orcidlogo};
\end{tikzpicture}
}{|}}}}

\usepackage{hyperref} 
\hypersetup{
    colorlinks=true,
    citecolor=blue,
    linkcolor=blue,
    urlcolor=blue,
    }

\definecolor{cadmiumred}{rgb}{0.89, 0.0, 0.13}

\definecolor{scc}{rgb}{0.13, 0.67,  0.8}
\definecolor{scc}{rgb}{0.54, 0.17, 0.89}
\definecolor{lars}{rgb}{0., 0.5,  0.}
\definecolor{ocol}{rgb}{1.0, 0.49, 0.0}
\definecolor{mpc}{rgb}{0.5, .3, 0.0}

\definecolor{scc}{rgb}{0.54, 0.17, 0.89}

\newcommand{\ULISSE}{\texttt{ULISSE}\xspace}
\newcommand{\ULISSEacro}{(aUtomatic Lightweight Intelligent System for Sky Exploration)\xspace}
\newcommand{\perc}{\,\%\xspace}
\newcommand{\comment}[1]{}

\makeatletter
\renewcommand*\aa@pageof{, page \thepage{} of \pageref*{LastPage}}
\makeatother

\begin{document}

\title{\ULISSE: A Tool for One-shot Sky Exploration and its Application to Active Galactic Nuclei Detection}

   \author{Lars Doorenbos\inst{1,\orcidicon{0000-0002-0231-9950}}, 
   Olena Torbaniuk\inst{2,3,\orcidicon{0000-0003-4465-2564}}, 
   Stefano Cavuoti\inst{4,5,\orcidicon{0000-0002-3787-4196}}, 
   Maurizio Paolillo\inst{2,4,5\orcidicon{0000-0003-4210-7693}}, \\
   Giuseppe Longo\inst{2,\orcidicon{0000-0002-9182-8414}}, 
   Massimo Brescia\inst{4,\orcidicon{0000-0001-9506-5680}}, 
   Raphael Sznitman\inst{1,\orcidicon{0000-0001-6791-4753}} \and 
   Pablo Márquez-Neila\inst{1,\orcidicon{0000-0001-5722-7618}}\\
   }
\titlerunning{\ULISSE: A Tool for One-shot Sky Exploration}
\authorrunning{L. Doorenbos et al. 2022}

   \institute{
   AIMI, ARTORG Center, University of Bern, Murtenstrasse 50, CH-3008 Bern, Switzerland\\
              \email{lars.doorenbos@unibe.ch}
    \and
    Department of Physics, University Federico II, Strada Vicinale Cupa Cintia, 21, 80126 Napoli, Italy
         \and
    Main Astronomical Observatory of National Academy of Sciences, 27 Akademika Zabolotnoho str., 03143 Kyiv, Ukraine
    \and
    INAF - Astronomical Observatory of Capodimonte, Salita Moiariello 16, I-80131 Napoli, Italy
    \and 
    INFN - Sezione di Napoli, via Cinthia 9, 80126 Napoli, Italy
             }

   \date{Received XXXX xx, 2022; accepted XXXX xx, 2022}

 
  \abstract
{Modern sky surveys are producing ever larger amounts of observational data, which makes the application of classical approaches for the classification and analysis of objects challenging and time-consuming. However, this issue may be significantly mitigated by the application of automatic machine and deep learning methods.}
{We propose \ULISSE, a new deep learning tool that, starting from a single prototype object, is capable of identifying objects sharing the same morphological and photometric properties, and hence of creating a list of candidate sosia. In this work, we focus on applying our method to the detection of AGN candidates in a Sloan Digital Sky Survey galaxy sample, since the identification and classification of Active Galactic Nuclei (AGN) in the optical band still remains a challenging task in extragalactic astronomy.}
{Intended for the initial exploration of large sky surveys, \ULISSE directly uses features extracted from the ImageNet dataset to perform a similarity search. The method is capable of rapidly identifying a list of candidates, starting from only a single image of a given prototype, without the need for any time-consuming neural network training.}
  %
{Our experiments show \ULISSE is able to identify AGN candidates based on a combination of host galaxy morphology, color and the presence of a central nuclear source, with a retrieval efficiency ranging from 21\perc to 65\perc (including composite sources) depending on the prototype, where the random guess baseline is 12\perc. We find \ULISSE to be most effective in retrieving AGN in early-type host galaxies, as opposed to prototypes with spiral- or late-type properties. }
{Based on the results described in this work, \ULISSE can be a promising tool for selecting different types of astrophysical objects in current and future wide-field surveys (e.g. Euclid, LSST etc.) that target millions of sources every single night.}


\maketitle
%
\input{01introduction}

\input{02method}

\input{03data}

\input{04experiments}
\input{05conclusion}

\begin{acknowledgements}
We acknowledge support from the European Union Horizon 2020 research and innovation programme under the Marie~Skłodowska-Curie grant agreement no. 721463 to the SUNDIAL Innovative Training Networks. L.D., R.S. and P.M.-N. acknowledge support from research grant 200021\_192285 `Image data validation for AI systems' funded by the Swiss National Science Foundation (SNSF). The work of O.T. was supported by the research grant number 2017W4HA7S `NAT-NET: Neutrino and Astroparticle Theory Network' under the program PRIN 2017 funded by the Italian Ministero dell'Universit\`a e della Ricerca (MUR). M.P. and O.T. also acknowledge financial support from the agreement ASI-INAF n.2017-14-H.O. M.B. acknowledges financial contributions from the agreement ASI/INAF 2018-23-HH.0, Euclid ESA mission - Phase D.  The authors would like to thank the anonymous referee for the comments and suggestions which helped us to improve the paper.

\end{acknowledgements}

\bibliographystyle{aa}
\bibliography{main} 
\onecolumn

\begin{appendix}

\input{06supplementary.tex}

\end{appendix}
\end{document}

%% file: 01introduction.tex
\section{Introduction}

In the last twenty years numerous digital surveys such as the Sloan Digital Sky Survey (SDSS, \citealt{York2000}), Kilo Degree Square Survey (KiDS, \citealt{deJong2015}), Panoramic Survey Telescope and Rapid Response System (Pan-STARRS, \citealt{Magnier2020}), Dark Energy Survey (DES, \citealt{Abbott2016}), Hyper Suprime-Cam Subaru Strategic Program (HSC SSP, \citealt{Aihara2019}) greatly improved our knowledge of the Universe by exploring deep and wide areas of the sky through multi-wavelength imaging campaigns.
In the coming years, new  multi-band  wide-field survey projects, such as the Vera C. Rubin Observatory Legacy Survey of Space and Time (Rubin-LSST, \citealt{Ivezic2019}), Euclid \citep{Scaramella2021}, Nancy Roman Telescope (formerly the Wide-Field Infrared Survey Telescope or WFIRST, \citealt{Green2012}), James Webb Space Telescope (JWST, \citealt{AlvarezMarquez2019}), will further increase by orders of magnitudes the amount of observational data. 
Most of these future  surveys will in fact produce photometric data for several millions of sources each night. However, since spectroscopic follow-ups for even a small fraction of the observed sources will be unfeasible, there is the need for algorithms capable to exploit photometric information to classify, or at least identify, interesting candidates to be further investigated. It comes as no surprise that, in recent years, much work has been devoted to implement and fine tune  fast and self-adaptive learning methods for prediction, classification, visualization (in other words, for data understanding) inducing the exploitation of astroinformatics solutions, for instance the paradigms of machine and deep learning \citep{Baron2019,Longo2019,Fluke2020,Lecun1998,Disanto2018,Schaefer2018}, replacing more classical methods, considered inefficient in the big data regime. 

We can roughly subdivide the machine learning (ML) algorithms in two broad classes. The first one, which is probably the most used, is called supervised. Such methods rely on the availability of a set of data for which we believe to possess some ground truth (labels) that is used to train the algorithm. The other possibility is to have an unsupervised model, working on the data without any or almost any a priori knowledge. In this case, the labels (if any) are used only a posteriori, to analyze and understand the results. It goes without saying that with a supervised approach the interpretation of the results is by far easier and that such methods can be more easily tailored to solve a specific problem. This explains why, so far, the number of works dealing with supervised methods is much larger (see a few examples: \citealt{Weir1995,Kim2011,Brescia2013,Disanto2018b,DelliVeneri2019,Schmidt2020,Kinson2021,Wenzl2021}) than those about unsupervised approaches. Nonetheless, there are several successful examples of unsupervised approaches in astrophysical problems (e.g. \citealt{Masters2015,Baron2017,FronteraPons2017,Mislis2018,CastroGinard2018,Razim2021,Ofman2022}).

Supervised ML methods imply the necessity of a training set, derived from real data or simulations. However, in the case of real data, multi-band photometric observations cannot provide a full  understanding of the physical processes at work  and even spectroscopic observations are seldom fully representative of the complexity of the parameter space describing our universe.  \citet{Masters2015}, for instance, have shown that there are always portions of the parameter space left under-sampled (if not unexplored) and this is even more true when dealing with rare objects for which there are very few labels (if compared with more common objects).

The vast it  of unsupervised solutions can be regarded as clustering or pre-clustering methods, such as those devoted to the reduction of dimensionality \citep{Bishop2006}. Under-sampled or rare objects usually are penalized in these kinds of representation, because they seldom succeed in creating a cluster on their own.

On the edge between the two approaches, lies the field of one-shot learning, where only a single labelled sample is available per class of interest \citep{wang2020generalizing}. We apply this paradigm here as it has the potential to combine the best of both worlds: it removes the need for the expensive and often unfeasible process of collecting a large labelled dataset, inherent to supervised methods, while at the same time removing the problem encountered with unsupervised methods, that have trouble with rare and under-sampled objects.

In this work we present \ULISSE \ULISSEacro: a one-shot method capable of retrieving objects closely related to a given input, and apply it directly to multi-band images\footnote{An example notebook is provided at \url{https://github.com/LarsDoorenbos/ULISSE}.}. 
The idea behind our method is quite intuitive: we transform the image of a given source (hereafter prototype) into a set of representative features, after which we use this information to look for other objects similar to the prototype in the feature space, which should translate to similarity in the astronomical sense.
The power of this method comes from the fact that even if we take a rare object as a prototype, the method allows us to search for similar objects in the dataset, thus bypassing the need for a large and well sampled training set, and actually provide a reliable list of candidates for follow-up observations, thus opening the way to the construction of reliable training sample for supervised methods.

In order to test the method, we apply it here to the detection of Active Galactic Nuclei (AGN). 

The identification of AGN in the optical band is not trivial due to the strong contamination from the host galaxy and obscuration by the circumnuclear or galactic dust. This imposes a whole set of problems, which can be resolved with the usage of multi-wavelength observations (from radio to X-ray) and the combination of different selection techniques. A proper AGN selection plays a crucial role in the study of the formation and evolution of supermassive Black Holes \citep{Brandt2005,Merloni2016, Hickox2018} and their feedback on the host galaxies \citep{Fabian2012, Kormendy2013, Heckman2014, Tracker2014}. The identification of AGNs will be an important task for all future surveys such as the Rubin-LSST, since the new data will allow for the study of the formation and co-evolution of supermassive black holes, their host galaxies, and their dark matter halos. Furthermore, the classification of AGNs is important also for other science cases, since they need to be handled in an independent way with respect to standard galaxies when deriving, for example, photometric redshifts \citep{Brescia2019,Desprez2020}. 

The search for candidate AGNs in surveys has been performed already in the past with ML models applied to photometric tabular data \citep{Cavuoti2013,Fotopoulou2018,Chang2021,Falocco2022} or to their variability \citep{Palaversa2013,Disanto2016,Sanchez-Saez2019,Faisst2019,DeCicco2021}. More recently there was also an attempt to use, for the same task, deep neural networks \citep{Chen2021}. 

A concurrent work in \cite{stein2021mining} presents a similar approach to ours for the detection of astronomical objects by performing similarity search on images, which is applied to the detection of strong gravitational lenses. Contrary to our approach, rather than pre-trained features, they make use of self-supervised pre-training.

This paper is structured as follows: In Sect.~\ref{sec:method} we present our method, while in Sect.~\ref{sec:data} we describe our data sample and present the studied prototypes. Section~\ref{sec:exp} is devoted to the presentation of the experiments and their outcome. Section \ref{sec:discussion} contains the discussion of the results together with the analysis of the method limitations and the possible improvements. Finally, in Sect. \ref{sec:conclusions} we present brief conclusions.

%% file: 02method.tex
\section{Method}\label{sec:method}
For \ULISSE we make use of the features extracted from a convolutional neural network (CNN), that has been pretrained on a general large-scale dataset. It then finds relevant objects through a nearest neighbor search. We describe each of the architecture components in the following paragraphs.

\subsection{Convolutional neural networks}
In the context of classification, convolutional neural networks usually consist of two parts \citep{Schmidhuber2015}: the first part, transforms the input image into a feature vector through a series of convolutional layers, pooling layers and activation functions (described below) and, in practice, operates as a feature extractor. 
The second part takes these features and uses them to perform the actual classification task. Usually this second part consists of a multi layer perceptron (MLP, \citealt{McCulloch1943,Rosenblatt1958}) neural network.

The convolutional layers shift a number of small windows (kernels) over the input, computing a weighted average of the local surroundings, obtaining the so-called feature maps as the output, which indicate how strongly a given location correlates with the window. The weights of these windows are learned by the model itself.

Pooling layers decrease the size of these feature maps, by replacing each location in the input with an aggregate statistic derived from its rectangular neighborhood. The size of this neighborhood is set by the user. Commonly used examples are average pooling, which outputs the average value within the window, and max pooling, which reduces a region to its maximum value.

\begin{figure*}  
\centering
\begin{tabular}{cccccccc} 
\hline
11 & 41 & 541 & 835 & 1073\\
\hline \\
\includegraphics[width=0.17\linewidth]{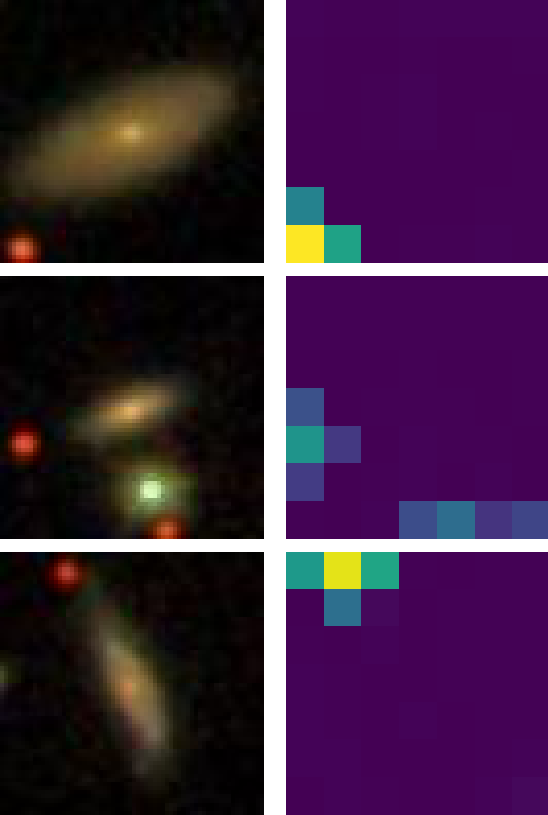} &
\includegraphics[width=0.17\linewidth]{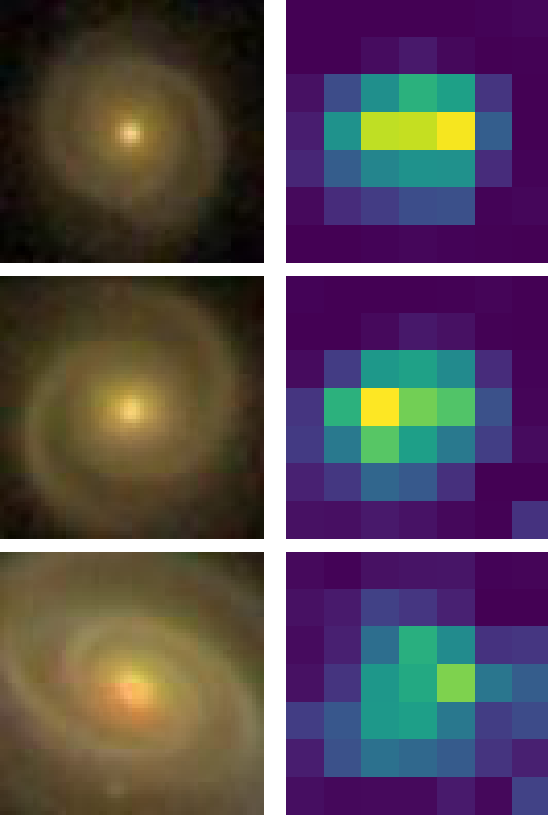} &
\includegraphics[width=0.17\linewidth]{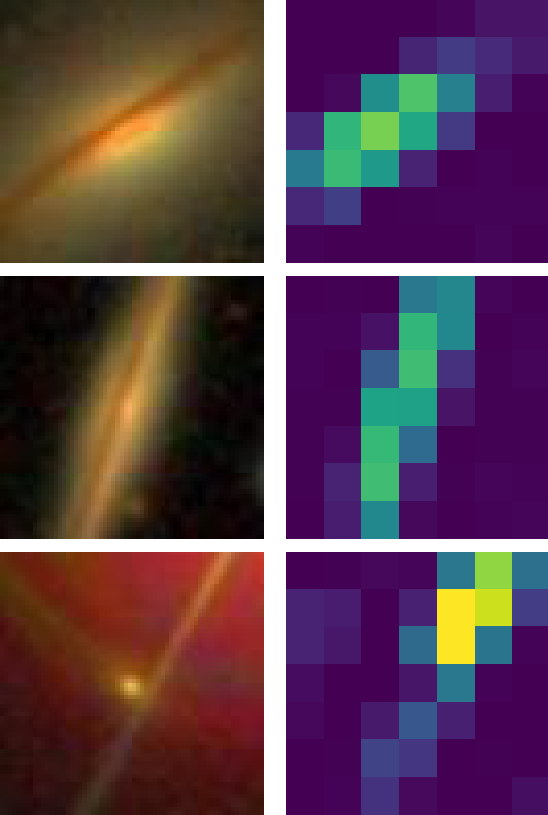} &
\includegraphics[width=0.17\linewidth]{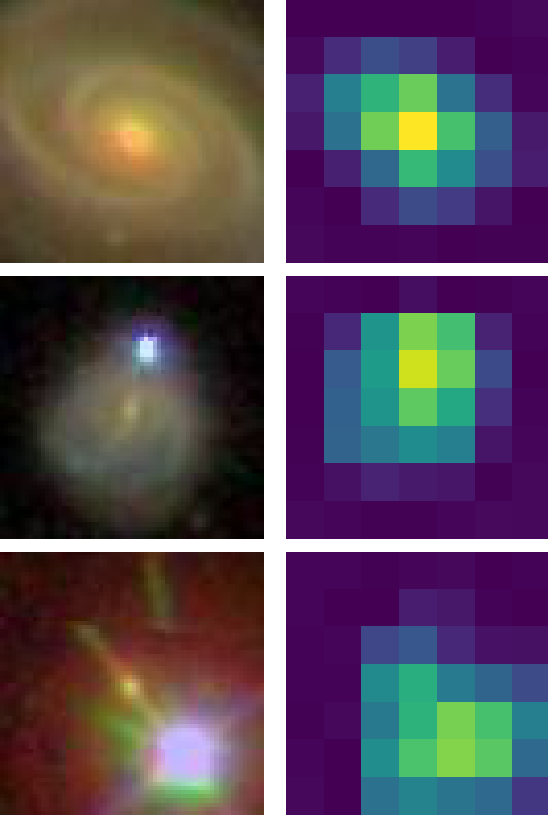} & 
\includegraphics[width=0.17\linewidth]{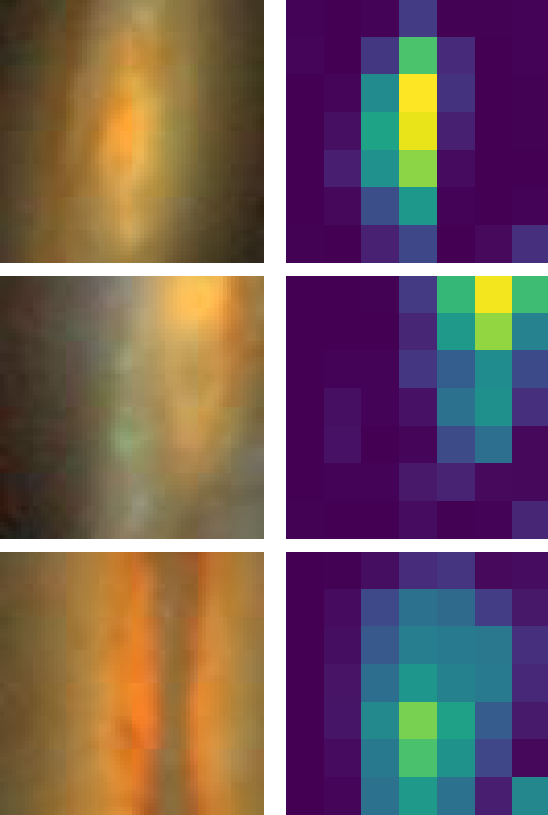} 
 \\
\hline
\hline
\end{tabular}
\caption{The three objects in our sample which most strongly activate features 11, 41, 541, 835 and 1073 (arbitrarily chosen for visualization), together with their feature maps. We provide these for all 1280 features at the  \url{http://dame.na.astro.it/ulisse}.}\label{tab:maps}

\end{figure*}

The activation functions are applied with the goal to introduce non-linearity in the feature representation. A common example is the ReLU (Rectified Linear Unit) activation function, which sets the negative part of its input to zero, $f(x)=max(0,x)$.

The feature extractor part of the algorithm is followed by an MLP, consisting of one or more fully connected layers, which performs the actual classification of the input, based on the features obtained in the previous part. In an MLP a fully connected layer connects every input neuron to every output neuron. The final layer outputs a vector with as many elements as the number of classes and the neuron with the highest value identifies the final decision.
The model is optimized with respect to a loss function end-to-end by forward propagation and error back-propagation \citep{Bishop2006, Goodfellow-et-al-2016}.

\subsubsection{Pretraining}
Training a CNN on a large and varied dataset allows it to learn a diverse and general set of features, which should be useful beyond the original task. These features are often used as the starting point for complex tasks in another domain (the target domain), with the aim of avoiding labelling the large amounts of data often needed to train a model from scratch \citep{pan2009survey}. This concept is known as transfer learning and has proven very successful in many domains (for instance, astronomy,  \citealt{Awang2020,Martinazzo2020}, malware classification, \citealt{Prima2020}, earth science, \citealt{Zou2018}, and medicine, \citealt{ding2019deep, esteva2017dermatologist,Kim2021tl, menegola2017knowledge}). 

The typical large-scale dataset of choice used for training is ImageNet \citep{deng2009imagenet} which contains around 1.3 million images, where the original task was to classify each image into one of 1000 classes.
In practice, when moving to the target domain, 
the second part of the above described architecture, namely the fully connected layers of the pre-trained network (the classifier) are discarded, as the new domain does not contain the same classes. 
The feature extracting part of the network is then used to tackle the target task. As an additional benefit of this approach, no fine-tuning to the target domain is needed, thus reducing the training time to almost zero and making it directly applicable to any new dataset.

\subsubsection{Feature extraction}
In order to obtain our features, we first extract the feature maps (7x7 pixels each) from the final convolutional layer of the pre-trained neural network. In order to reduce their dimensionality, we then average over the spatial dimensions. As a result, the features represent image-level properties.

In deep learning, it is a common understanding that the deeper the layers are in the network, the higher is the level of abstraction of the extracted features \citep{Goodfellow-et-al-2016}. Our approach is therefore based on the assumption that objects whose image share many deep features with the prototype (i.e. that are close together in this feature space), have also the same morphological properties. 
We wish to emphasize that, as it will be further discussed in the coming sections, since we are working with multiband images, in this context morphology must be intended in a broader sense, since we also take implicitly into account variations in the color distribution.

Throughout this work we use an EfficientNet-b0, a specific type of CNN architecture \citep{tan2019efficientnet} that was trained for classification on ImageNet as the CNN from which we obtain the features. Its penultimate layer consists of 1280 channels, leading to a 1280-dimensional feature descriptor for each image.

Note that the features are extracted from the model, and were derived from natural images (i.e. heterogeneous images of everyday objects or scenes such as cats, cars, rivers and so on), rather than astronomical ones. Hence, they are not directly interpretable. Nonetheless, we can get an idea of the patterns individual features are looking for, by looking at the images in our dataset which most strongly activate them. We show this for the five cases in Fig.~\ref{tab:maps}, where it becomes clear that different features are focusing on different aspects of the image. We provide these visualizations for each of the 1280 features at \url{http://dame.na.astro.it/ulisse}. 
With reference to Fig.~\ref{tab:maps}, we need to stress that even though the similarity of the objects is defined through a complex combination of features, in some cases it is possible to recognize specific patterns which are associated to a given feature.
For instance, feature 41 seems to be activated by extended objects with a bright nucleus, while feature 541 responds most strongly to narrow objects with some sort of bulge.

\subsection{\ULISSE}

\begin{figure}
    \centering
    \includegraphics[width=.49\textwidth]{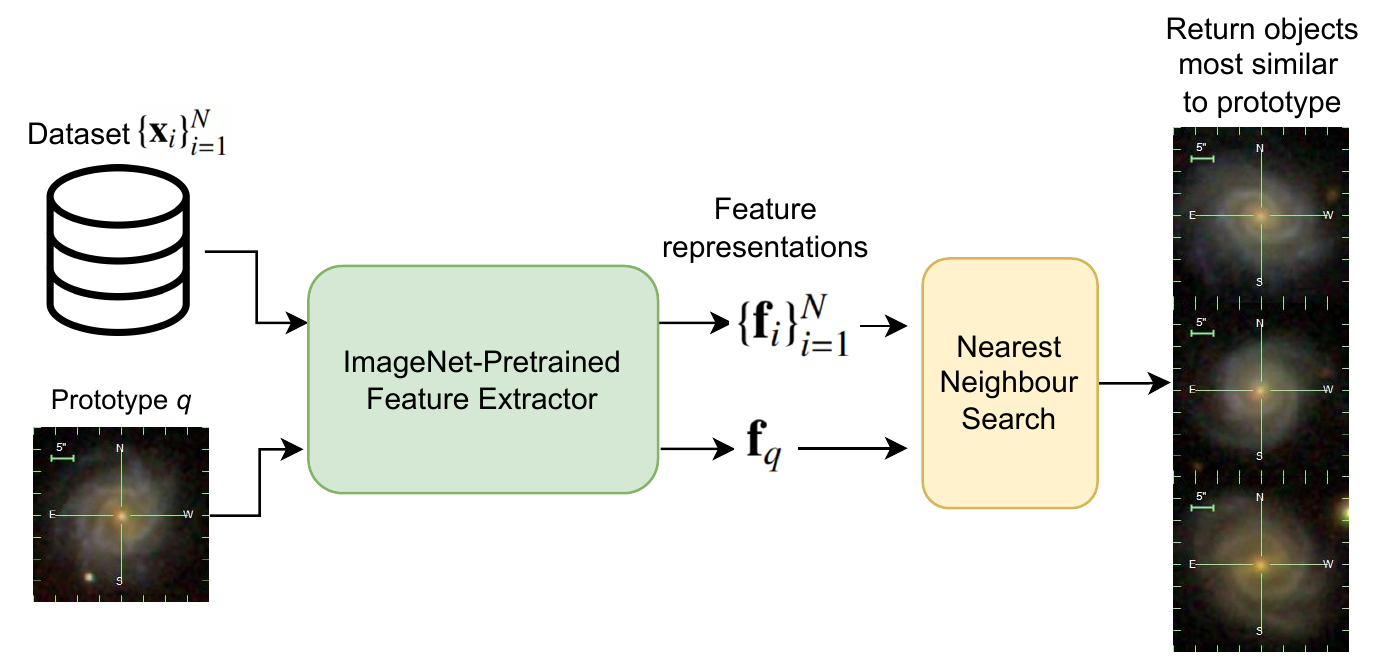}
    \caption{Overview of \ULISSE.}
    \label{fig:ULISSE}
\end{figure}

These pretrained features are used by \ULISSE to identify objects with similar properties. This is done by performing a similarity search in the feature space. 
Given a prototype object, the closest objects in this feature space 
provide a list of candidate lookalikes for the prototype of interest. A schematic of the proposed method is shown in Fig.~\ref{fig:ULISSE}.

Formally, given a prototype image $\textbf{x}_q$, we wish to retrieve its nearest neighbors from a dataset $\{\textbf{x}_i\}_{i=1}^N$. 
For an image $\textbf{x}_i$, we denote its pretrained feature representation as $\mathbf{f}(\textbf{x}_i) \equiv \textbf{f}_i$. 
We find the nearest neighbors by their Euclidean distance in this pretrained feature space \citep{hastie2009elements}. Hence, we find those objects $\textbf{x}_i$ that minimize $d(\textbf{x}_q, \textbf{x}_i) = ||\textbf{f}_{q} - \textbf{f}_i||^2$. With the use of acceleration structures such as \textit{k}-d trees, which allow for the efficient computation of nearest neighbor searches, these lookalikes are extremely fast to find after their initial construction.

If no validation data is available, \ULISSE can be applied by simply returning the $n$ closest objects in the dataset as a list of candidates (measured in feature space, not in terms of astronomical distance), where $n$ is set by the user. However, if we do have access to a validation set (even if it is limited), we can use this to measure our desired performance metric as a function of the distance of the lookalike from the prototype, with which we can determine a threshold on the distance for use on new data. For instance, if our aim is to retrieve AGN, we can determine until which distance from our prototype the fraction of AGN among the retrieved objects from the validation set remains high, and select all objects that fall within this distance from our target dataset.

Concretely, one can choose the furthest object $\textbf{x}_f$ at which point the selected sample still gives good performance, with distance $t = ||\textbf{f}_{q} - \textbf{f}_f||^2$. Then, on the target dataset, all objects from $\{\textbf{x}_i\}_{i=1}^N$ satisfying $d(\textbf{x}_q, \textbf{x}_i) \leq t$ are selected.

%% file: 03data.tex
\section{Data}\label{sec:data}

In order to assess the result of this method we need a sample with well-known object classification and for this purpose, we choose to retrieve thumbnails for the sample used in \cite{Torbaniuk2021}, which is based on the {\it galSpec} catalog of galaxy properties\footnote{\url{https://www.sdss.org/dr12/spectro/galaxy_mpajhu/}} produced by the MPA-JHU group as a subsample from the main galaxy catalog of the 8th Data Release of the Sloan Digital Sky Survey (SDSS DR8, \citealt{Brinchmann2004}).
This sample has been cleaned by removing duplicates and objects with  low-quality photometry using the basic photometry processing flags recommended by SDSS\footnote{\url{https://www.sdss.org/dr16/algorithms/photo_flags_recommend/}}. 
The objects were classified using the so-called BPT-diagrams \citep[from the names of the proposers: Baldwin, Phillips \& Terlevich,][]{Baldwin1981} as: star-forming galaxies `SFG', `AGN' and `Composite' according to the ratios of four specific emission lines in their optical spectra \citep{Kauffmann:03c,Kewley:06}. \citet{Brinchmann2004} expanded this selection criteria and add two additional classes of objects with low signal-to-noise lines (low S/N SFG and AGN). In our work, we used the naming `SFG' for both low S/N SFG and SFG, and `AGN' for both AGN and low S/N AGN. At the same time, our sample also contains a significant fraction of objects with weak or no emission lines, which remain unclassified due to limitations of the BPT-diagram (see Table\,\ref{tab:proportions}). Moreover, some SFGs may contain a low-luminosity AGN, which may not be classified by the BPT-diagram since the emission of the star-forming processes in the host galaxy will dominate the optical spectrum. To address these problems and improve the selection, we added X-ray detections from the XMM-Newton Serendipitous Source Catalog (3XMM-DR8, \citealt{Rosen2016}). The total number of objects in our optical sample observed by {\it XMM-Newton} and their fraction in each class are presented in Table\,\ref{tab:proportions} (X-ray MOC subsample). X-ray AGN were classified according to the X-ray selection criteria described in detail in \citet{Torbaniuk2021}, that is large X-ray luminosity or X-ray/optical ratio. 

\begin{table}
\caption{The summary of the different datasets studied in this work. The fraction represents the percentage of objects in each dataset classified as AGN, SFG or composite according to the optical BPT-diagram or X-ray AGN/non-AGN by X-ray selection criteria (see details in the text). Unknown class indicates the fraction of objects which have not been observed by {\it XMM-Newton}.}
\begin{center}
\begin{tabular}{cclrrr}
\hline\hline
\multicolumn{3}{c}{\multirow{2}{*}{Sample}} & \multicolumn{3}{c}{Fraction} \\
\cline{4-6}
\multicolumn{3}{l}{} & \multicolumn{1}{c}{Entire} & \multicolumn{1}{c}{X-ray MOC} & \multicolumn{1}{c}{Random} \\
\hline\hline\\[-1.5ex]
\multirow{7}{*}{\rotatebox[origin=c]{90}{\centering Criteria}} & \multirow{4}{*}{\rotatebox[origin=c]{90}{\centering BPT}} & AGN & 12.0\perc & 11.8\perc\hphantom{0} & 12.0\perc\\
& & Composite & 5.8\perc & 5.5\perc\hphantom{0} & 5.8\perc \\
& & SFG & 44.0\perc & 41.5\perc\hphantom{0} & 44.1\perc\\
& & Unclassified & 38.2\perc & 41.2\perc\hphantom{0} & 38.1\perc\\
\cline{2-6}\\[-1.5ex]
& \multirow{3}{*}{\rotatebox[origin=c]{90}{\centering X-ray}} & AGN & 0.2\perc & 4.0\perc\hphantom{0} & 0.2\perc \\
& & non-AGN & 5.6\perc & 96.0\perc\hphantom{0} & 5.6\perc\\
& & Unknown & 94.2\perc & ---\hphantom{0000} & 94.2\perc\\
\hline\\[-1.5ex]
\multicolumn{3}{l}{Number of objects} & 703\,422 & 40\,889\hphantom{00} & 99\,991\\
\hline
\hline
\end{tabular}
\end{center}
\label{tab:proportions}
\end{table}

For each object in our sample, we extracted thumbnails from the Sloan Digital Sky Survey Data Release 16 (SDSS DR16, \citealt{Eisenstein2011,Blanton2017,Ahumada2020,York2000}) using the Image Cutout service\footnote{\url{http://skyserver.sdss.org/dr16/en/help/docs/api.aspx}}. This service allows the retrieval of \texttt{JPEG} images composed of the three inner SDSS bands (\textit{g}, \textit{r} and \textit{i}) for any portion of the sky observed by SDSS just based on its coordinates. 
In this work, we decided to retrieve $64\times64$ pixel thumbnails. This is based on initial experiments with $48\times48$, $55\times55$, $64\times64$ and $73\times73$ pixel thumbnails, where we found that $64\times64$ gave the best results. As the images are reduced to $56\times56$ pixels after center cropping, they correspond to a size of $\sim 22.2\times22.2$ arcseconds (the SDSS pixel scale is $0.396$\,arcseconds per pixel).
The choice of the thumbnail size is important because a small size may lead to cutting the edges of an extended object, while a uselessly large thumbnail could contain more than one object, thus affecting the characteristic extracted by our method. Recall that as we average our features over the spatial dimensions, they describe image-wide properties.

There is no proper way to completely eliminate both effects, but considering our preliminary experiments and the reference objects themselves (see Sect.\,\ref{sec:ref-obj}) we decided on $22.2\times22.2$ arcseconds as a good compromise.
Although the spectroscopic information comes from an earlier data release (DR8) we decided to retrieve the images from the latest version of SDSS in order to benefit from any possible improvements or bug removals
of the pipeline. Considering the redshift range covered by our sample, these thumbnails correspond to physical sizes ranging from $\sim 50$ pc up to $\sim 110$ kpc on a side.

\begin{table*}  
\renewcommand{\arraystretch}{1.4}
\caption{Summary of the thumbnails used as AGN prototypes in our work. BPT class for each prototype was taken from the {\it galSpec} catalog (see details in \citealt{Brinchmann2004}). X-ray classification is from \citet{Torbaniuk2021} based on 3XMM-DR8 data \citep{Rosen2016}. }\label{tab:reference}
\centering
\setlength\tabcolsep{1 pt}
\begin{tabular}{lccccc} 
\hline\hline   
\multicolumn{1}{c}{\#} & 1 & 2 & 3 & 4\\
\hline\hline   
SDSS & J032525.36-060837.8 & J164607.00+422737.4 & J153621.30+222913.6 & J133548.24+025956.1 \\
RA & 51.35569 & 251.52917 & 234.08879 & 203.95103 \\
DEC & $-6.14386$ & 42.46041 & 22.48712 & 2.99892 \\
Redshift & 0.034 &0.049 & 0.089 & 0.022 \\
Thumbnails &  \parbox{3.1cm}{\includegraphics[width=3.1cm]{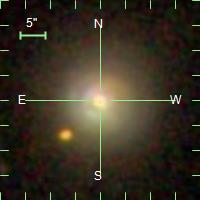}} & \parbox{3.1cm}{\includegraphics[width=3.1cm]{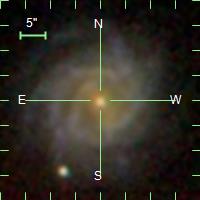}} & 
\parbox{3.1cm}{\includegraphics[width=3.1cm]{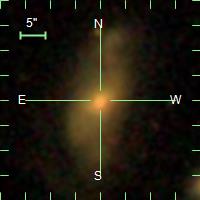}} &  \parbox{3.1cm}{\includegraphics[width=3.1cm]{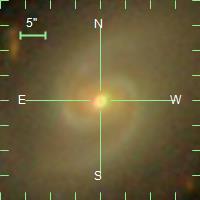}} \\ 
SDSS class & Galaxy & Galaxy & QSO & Galaxy \\
SDSS subclass & AGN & Starforming & AGN Broadline & AGN \\
BPT class & AGN & SFG & AGN & SFG \\
X-ray class & AGN & AGN & AGN & AGN\\
\hline\hline   
\multicolumn{1}{c}{\#}  & 5 & 6 & 7 & 8\\
\hline\hline   
SDSS & J084002.36+294902.6 & J125754.36+272926.2 & J075219.80+174210.3 & J110511.08+382129.3 \\
RA & 130.00986 & 194.47651 & 118.08252 & 166.29618 \\
DEC & 29.81740 & 27.49063 & 17.70287 & 38.35815 \\
Redshift & 0.065  & 0.017  &  0.097 &  0.046 \\
Thumbnails &  \parbox{3.1cm}{\includegraphics[width=3.1cm]{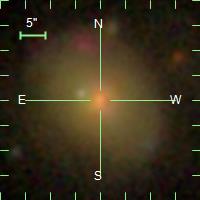}} & \parbox{3.1cm}{\includegraphics[width=3.1cm]{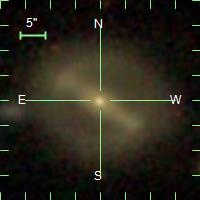}} &  \parbox{3.1cm}{\includegraphics[width=3.1cm]{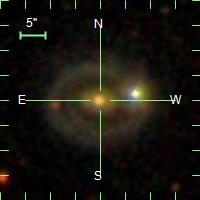}} & \parbox{3.1cm}{\includegraphics[width=3.1cm]{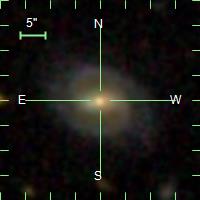}}  \\ 
SDSS class & Galaxy & Galaxy & Galaxy & Galaxy \\
SDSS subclass & AGN Broadline & Starforming & --- & Starforming \\
BPT class & AGN & SFG & AGN & Composite \\
X-ray class & AGN & AGN & AGN & AGN\\
\hline\hline   
\end{tabular}
\end{table*}

As our goal is to show the validity of our method, rather than to obtain results for the whole sample, we create a smaller subset of the data for efficient computation. To this end, we randomly shuffle the coordinates, and take the first 100 000 objects. In some cases, the Image Cutout service fails to retrieve a thumbnail, hence we are left with 99 991 images as our dataset. This is denoted with \textit{Random} in Table\,\ref{tab:proportions}, where it can be seen its proportions are practically equal to that of the whole sample.
The percentage of sources  classified as AGN according to the BPT selection in the \textit{Random} sample ia 12\perc (see Table\,\ref{tab:proportions}) and will be used as the random guess baseline to test our method.

\begin{table*}  
\renewcommand{\arraystretch}{1.4}
\caption{Summary of the thumbnails used as non-AGN prototypes in our work. BPT and X-ray classification are the same as in Table\,\ref{tab:reference-non-agn}.}\label{tab:reference-non-agn}
\centering
\setlength\tabcolsep{1 pt}
\begin{tabular}{lccccc} 
\hline\hline   
\multicolumn{1}{c}{\#}  & 9 & 10 & 11 & 12 & 13 \\
\hline\hline   
SDSS & J115928.62+423542.8 & J151121.53+072250.6 & J134059.80+302058.0 & J083114.54+524224.8 & J151105.13+053112.7\\
RA & 179.86926 & 227.83972 & 205.24919 & 127.81060 & 227.77139 \\
DEC & 42.59522 & 7.38073 & 30.34947 & 52.70690 & 5.52020\\
Redshift & 0.114  & 0.044  &  0.040 & 0.064 & 0.035 \\
Thumbnails &  \parbox{3.1cm}{\includegraphics[width=3.1cm]{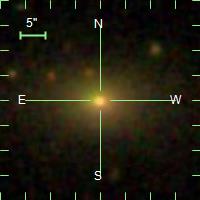}} & \parbox{3.1cm}{\includegraphics[width=3.1cm]{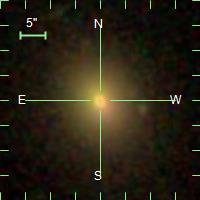}} &  \parbox{3.1cm}{\includegraphics[width=3.1cm]{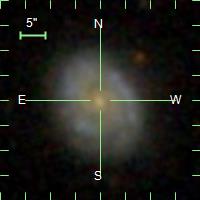}} & \parbox{3.1cm}{\includegraphics[width=3.1cm]{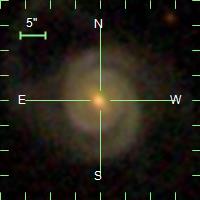}} & \parbox{3.1cm}{\includegraphics[width=3.1cm]{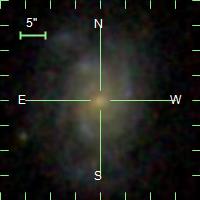}}\\ 
SDSS class & Galaxy & Galaxy & Galaxy & Galaxy & Galaxy \\
SDSS subclass & --- & --- & Starforming & Starforming & Starforming \\
BPT class & Unclassified & Unclassified & SFG & SFG & SFG \\
X-ray class & non-AGN & non-AGN & non-AGN & non-AGN & non-AGN \\
 \hline \hline
\end{tabular}
\end{table*}

\subsection{Prototypes}\label{sec:ref-obj}

In order to detect candidate AGN we need to select a set of prototypes that contains `true' AGN and non-AGN confirmed by some reliable criteria. Since X-ray emission is a good (and least contaminated) tracer of accretion processes, we selected the two groups of prototypes (AGN and non-AGN) within the X-ray MOC subsample in Table \ref{tab:proportions}, based on X-ray criteria. Furthermore, numerous studies show that AGN may be detected in galaxies of different morphological types, however they show some preference for star-forming galaxies (probably due to larger reservoir of cold gas available for SMBH accretion in star-forming galaxies compared to quiescent ones, see \citealt{Lutz2010, Mendez2013, Rosario2013, Stemo2020}). We thus selected prototypes that represent a variety of morphological types. The list of eight AGN prototypes according to the X-ray criteria is presented in Table~\ref{tab:reference}, while five non-AGN prototypes are presented in Table\,\ref{tab:reference-non-agn}. For the AGN prototypes we selected a couple of blue spiral galaxies (i.e. with younger stellar population and ongoing star-formation processes,\,\#2,\,8), elliptical galaxies with red color (i.e. older stellar population and quenched star-formation,\,\#3,\,5), a transition galaxy (\#4) with red color and spiral structures, a peculiar galaxy with possible arms or shell-like features which may be caused by past galaxy interactions (\#1), ring (\#7) and bar (\#6) galaxies. 
In addition to X-ray AGN identification, these prototypes have also been classified using the BPT-diagram. Prototypes\,\#1,\,3,\,5,\,7 are identified as AGN (i.e the AGN signatures dominate the optical spectrum,\,\#2,\,4,\,6 are SFGs according to the BPT-diagram (i.e the star-forming signatures dominate the optical spectrum). At the same time, the prototype\,\#8 is marked as a ‘composite’ displaying the contribution from both an AGN and star-forming processes. Since the \ULISSE performance may depend on the morphology of the host galaxy rather than the presence of an AGN, we also added non-AGN prototypes to our study. The selection of non-AGN prototypes was performed by a similar approach applied to AGN-prototypes selection: we chose elliptical (\#9,\,10) and spiral galaxies with blue (\#11,\,13) and red color (\#12). In this case, three spiral galaxies are classified as SFG according to the BPT-diagram, while the ellipticals remain unclassified due to the absence or weakness of emission lines in their optical spectra. 

It should be noticed that in Table~\ref{tab:reference} and~\ref{tab:reference-non-agn} we used the $100 \times 100$ explorer images, while our experiments were performed based on $64\times64$ thumbnails.

%% file: 04experiments.tex
\section{Experiments and results}\label{sec:exp}

In this section, we present our results for the one-shot AGN detection, along with an analysis of the morphology and color of each prototype object and their contribution to the overall AGN fraction (Sect.~\ref{sec:color-res}). In addition, in Sect.~\ref{sec:recursive} we introduce a recursive application of our method as a promising technique to retrieve a purer sample with a larger number of candidates. 

\subsection{AGN detection results}\label{sec:results}

As it was mentioned in the previous section, for each object in our sample we have labels according to the BPT diagram (see Table~\ref{tab:proportions}), while the labels from the X-ray classification are available only for the X-ray MOC subsample. To begin our experiment we used the $g,r,i$ color-composite thumbnails. The results corresponding to each prototype are visualized as the number of objects of different classes (AGN, SFG etc.) selected by our method versus the distance from our prototype (see Table~\ref{tab:refs},~\ref{tab:refs2} for AGN prototypes and Table~\ref{tab:refs3} for non-AGN prototypes). As no validation set is available, we set the number of objects $n$ retrieved by \ULISSE to 300 nearest neighbors. We chose this number in order to study the variation of AGN fraction with the distance. However, in practice, the choice of $n$ depends on the purpose of the user and goals of the study. 

The experiments for the 8 ``AGN'' prototypes retrieve on average 34.0\perc objects also labelled as AGN over all 300 nearest neighbors. However, this fraction varies from object to object. For instance, for prototypes\,\#1,\,3 and 5 our method retrieved $\sim40$\perc of AGN (on average) within the 300 nearest neighbors. Taking into account also composite objects (i.e. with spectral signatures both from the AGN and star-formation) the resulting fraction of AGN+Composite reaches  $\sim 55\perc$. The rest of the sources belong to the SFG and Unclassified classes (see details in Table~\ref{tab:fractions}). In the case of prototype\,\#4 the fraction of AGN reached 53\perc (up to 65\perc for AGN+Composite). On the other hand, \ULISSE retrieved a relatively lower fraction of AGN (21-28\perc) for prototypes\,\#2,\,6,\,8 (close to 40\perc for AGN+Composite, see Table~\ref{tab:fractions}). In general, however, the retrieved AGN fraction for all AGN prototypes significantly exceeded the value expected by the random guess baseline, which is 12\perc according to the BPT selection presented in Table~\ref{tab:proportions}.

In the case of non-AGN prototypes (objects\,\#9-13 in Table\,\ref{tab:reference-non-agn}) we expect a smaller number of AGN to be retrieved. For instance, for prototypes\,\#9, 10 labeled as `unclassified' according to BPT-diagram, \ULISSE found only 21\perc of AGN, while the rest of resulting objects is labelled as `unclassified'. Similar result were obtained for  prototypes\,\#11,\,13 labeled as `SFG', where \ULISSE retrieved only 14.3\perc and 1.7\perc of AGN. In the case of prototype\,\#12 (SFGs according to BPT-diagram) the method retrieved relatively large fraction of AGN compared to other non-AGN prototypes mentioned above, which is also comparable to the resulting percentage of SFGs. 

The BPT classification is not always able to detect the presence of an AGN and disentangle nuclear from star-formation activity, especially for low S/N spectra. For this reason, we decided to repeat our experiments using the X-ray MOC subsample of sources observed by {\it XMM-Newton} (Table\,\ref{tab:proportions}) and the X-ray classification criteria. The resulting AGN/non-AGN fraction as a function of distance are presented in Table\,\ref{tab:refs},\,\ref{tab:refs2} (for AGN prototypes\,\#1-8) and Table\,\ref{tab:refs3} (non-AGN prototypes\,\#9-13). The percentage of AGN/non-AGN obtained by \ULISSE for all studied prototypes are presented in Table\,\ref{tab:fractions}. The average retrieved AGN fraction ranges between 8\perc and 12\perc, which again exceed the percentage of AGN expected by the random guess baseline for X-ray MOC sample (4\perc; see Table\,\ref{tab:proportions}). The same test with the X-ray MOC sample, again yields lower AGN fractions for non-AGN prototypes (see Table\,\ref{tab:fractions}). 

In Appendix~\ref{sec:appendix} we present the closest 25 neighbors returned by \ULISSE for each AGN/non-AGN prototypes (see Fig.~\ref{fig:nns1}-\ref{fig:nns12}).

\subsection{Disentangling morphology and color}\label{sec:color-res}

Traditionally, AGN identification relies on different selection criteria based on their photometric and spectroscopic information (such as color-color diagrams, \citep{Richards2002,Richards2005,Schneider2007,Schneider2010,Chung2014}, spectral lines ratios, \citealt{Heckman1980,Kauffmann:03c,Kewley:06}. As explained above, our experiments used color-composite $g,r,i$ thumbnails to recognize the different classes of sources. To assess whether \ULISSE's results are mainly based on the morphological features of the prototypes or on their colors, we performed an additional set of tests using only single optical bands.

In Table\,\ref{tab:fract-color} we present the fractions of AGN, SFG, composite and unclassified objects based on single ($g$, $r$ or $i$-band) band. Considering the fraction averaged over the 8 AGN-prototypes, we found that \ULISSE returned a smaller fraction of AGN and composite using only single-band images. Obviously, the lower efficiency implies an average increase in the fraction of SFG and unclassified classes among the sources retrieved with the use of a single band.
In particular, the $g$-band showed the lowest AGN retrieval efficiency (23.9\perc) among the three bands (30.3\perc and 29.4\perc for $r$ and $i$ bands, respectively).  
For non-AGN prototypes we observed similar trends (see Table\,\ref{tab:fract-color}).

In Table\,\ref{tab:morphcolor} we present fractions of AGN, SFG, composite and unclassified objects as a function of the distance for prototypes\,\#2,\,3,\,6 obtained based on a single band or on their combination.

\begin{table*}
\caption{Fractions of SFG, Composite, AGN and Unclassified objects for each of the 13 prototypes over their 300 nearest neighbors. The random guess baselines of our method for each class of objects are presented in parentheses (see details also in Table\,\ref{tab:proportions}).}
\begin{center}
\begin{tabular}{@{\extracolsep{4pt}}cccccccc@{}}
\hline\hline
\multicolumn{2}{c}{\multirow{3}{*}{Prototype}} & \multicolumn{4}{c}{BPT} & \multicolumn{2}{c}{X-ray}\\
\cline{3-6}\cline{7-8}
& & AGN & Composite & SFG & Unclassified  & AGN & non-AGN\\
& & (12\perc) & (5.8\perc) & (44.1\perc) & (38.1\perc) & (4.0\perc) & (96.0\%)\\
\hline\hline
\multirow{9}{*}{\rotatebox[origin=c]{90}{\centering AGN}} & 1 & 41.2\perc & 14.3\perc & 16.9\perc & 27.6\perc & 16.0\perc & 84.0\perc\\
& 2 & 28.3\perc & 18.6\perc & 48.8\perc & 4.3\perc & 12.3\perc & 87.7\perc\\
& 3 & 36.5\perc & 19.3\perc & 29.6\perc & 14.6\perc & 7.7\perc & 92.3\perc\\
& 4 & 53.0\perc & 12.4\perc & 23.3\perc & 11.3\perc & 17.3\perc & 82.7\perc\\
& 5 & 42.9\perc & 9.0\perc & 12.9\perc & 35.2\perc & 16.3\perc & 83.7\perc \\
& 6 & 21.3\perc & 17.6\perc & 57.5\perc & 3.6\perc & 9.3\perc & 90.7\perc\\
& 7 & 24.3\perc & 14.9\perc & 26.2\perc & 34.6\perc & 9.7\perc & 90.3\perc \\
& 8 & 24.9\perc & 16.6\perc & 48.5\perc & 10.0\perc & 6.0\perc & 94.0\perc\\
\cline{3-6}\cline{7-8}\\[-1.5ex]
& {Average} & 34.1\perc & 15.3\perc & 32.9\perc & 17.7\perc & 11.8\perc & 88.2\perc \\
\hline\hline\\[-1.5ex]
\multirow{6}{*}{\rotatebox[origin=c]{90}{\centering non-AGN}} & 9 & 21.9\perc & 4.3\perc & 8.7\perc & 65.1\perc & 8.3\perc & 91.7\perc\\
& 10 & 21.3\perc & 7.3\perc & 11.7\perc & 59.7\perc & 7.7\perc & 92.3\perc \\
& 11 & 14.3\perc & 14.0\perc & 69.4\perc & 2.3\perc & 10.7\perc & 89.3\perc \\
& 12 & 36.7\perc & 16.3\perc & 37.3\perc & 9.7\perc & 11.0\perc & 89.0\perc \\
& 13 & 1.7\perc & 1.3\perc & 95.3\perc & 1.7\perc & 2.7\perc & 97.3\perc \\
\cline{3-6}\cline{7-8}\\[-1.5ex]
& {Average} & 19.2\perc & 	8.6\perc & 	44.5\perc & 	27.7\perc & 	8.1\perc & 	91.9\perc\\
\hline\hline

\end{tabular}
\end{center}

\label{tab:fractions}
\end{table*}

\begin{table*}
\caption{Fractions of SFG, Composite, AGN and Unclassified objects for each of the 13 prototypes over their 300 closest neighbors obtained based on a single band SDSS image ($g$, $r$ or $i$ band). Arrows show the decrease ($\downarrow$), increase ($\uparrow$) or no change (=) of the fraction obtained by our method based on one single band relative to the corresponding fraction obtained in three bands together (see Table\,\ref{tab:fractions}).}
\begin{center}
\setlength{\tabcolsep}{2pt}\resizebox{\textwidth}{!}{  

\begin{tabular}{@{\extracolsep{3pt}}cccccccccccccc@{}}
\hline\hline
\multicolumn{2}{c}{\multirow{2}{*}{Prototype}} & \multicolumn{3}{c}{AGN} & \multicolumn{3}{c}{Composite} & \multicolumn{3}{c}{SFG} & \multicolumn{3}{c}{Unclassified}  \\
\cline{3-5}\cline{6-8}\cline{9-11}\cline{12-14}
& & $g$ & $r$ & $i$ & $g$ & $r$ & $i$ & $g$ & $r$ & $i$ & $g$ & $r$ & $i$    \\
\hline\hline\\[-1.5ex]
\multirow{9}{*}{\rotatebox[origin=c]{90}{\centering AGN}} & 1 & $\downarrow$\,10.3\perc  & $\downarrow$\,35.2\perc & $\downarrow$\,35.5\perc &  $\downarrow$\,\hphantom{0}9.6\perc  & $\downarrow$\,\hphantom{0}6.6\perc & $\downarrow$\,\hphantom{0}6.0\perc & $\uparrow$\,63.1\perc & $\downarrow$\,\hphantom{0}8.6\perc & $\downarrow$\,\hphantom{0}8.6\perc &  $\downarrow$\,16.9\perc & $\uparrow$\,49.5\perc & $\uparrow$\,49.8\perc\\
& 2 & $\downarrow$\,12.6\perc & $\downarrow$\,25.6\perc & $\downarrow$\,24.6\perc & $\downarrow$\,10.6\perc & $\downarrow$\,15.0\perc & $\downarrow$\,13.6\perc & $\uparrow$\,73.1\perc & $\uparrow$\,52.2\perc & $\uparrow$\,54.2\perc & $\downarrow$\,\hphantom{0}3.7\perc & $\uparrow$\,\hphantom{0}7.3\perc & $\uparrow$\,\hphantom{0}7.6\perc\\
& 3 & $\downarrow$\,28.2\perc & $\downarrow$\,16.3\perc & $\downarrow$\,26.6\perc & $\downarrow$\,\hphantom{0}8.6\perc & $\downarrow$\,\hphantom{0}8.3\perc & $\downarrow$\,14.0\perc & $\downarrow$\,21.3\perc & $\uparrow$\,68.1\perc & $\uparrow$\,31.2\perc & $\uparrow$\,41.9\perc & $\downarrow$\,\hphantom{0}7.3\perc & $\uparrow$\,28.2\perc \\
& 4 & $\downarrow$\,31.0\perc & $\downarrow$\,49.0\perc & $\downarrow$\,45.7\perc & $\downarrow$\,10.0\perc & $\uparrow$\,12.7\perc & $\downarrow$\,\hphantom{0}9.0\perc & $\uparrow$\,36.3\perc & $\uparrow$\,24.3\perc & $\downarrow$\,16.0\perc & $\uparrow$\,22.7\perc & $\uparrow$\,14.0\perc & $\uparrow$\,29.3\perc\\
& 5 & $\downarrow$\,36.2\perc & $\downarrow$\,40.2\perc & $\downarrow$\,34.9\perc & $\uparrow$\,\hphantom{0}9.3\perc & $\uparrow$\,11.3\perc & $\downarrow$\,\hphantom{0}8.6\perc & $\uparrow$\,15.6\perc & $\uparrow$\,16.6\perc & $\downarrow$\,\hphantom{0}9.6\perc & $\uparrow$\,38.9\perc & $\downarrow$\,31.9\perc & $\uparrow$\,46.8\perc\\
& 6 & $\uparrow$\,32.9\perc & $\uparrow$\,23.3\perc & $\downarrow$\,20.6\perc & $\downarrow$\,11.6\perc & $\downarrow$\,14.0\perc & $\downarrow$\,12.0\perc & $\downarrow$\,23.9\perc & $\uparrow$\,59.5\perc & $\uparrow$\,62.2\perc & $\uparrow$\,31.6\perc & $\downarrow$\,\hphantom{0}3.3\perc & $\uparrow$\,\hphantom{0}5.3\perc\\
& 7 & $\downarrow$\,21.6\perc & $\uparrow$\,27.9\perc & $\downarrow$\,23.3\perc & $\downarrow$\,\hphantom{0}9.6\perc & $\uparrow$\,15.9\perc & $\downarrow$\,\hphantom{0}8.6\perc & $\downarrow$\,20.3\perc & $\downarrow$\,21.3\perc & $\downarrow$\,20.6\perc & $\uparrow$\,48.5\perc & $\uparrow$\,34.9\perc & $\uparrow$\,47.5\perc\\
& 8 & $\downarrow$\,18.6\perc & $\downarrow$\,24.6\perc & $\downarrow$\,24.3\perc & $\downarrow$\,17.3\perc & $\uparrow$\,17.9\perc & $\downarrow$\,12.6\perc & $\uparrow$\,61.5\perc & $\downarrow$\,39.2\perc & $\downarrow$\,25.2\perc & $\downarrow$\,\hphantom{0}2.7\perc & $\uparrow$\,18.3\perc & $\uparrow$\,37.9\perc\\
\cline{3-5}\cline{6-8}\cline{9-11}\cline{12-14}\\[-1.5ex]
& {Average} & $\downarrow$\,23.9\perc & $\downarrow$\,30.3\perc & $\downarrow$\,29.4\perc & $\downarrow$\,10.9\perc & $\downarrow$\,12.7\perc & $\downarrow$\,10.6\perc & $\uparrow$\,39.4\perc & $\uparrow$\,36.2\perc & $\downarrow$\,28.5\perc & $\uparrow$\,25.9\perc & $\uparrow$\,20.8\perc & $\uparrow$\,31.6\perc \\
\hline\hline\\[-1.5ex]
\multirow{6}{*}{\rotatebox[origin=c]{90}{\centering non-AGN}} & 9 & $\downarrow$\,17.9\perc & $\downarrow$\,20.9\perc  & $\downarrow$\,21.6\perc  & $\downarrow$\,\hphantom{0}3.7\perc & $\downarrow$\,\hphantom{0}3.3\perc  & $\uparrow$\,\hphantom{0}6.3\perc & $\uparrow$\,\hphantom{0}9.0\perc &  $\uparrow$\,11.3\perc & $\uparrow$\,12.6\perc & $\uparrow$\,69.4\perc &  $\downarrow$\,64.5\perc & $\downarrow$\,59.5\perc\\
& 10 & $\uparrow$\,22.3\perc & $\uparrow$\,23.3\perc  & $\uparrow$\,28.0\perc & $\uparrow$\,\hphantom{0}8.7\perc & $\downarrow$\,\hphantom{0}6.7\perc  & $\uparrow$\,\hphantom{0}8.0\perc & $\uparrow$\,13.0\perc & $\downarrow$\,10.3\perc & $\downarrow$\,11.0\perc & $\downarrow$\,56.0\perc & =\,59.7\perc & $\downarrow$\,53.0\perc\\
& 11 & $\downarrow$\,\hphantom{0}3.0\perc &  $\downarrow$\,10.6\perc & $\uparrow$\,17.3\perc & $\downarrow$\,\hphantom{0}4.7\perc & $\downarrow$\,10.3\perc  & $\downarrow$\,12.3\perc & $\uparrow$\,91.7\perc &  $\uparrow$\,77.4\perc & $\downarrow$\,67.8\perc & $\downarrow$\,0.7\perc & $\downarrow$\,\hphantom{0}1.7\perc & $\uparrow$\,\hphantom{0}2.7\perc \\
& 12 & $\uparrow$\,39.7\perc & $\downarrow$\,29.3\perc  & $\downarrow$\,31.7\perc & $\downarrow$\,12.0\perc & $\uparrow$\,16.7\perc  & $\downarrow$\,14.0\perc & $\downarrow$\,32.7\perc & $\uparrow$\,39.0\perc  & $\uparrow$\,43.7\perc & $\uparrow$\,15.7\perc & $\uparrow$\,15.0\perc & $\uparrow$\,10.7\perc \\
& 13 & $\uparrow$\,21.9\perc & $\uparrow$\,5.6\perc & $\uparrow$\,4.3\perc & $\uparrow$\,11.0\perc & $\uparrow$\,2.0\perc & $\uparrow$\,1.7\perc & $\downarrow$\,62.1\perc & $\downarrow$\,85.7\perc & $\downarrow$\,88.7\perc & $\uparrow$\,5.0\perc & $\uparrow$\,6.6\perc & $\uparrow$\,5.3\perc  \\
\cline{3-5}\cline{6-8}\cline{9-11}\cline{12-14} \\[-1.5ex]
& {Average} & $\uparrow$\,21.0\perc & 	$\downarrow$\,17.9\perc & 	$\uparrow$\,20.6\perc & 	$\downarrow$\,\hphantom{0}8.0\perc & 	$\downarrow$\,\hphantom{0}7.8\perc & 	$\downarrow$\,\hphantom{0}8.5\perc & 	$\downarrow$\,41.7\perc & 	$\uparrow$\,44.7\perc & 	$\uparrow$\,44.8\perc & 	$\uparrow$\,29.4\perc & 	$\uparrow$\,29.5\perc & 	$\uparrow$\,26.2\perc\\
\hline\hline

\end{tabular}
}
\end{center}

\label{tab:fract-color}
\end{table*}

\subsection{Recursive application}\label{sec:recursive}

While \ULISSE can return hundreds of candidates for a single prototype, the success rate typically drops with the number of returned objects, that is with the distance from the prototype in the feature space. Due to this effect, in Sect.~\ref{sec:results}, we find that retrieval of AGN is most effective using the closest neighbors, typically the nearest 20-30 objects returned by our method. One possible way to keep the purity high while increasing the number of sources is to apply \ULISSE in a recursive way. We do this by using one of the resulting candidates as a prototype for the next step of the selection, and in this section verify if this recursive technique returns a sample with equal or higher success rate, compared to our initial approach described in Sect.~\ref{sec:results}.

Concretely, we select the closest object in feature space to our reference prototype and apply it as the prototype for the next step. As this closest object likely shares the most morphological and photometric properties with our reference prototype, it should also be the most likely to be an AGN. Then, at each next step we repeat the same procedure choosing the closest object to the reference object used in the previous step, but excluding objects considered in earlier steps. We set $n = 25$. Results can be found in Table~\ref{tab:recursive}, where find the recursive version has a slightly higher purity, and a lower standard deviation compared to the normal version, suggesting it might be more robust to the choice of prototype. 

Using this set-up, for the relatively unsuccessful prototype\,\#2, we obtained a total of 89 objects excluding duplicates in five iterations, of which 38 are AGN (42.7\perc, see Table~\ref{tab:recursiveLowest}). In contrast, the resulting AGN fraction setting $n = 89$ directly for prototype\,\#2 would only result in 25.8\perc AGN.

We found this mostly beneficial when the chosen prototype did not score exceptionally well. For example, in the case of prototype\,\#4, we obtain 87 objects of which 47.1\perc are AGN with 5 iterations of our recursive method, yet this is 63.2\perc if we simply set $n=87$.

\begin{table}
\caption{Results obtained after five iterations of the recursive application of \ULISSE with $n = 25$. $N_{total}$ is the total number of found objects without duplicates.}
\begin{center}
\begin{tabular}{@{\extracolsep{4pt}}ccccc@{}}
\hline\hline
\multicolumn{2}{c}{\multirow{2}{*}{Prototype}} & \multirow{2}{*}{$N_{total}$} & \multicolumn{2}{c}{Fraction of AGN} \\
\cline{4-5}
& & & Recursive & Normal\\
\hline\hline
\multirow{10}{*}{\rotatebox[origin=c]{90}{\centering AGN}}  & 1 & 	84 & 	38.1\perc & 	45.2\perc \\
 & 2 & 	89 & 	42.7\perc & 	25.8\perc \\
 & 3 & 	93 & 	43.0\perc & 	43.0\perc \\
 & 4 & 	87 & 	47.1\perc & 	63.2\perc \\
 & 5 & 	71 & 	47.9\perc & 	40.8\perc \\
 & 6 & 	88 & 	44.3\perc & 	30.7\perc \\
 & 7 & 	84 & 	34.5\perc & 	25.0\perc \\
 & 8 & 	91 & 	23.1\perc & 	25.3\perc \\
\cline{2-5}\\[-1.5ex]
& \multicolumn{2}{l}{Average} &  40.1\perc & 37.4\perc\\
& \multicolumn{2}{l}{Std deviation} & 8.2 & 13.4 \\
\hline\hline\\[-1.5ex]

\end{tabular}
\end{center}

\label{tab:recursive}
\end{table}

%% file: 05conclusion.tex
\section{Discussion}\label{sec:discussion}

In Sect.\,\ref{sec:exp} we presented the AGN fraction retrieved by \ULISSE for AGN and non-AGN prototypes, and its correlation with the color and the morphological type of studied prototypes. Furthermore, we analyzed the possibility to improve our results with usage of a recursive technique. In this section, we discuss and interpret our results, \ULISSE limitations and possible future improvements.

\subsection{The correlation between prototype properties and AGN fraction retrieved by \ULISSE}

The traditional optical AGN selection criteria are based on photometric and spectroscopic properties such as magnitudes, colors, the ratios of high- and low-excitation emission lines etc. \citep{Kauffmann:03c, Hickox2018, Zhang2018}. However, AGN identification in the UV/optical band is non-trivial and faces several limitations. For instance, the so-called color-color diagrams can be used to distinguish distant bright quasars from stars in our Galaxy, because of the bluer color of quasars compared to stars  (i.e. the peak of emission from AGN accretion disk is located in UV range; \citealt{Shakura1973}). At the same time, such color-color criteria do not work well for nearby, low-luminosity AGN, because the stellar emission of the host galaxy begins to dominate on the AGN emission in the optical range. This issue may be partially resolved by a spectroscopic approach, for instance by the BPT-diagram, where the AGN identification relies on the ratios of emission lines in the optical spectrum \citep{Kauffmann:03c, Kewley:06}, but even in this case we lose a large fraction of AGN due to the fact that the star formation processes also contribute to the optical emission lines. In addition, all UV/optical selection criteria are inefficient to identify AGN obscured by circumnuclear and galactic dust \citep{Hickox2018, Zhang2018, Ji2022}. Therefore, a more efficient AGN identification usually requires use of  multi-wavelength observations including radio, IR and X-ray bands \citep{Mateos2012, Trump2013, Heinis2016, Agostino2019}.  

In Sect.~\ref{sec:exp} we mentioned that the total fraction of AGNs in our sample according to the BPT diagram is only 12\perc (i.e. the random guess baseline of our method), while \ULISSE appears to be more effective than random selection in identifying AGN, yielding on average 34\perc of confirmed AGN for the 8 prototypes studied here (see Table~\ref{tab:fractions}). At the same time, a detailed analysis of each separate prototype showed that this AGN fraction varies depending on the prototypes properties: in general, the AGN selection efficiency for prototypes which visually belong to late-type galaxies (spiral morphology) is lower than for early-type galaxies (elliptical morphology).

For instance, for prototypes\,\#2,\,6,\,8 \ULISSE found on average only $\sim 25$\perc of AGN (see Table~\ref{tab:fractions}). Such relatively low efficiency in selecting AGN can be due to several causes. Firstly, as we mentioned in Sect.\,\ref{sec:data} the BPT diagram separates AGN from star-forming galaxies based on the ratio of emission lines and therefore the source identification depends on whose process (AGN or star formation) dominates the spectrum. The majority of spiral galaxies have strong ongoing star-formation whose emission can outshine less powerful AGN, which would then be classified as SFGs according to the BPT-diagram. 

A low retrieval fraction is observed instead for\,\#7 at small distances from the prototype; this could be due to the peculiar nature of this source both for its ring-like morphology and for the presence of a bright nearby star. However, overall the average efficiency settles on $\sim 24$\perc, not far from other late-type galaxies.

At the same time, our sample contains a sizeable fraction of so-called composite objects, where AGN and star-forming emission contribute equally to the optical spectrum and can not be easily separated \citep{Kewley:06, Kauffmann:03c}. This fact allows us to combine the fraction of AGN and composite objects obtained by \ULISSE, which gives us an average AGN content of $\sim 42$\perc for prototypes\,\#2,\,6-8.

For prototypes\,\#3,\,5 with early-type morphology \ULISSE identified on average $\sim 40$\perc of confirmed AGN (54\perc in the case of AGN+Composite, see Table~\ref{tab:fractions}). The larger  AGN retrieval efficiency obtained for prototypes with early-type morphology can be explained by the  properties of passive galaxies, which typically have  partially or completely quenched star-formation \citealt{Thomas2002, Thom2012}; this allows more easily detect the AGN emission with negligible contamination from the host galaxy. 

The highest AGN fraction among 8 AGN prototypes was obtained by \ULISSE for prototype\,\#4 (53.0\perc, and 65.4\perc of AGN+Composite). This prototype shows the presence of bright red nucleus, which means that \ULISSE tends to select sources with a high nuclear luminosity and possibly, due to the known correlation between bulge and BH mass, suggests a stronger AGN activity \citep{Haring2004, Kormendy2013, McConnell2013}.

As it was mentioned above, an additional issue that may explain the differences in our AGN identification efficiency is the partial or complete obscuration of the nucleus, in the optical band, by circumnuclear and/or galactic dust. This effect is more significant for star-forming galaxies, because they usually show the presence of strong dust component in the disk, which plays important role in star forming processes helping to cool the cold gas \citep{Byrne2019,Lianou2019}. However, AGN obscuration can be also true for some elliptical galaxies, which show peculiar morphology and unusual dust lanes \citep{Goudfrooij1995,Hirashita2015}. In such cases, AGN detection is possible only at less affected wavelengths, such as IR or X-rays. 

In discussing the results obtained for different prototypes, we may wonder whether the \ULISSE performance depends mainly on the morphology of host galaxy more than on the presence of an AGN. To understand if this is the case, we performed additional test with non-AGN prototypes. As we explained in Sect.\,\ref{sec:ref-obj}, we chose several galaxies with different morphology (prototypes\,\#9-13 in Table\,\ref{tab:reference-non-agn}). Table\,\ref{tab:fractions} shows that the resulting AGN fraction obtained by our method is smaller on average (19.2\perc) compared to the results for AGN-prototypes and there is a similar correlation with the properties of the host galaxy (see Table\,\ref{tab:refs3}). At the same time, \ULISSE was very effective in retrieving objects with a BPT class similar to the studied prototypes, specifically `unclassified' and `SFG' for early-type and late-type systems, respectively. 

Overall, the lower AGN fraction for non-AGN prototypes indicates that \ULISSE is able to retrieve not only galaxies with a similar morphology and color to the studied prototype, but also to detect AGN with some level of reliability. However, we should also mention that prototype\,\#12 (see Table\,\ref{tab:refs3}) produced a relatively large AGN fraction (36.7\perc) compared to other non-AGN prototypes, and comparable to some of the AGN prototypes. In this case, the \ULISSE performance could be driven by the red color and bright central bulge of the spiral galaxy, similar to the AGN-prototype\,\#4 (see Table\,\ref{tab:refs2} and the discussion above). 

Using X-ray MOC sample (i.e. the sample of SDSS galaxies fall within {\it XMM-Newton} footprint, see \citealt{Torbaniuk2021}) we performed several additional tests aimed at avoiding the limitations in the BPT classification method described above. In Table\,\ref{tab:fractions} we present AGN fraction obtained by \ULISSE based on X-ray MOC sample. The average AGN fraction for AGN-prototypes\,\#1-8 is $\sim 12$\perc while for non-AGN prototypes\,\#9-13 it is $\sim 8$\perc; on the other hand the random guess baseline is 4\perc. 

For the individual prototypes we observe similar trends as for the BPT validation, that is prototypes with early-type morphology (\#3,\,5) produce in general higher AGN percentages (12\perc) compared to prototypes\,\#2,\,6-8 with spiral morphology (9.3\perc). While this result may seem to contradict recent studies which found that X-ray selected AGN prefer to reside in gas-rich galaxies with active star formation \citep{Lutz2010, Mullaney2012a, Mendez2013, Rosario2013, Shimizu2015, Birchall2020, Stemo2020, Torbaniuk2021}, in fact there is no disagreement here since we are selecting based on combined optical and X-ray data. The tests for non-AGN prototypes, using the X-ray MOC sample show again an average lower percentage of retrieved AGN; specifically for prototypes\,\#9,\,10 with elliptical morphology we obtain $\sim 7-8$\perc, while for prototypes\,\#11-13 with spiral morphology varies from 2.7\perc to 11\perc.

In Sect.\,\ref{sec:data} we mentioned that \ULISSE uses SDSS images in three bands ($g$, $r$ and $i$). 
To assess if the color information actually improves the \ULISSE efficiency or the results are only driven by the source morphology, we performed a further set of tests based on a single band images (see Sect. \,\ref{sec:color-res}). Using single $g$, $r$ or $i$-band we found a decrease in the retrieved AGN and composite fractions for almost all AGN prototypes compared to those obtained using the color information as well (the only exception is the peculiar ring-like galaxy, see Table\,\ref{tab:fract-color}). 

Based on this set of experiments with single bands, we can say that despite the presence of peculiar cases, \ULISSE actually exploits the available information from the different bands to improve the efficiency in detecting objects with similar physical properties, although the morphology has a dominant role. 

\subsection{Possible ways to increase AGN fraction}

There are several possible ways to increase the fraction of AGN identified by \ULISSE. We have already mentioned that the absolute number of sources $n$ retrieved by \ULISSE depends on the goal of the study. In our work, we set $n = 300$ to explore the dependence of the AGN fraction on the `distance' from the template, but as we saw in Sect.\,\ref{sec:results} the success rate is usually larger using the closest neighbors (see Table\,\ref{tab:refs},\,\ref{tab:refs2}). 
In the case of prototypes\,\#3,\,5 with elliptical morphology AGN fraction can reach 80\perc for the nearest 10 objects (i.e. 8 of 10 objects retrieved by \ULISSE are AGN). At the same time, for prototype\,\#4 \ULISSE was able to retrieved AGN with 100\perc efficiency for the nearest 10 objects. A similar performance at low distances is also visible for prototypes\,\#2,\,6,\,8 with spiral morphology; however, comparing to prototypes with elliptical morphology, we see a relatively smaller efficiency (near 40-60\perc for the nearest 10 objects). Furthermore, `spiral' prototypes seem to suggest a sharper decrease of AGN fraction with increasing distance to the reference prototype than prototypes with elliptical morphology. A different result was obtained only for the peculiar ring-like prototype\,\#7, which showed no AGN in the nearest 10 objects (the number of AGN reached 20\perc only in the next 10-15 objects, see Table\,\ref{tab:refs2}). 
Therefore, on average, reducing the number of sources $n$ would result in an increase in the AGN identification efficiency, but as a result we would obtain a smaller number of objects in the returned sample.

Moreover, the purity of the returned sample can be increased by applying our method in a recursive way. This makes use of the higher efficiency found among the closest neighbors, by iteratively choosing the next prototype as the closest new object, instead of simply enlarging the number of returned objects using one reference-prototype, therefore producing more robust results. In our work, we made a test with the application of this recursive technique (see details in Sect.\,\ref{sec:recursive}), which resulted in a higher AGN fraction together with a lower variance compared to our general method. In this way, the application of a recursive technique may be the preferential path to explore to increase the total number of objects while preserving a high AGN retrieval efficiency. 

On the other end, as we discussed in the previous section, our results are constrained by the usage of optical images and optical BPT selection criteria. Thus, the resulting AGN fraction obtained by \ULISSE could be underestimated due to, e.g, dust obscuration and/or contamination from star-forming processes in the host galaxy. To avoid or reduce this effect, we could try to extend our method using images at additional wavelengths such as IR or UV.

\subsection{Computational time}

Our method is based on two main steps. In the first step, all images are run through the pretrained neural network to extract the features. This step has to be done only once per dataset. The local run on a single NVIDIA GeForce\textsuperscript{\texttrademark} RTX~2060\footnote{The NVIDIA GeForce\textsuperscript{\texttrademark} RTX~2060 is powered by the  Turing architecture and has 1920 Cuda Cores,  6GB of RAM, and it is capable of 7.2 TFLOPS, for further details see: \url{https://www.nvidia.com/en-me/geforce/graphics-cards/rtx-2060/}} processes 100k images in 3-4 minutes. In the second step, we search for similarities to the chosen prototype within the entire dataset, comparing the features extracted in the previous stage. Using a $k$-d tree algorithm with $k=300$, we can pre-compute a structure that significantly speeds up later searches. Building the tree takes around 25-30 seconds on the same machine, results for any given prototype are then retrieved within one second.

Hence, starting from scratch, the closest 300 objects to a given prototype are retrieved, from a dataset of 100k images, in approximately 5 minutes; afterwards the process speeds up significantly due to the availability of the features and $k$-d tree, and 300 objects for any subsequent prototype are returned within a matter of seconds.

In addition, we should mention that SDSS thumbnails used in this work are of size $64 \times 64$ pixels, while our method had to rescale them to the size expected by the pretrained network ($224 \times 224$ pixel). Therefore, using images with higher resolution can be a natural way to improve \ULISSE performance without loss of computational speed.

\section{Conclusions}\label{sec:conclusions}

In this work, we present a new deep learning tool called \ULISSE \ULISSEacro for the exploration of sky surveys. The core of our method is to extract a set of representative features for each object in the sample under investigation, thus creating a common `feature space' where to search for objects with properties similar to a chosen prototype.

 Our method relies on only a single image of the requested object-prototype, making use of the first portion of a pretrained convolutional neural network, which transforms images into a set of representative features without requiring any specific astrophysical information. \ULISSE sorts all objects in the studied dataset according to the distance in this feature space (i.e. from the most similar to the least similar to the reference prototype). To verify \ULISSE's efficiency, we applied it to an extremely challenging task: the selection of AGN candidates using SDSS images of the {\it galSpec} catalogue. Based on the results obtained running \ULISSE on 8~AGN and 5~non-AGN prototypes with different host galaxy morphology and spectral properties, we arrived at the following conclusions:
\begin{itemize}
\setlength\itemsep{1ex}
    \item Our method is effective in identifying  AGN candidates, being able to retrieve galaxy samples with an AGN content that varies from 21\perc to 53\perc for different prototypes, significantly larger than the average AGN content of 12\perc in our full sample (according to the BPT classification). Including `composite' sources (which also host an AGN by definition) the retrieved AGN fraction raises to 65\perc;

    \item Our tests show that the \ULISSE performance is based on a combination of on host galaxy morphology, color and the presence of a central nuclear source. In fact, the retrieved AGN fraction for AGN prototypes is significantly higher on average than for non-AGN prototypes; 
    
    \item Despite being capable of obtaining reliable results even using one single band, \ULISSE is capable of combining the information coming from different bands, thereby increasing its efficiency in identifying objects sharing similar physical properties;
    
    \item We find \ULISSE is more effective in retrieving AGN in early-type host galaxies compared to prototypes with spiral/late-type properties;
    
    \item  The AGN retrieval efficiency is generally larger for closer neighbors to the prototype (i.e. the first 20-30). Thus, depending on the purpose of the study, \ULISSE can be used to retrieve either a higher percentage of AGN in a smaller sample, or a larger sample with lower AGN content. This dichotomy can be reduced using the recursive approach.
\end{itemize}

Based on the results described in this work and the high computational performance of the method, \ULISSE can be a promising tool for selecting various types of objects in current and future wide-field surveys (e.g. Euclid, LSST etc.) that target millions of sources every single night. For future work, the application of explainable artificial intelligence algorithms \cite{Goebel2018} might provide new insights into which of the features are most relevant for AGN detection, as well as to gain new insight on which observables better trace the physical properties of the sources under investigation. 

\subsection{LSST AGN data challenge}
We participated in the LSST AGN Data Challenge\footnote{\url{https://community.lsst.org/t/lsst-agn-science-collaboration-2021-data-challenge/5627}} (\citealt{Yu2021Challenge}) which is part of the LSST Enabling Science Program Awards\footnote{\url{https://www.lsstcorporation.org/enabling-science}}.
The dataset included sources coming from two main survey regions: SDSS Stripe 82 (S82) and XMM-LSS\footnote{\url{https://github.com/RichardsGroup/AGN_DataChallenge}}. 
The challenge did not provide a single specific task (although all are specifically tuned to AGNs selection and characterization) and the participants were free to address any possible AGN-related problem.
We decided to apply \ULISSE for the detection of AGN also to this specific dataset and were awarded the second place\footnote{\url{https://www.lsstcorporation.org/enabling-science/AGN-Data-Challenge}}. 
Although most experiments in the Data Challenge focused on Machine Learning applications, since each team was free to address any AGN-related problem, it is difficult to perform a fair comparison between the different results. For instance, Savi\'c et al. (in preparation) used support vector machines, random forests, and extreme gradient boosting, reaching classifying accuracies $>98\%$, which are far higher than ours. On the other hand, when using deep artificial neural networks that utilize pixel-level information, Savi\'c and collaborators did not observe any improvement. However, there are two main differences in these approaches: first, all these methods used tabulated (astrometric, photometric, color, morphological and variability) features, thus requiring a preliminary feature extraction  (and thus optimization) phase, while ULISSE works directly on images without any pre-processing steps. Second, the data challenge dataset was different from the one discussed here, and heavily skewed toward quasar-like sources and bright AGNs while in this work we extended the method to a more general dataset of low-luminosity AGNs with well resolved host-galaxy features. Thus, these approaches should be considered complementary, and their usage tailored to the specific scientific goal in mind.

%% file: 06supplementary.tex
\section{Fractions of different object classes as function of distance for AGN and non-AGN prototypes obtained by \ULISSE}
\begin{table*}[!hb]
\caption{The obtained results for AGN prototypes\,\#1-4 presented in Table~\ref{tab:reference}.}\label{tab:refs}
\centering
\resizebox{\textwidth}{!}{  
\setlength\tabcolsep{4 pt}
\begin{tabular}{lccc} 
\hline\hline             
\,\# & Thumbnail & Random & X-ray MOC \\
\hline \hline\\
 1 & \parbox{3.5cm}{\includegraphics[width=3.5cm]{images/queries/q1.jpeg}}  &  
     \parbox{8cm}{\centering\includegraphics[width=1\linewidth]{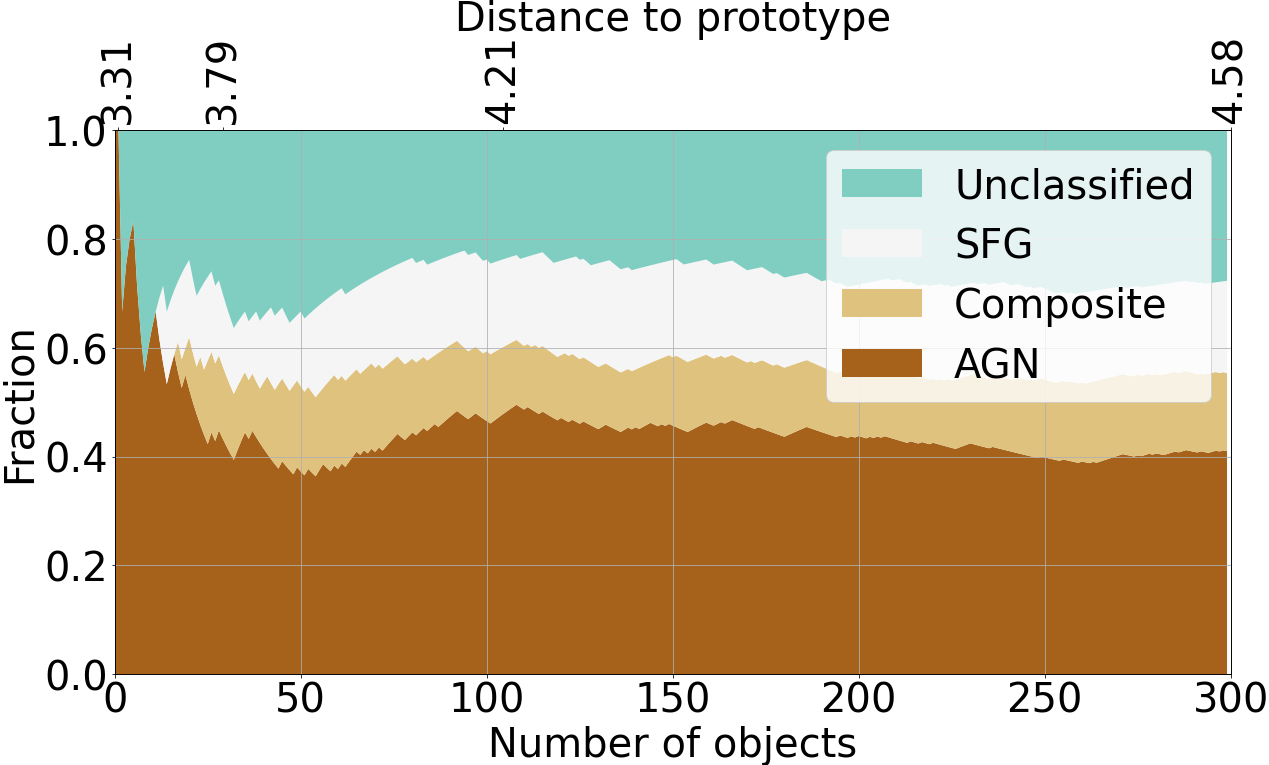}} & \parbox{8cm}{
      \centering\includegraphics[width=1\linewidth]{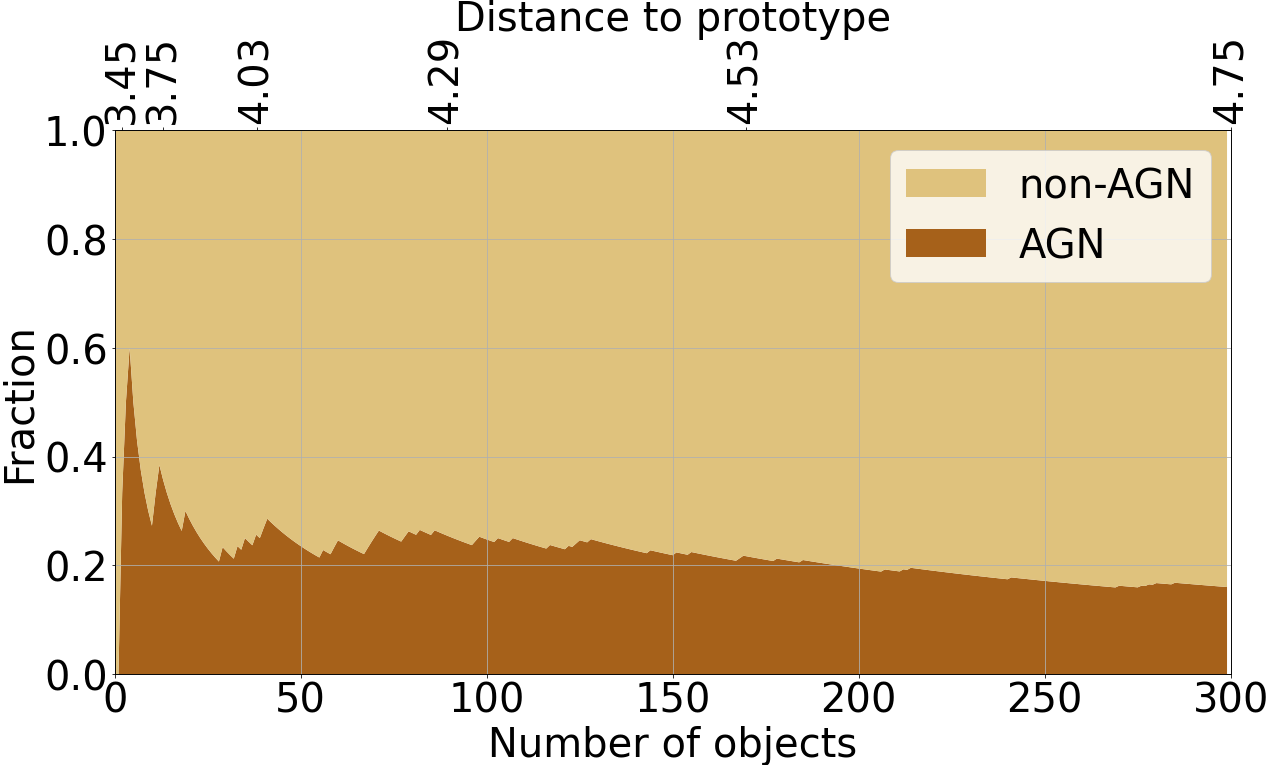}} \\
 2 & \parbox{3.5cm}{\includegraphics[width=3.5cm]{images/queries/olenaq2.jpeg}}  &  \parbox{8cm}{\centering\includegraphics[width=1\linewidth]{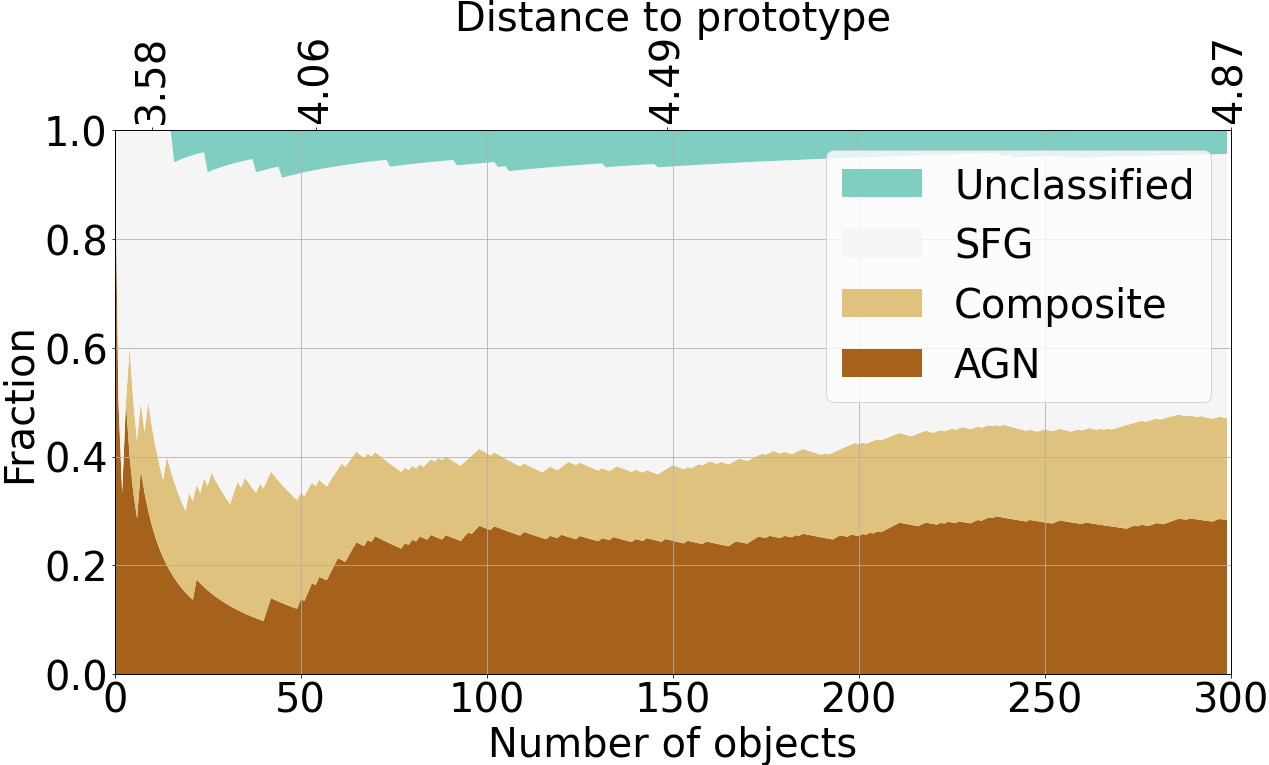}} &
  \parbox{8cm}{\centering\includegraphics[width=1\linewidth]{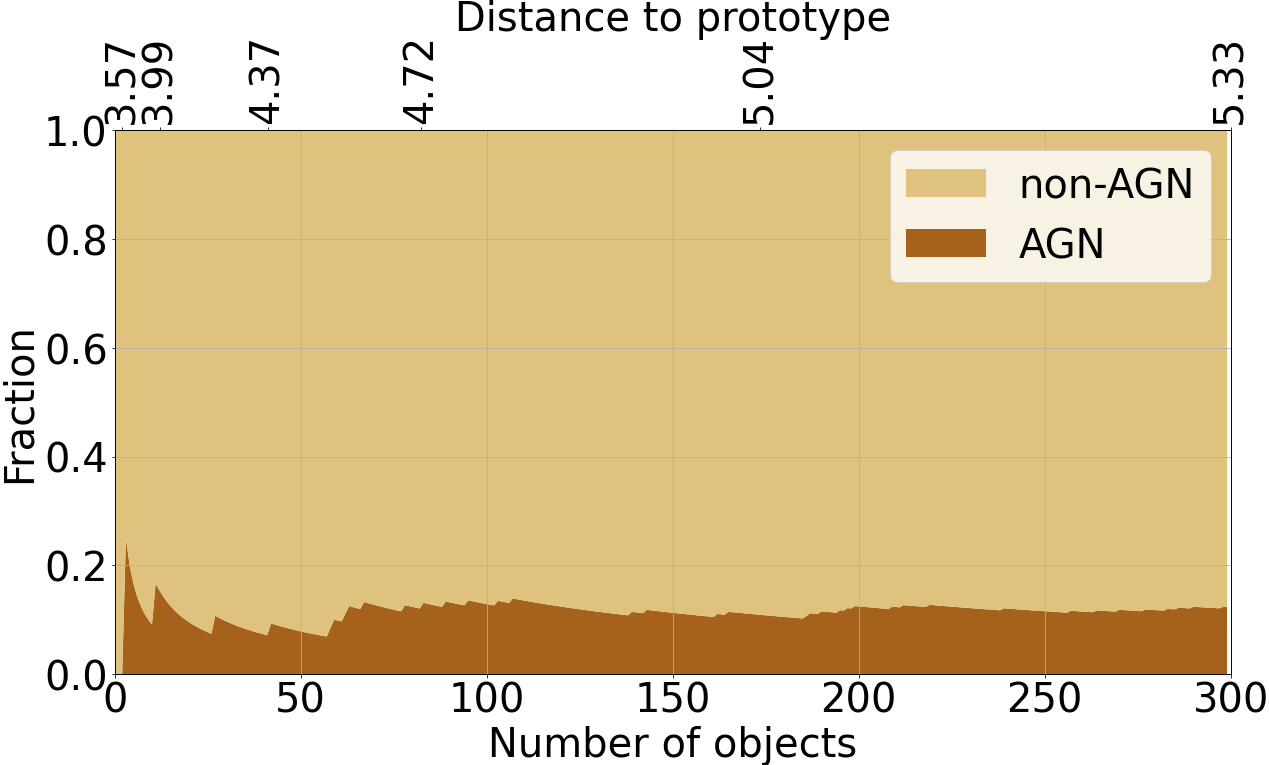}}
  \\
  3 & \parbox{3.5cm}{\includegraphics[width=3.5cm]{images/queries/olenaq3.jpeg}}  &  \parbox{8cm}{\centering\includegraphics[width=1\linewidth]{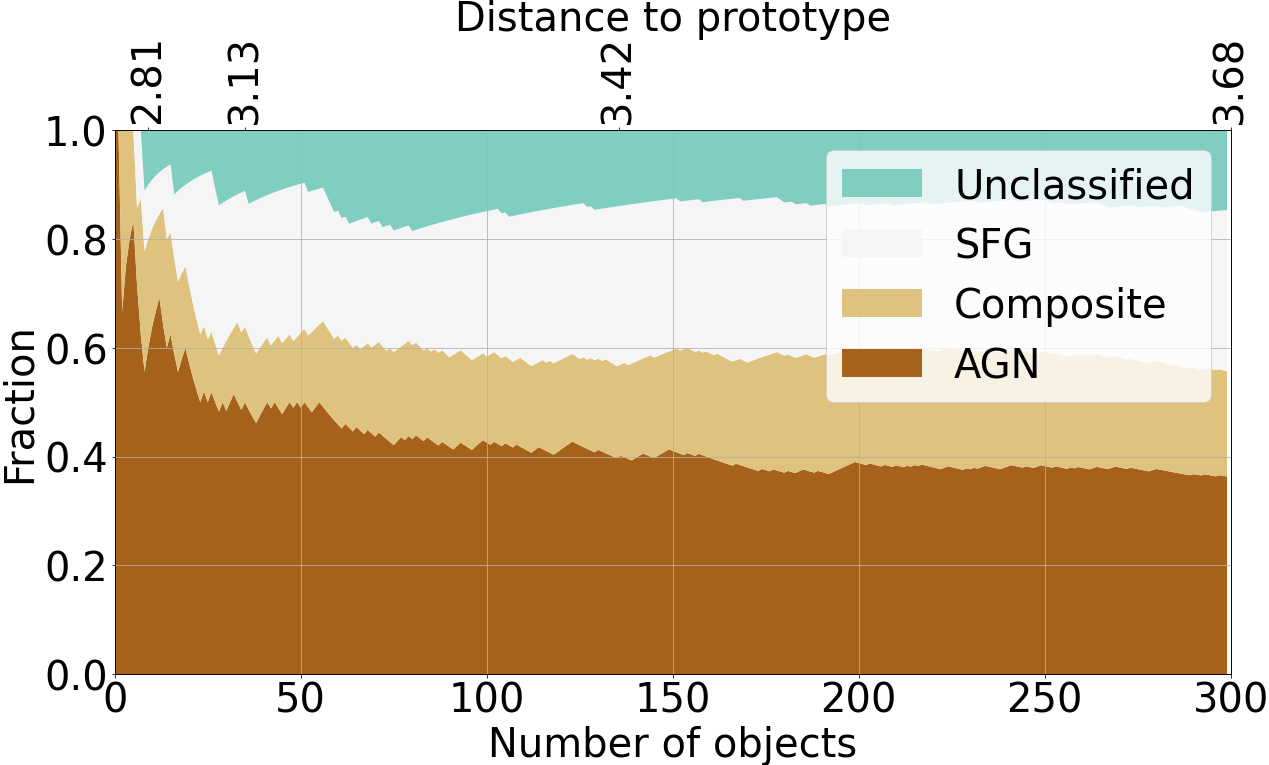}} &
  \parbox{8cm}{\centering\includegraphics[width=1\linewidth]{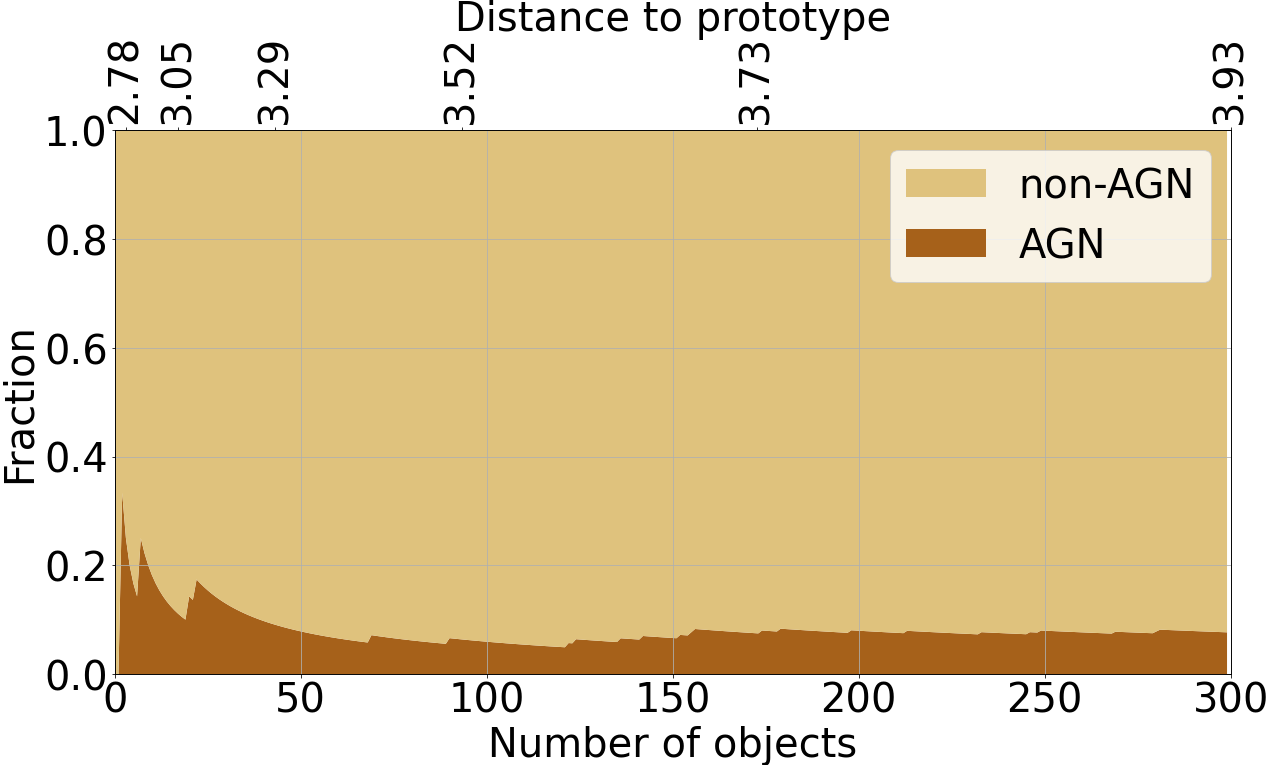}}
  \\
  4 & \parbox{3.5cm}{\includegraphics[width=3.5cm]{images/queries/olenaq4.jpeg}}  &  \parbox{8cm}{\centering\includegraphics[width=1\linewidth]{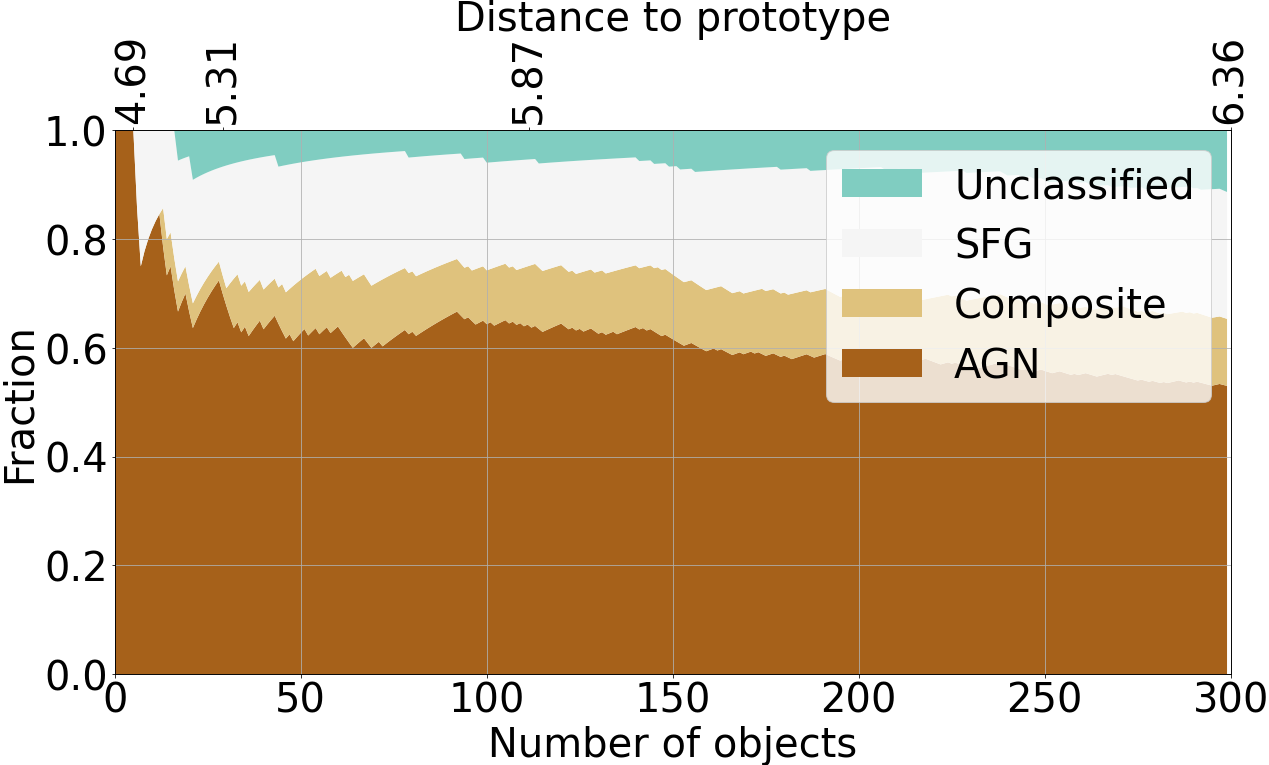}} &
  \parbox{8cm}{\centering\includegraphics[width=1\linewidth]{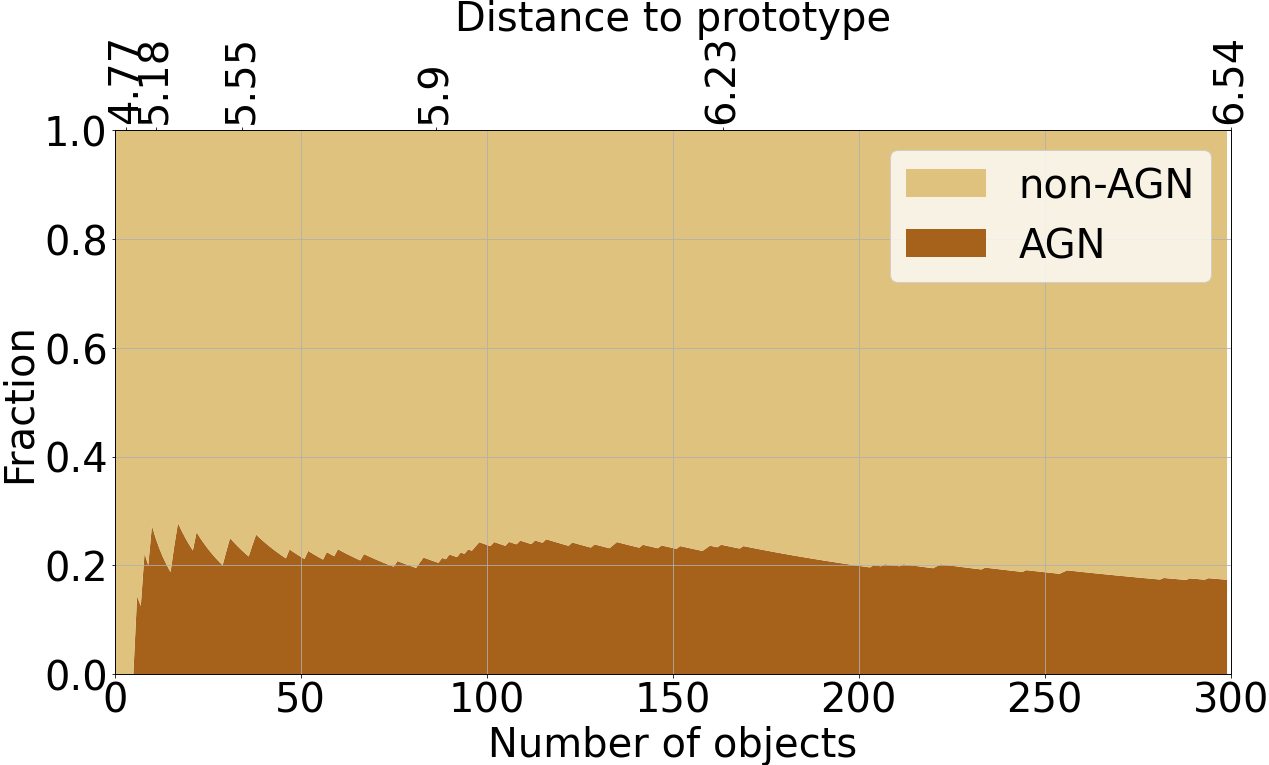}}
  \\\\
  
\hline \hline
\end{tabular}
}
\end{table*}

\begin{table*}  
\caption{The obtained results for AGN prototypes\,\#5-8 presented in Table~\ref{tab:reference}. }\label{tab:refs2}
\centering
\resizebox{\textwidth}{!}{  
\setlength\tabcolsep{4 pt}
\begin{tabular}{cccc} 
\hline\hline             
\,\# & Thumbnail & Random & X-ray MOC \\
\hline \hline \\
5 & \parbox{3.5cm}{\includegraphics[width=3.5cm]{images/queries/olenaq5.jpeg}}  &  
     \parbox{8cm}{\centering\includegraphics[width=1\linewidth]{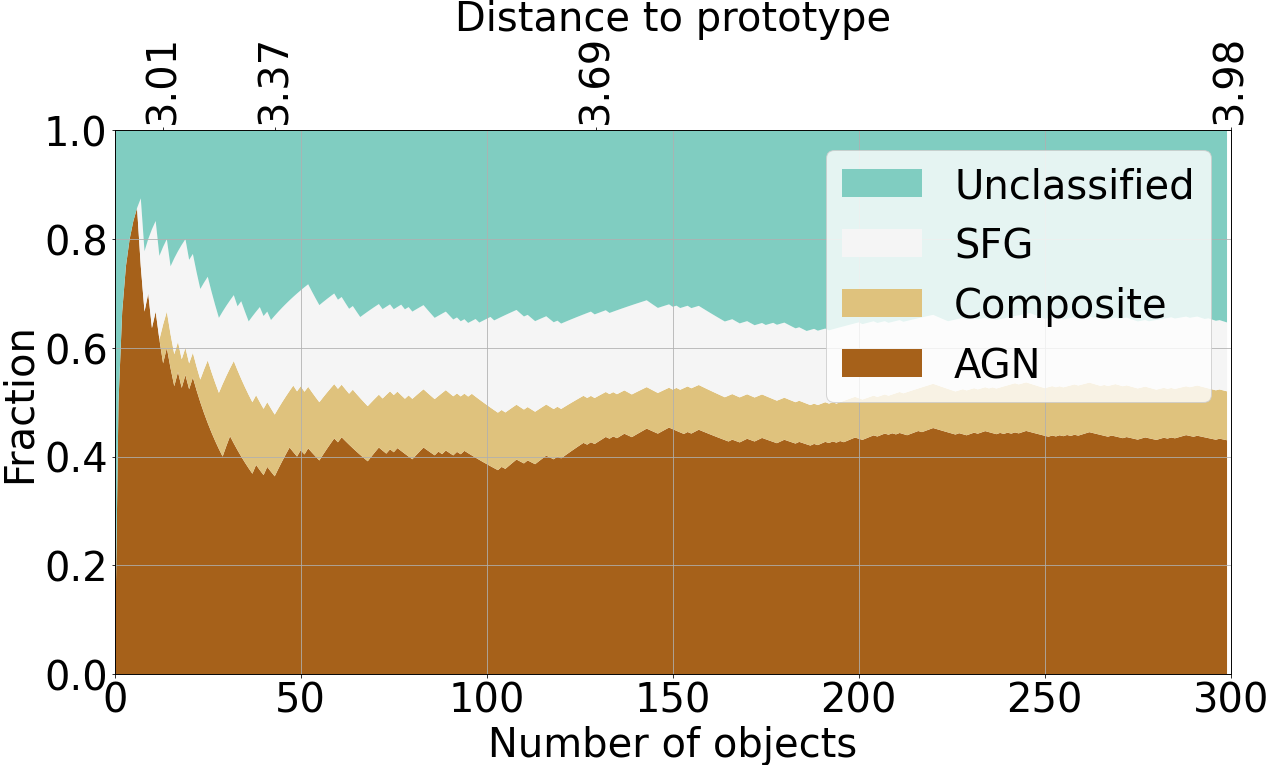}} & \parbox{8cm}{
      \centering\includegraphics[width=1\linewidth]{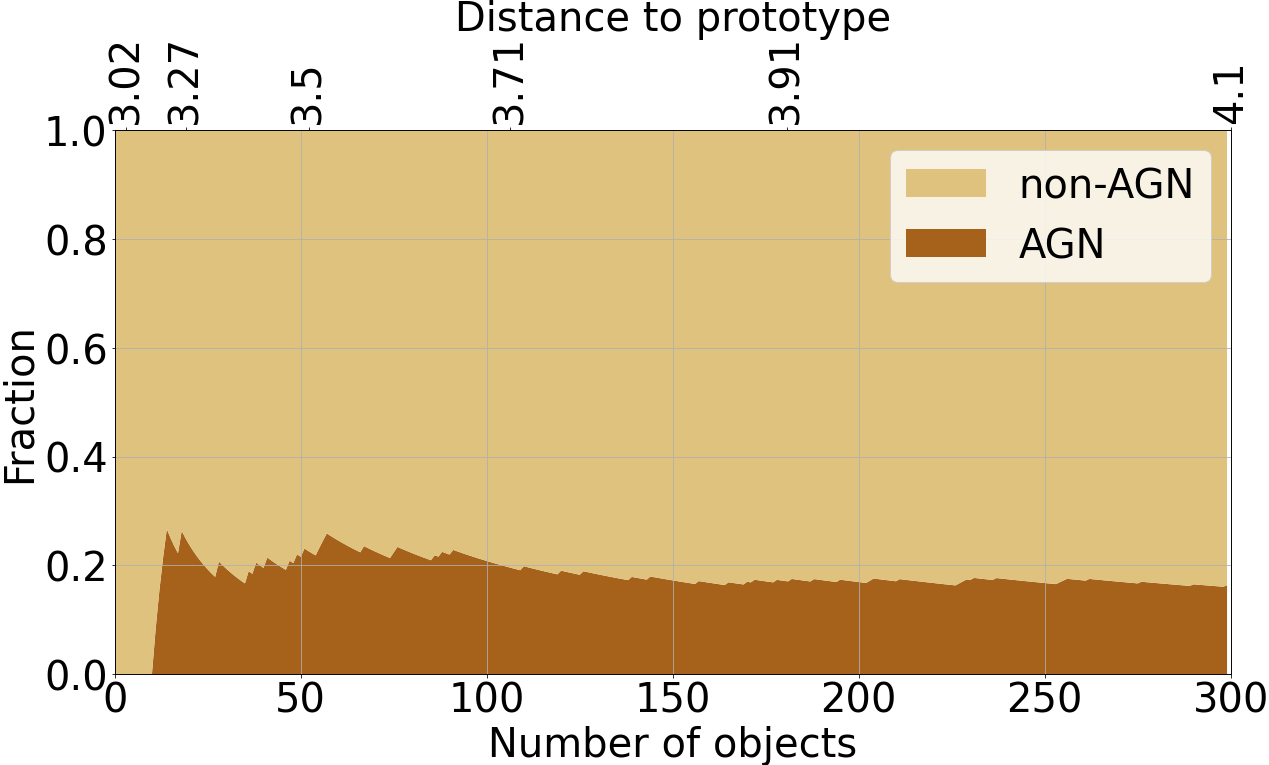}} \\
 6 & \parbox{3.5cm}{\includegraphics[width=3.5cm]{images/queries/olenaq7.jpeg}}  &  \parbox{8cm}{\centering\includegraphics[width=1\linewidth]{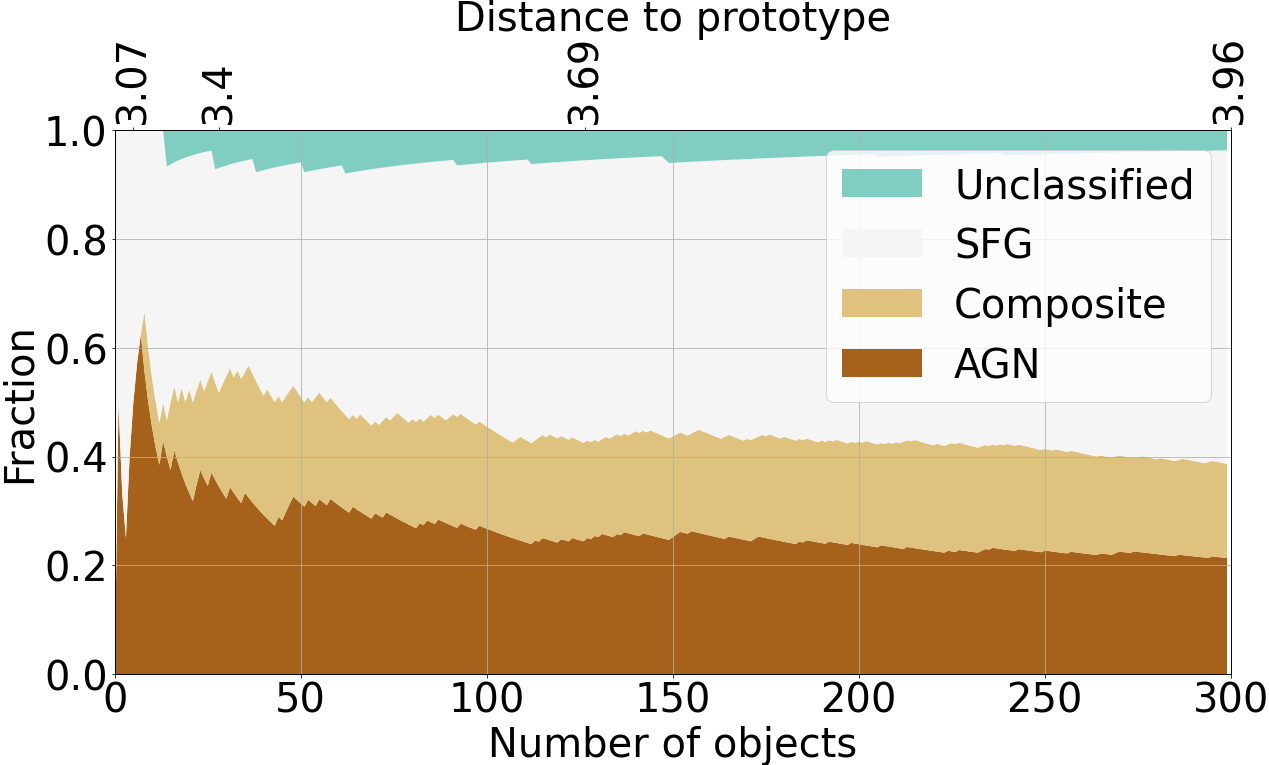}} &
  \parbox{8cm}{\centering\includegraphics[width=1\linewidth]{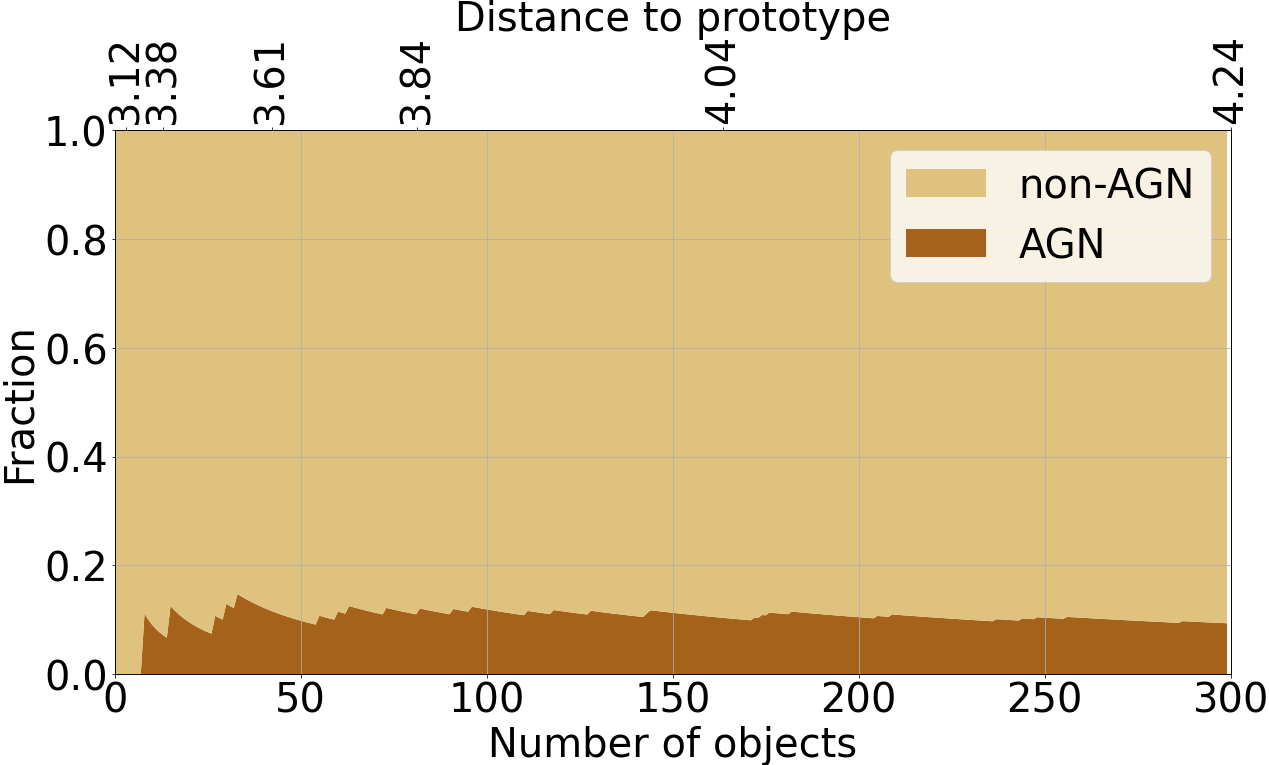}}
  \\
 7 & \parbox{3.5cm}{\includegraphics[width=3.5cm]{images/queries/olenaq8.jpeg}}  &  \parbox{8cm}{\centering\includegraphics[width=1\linewidth]{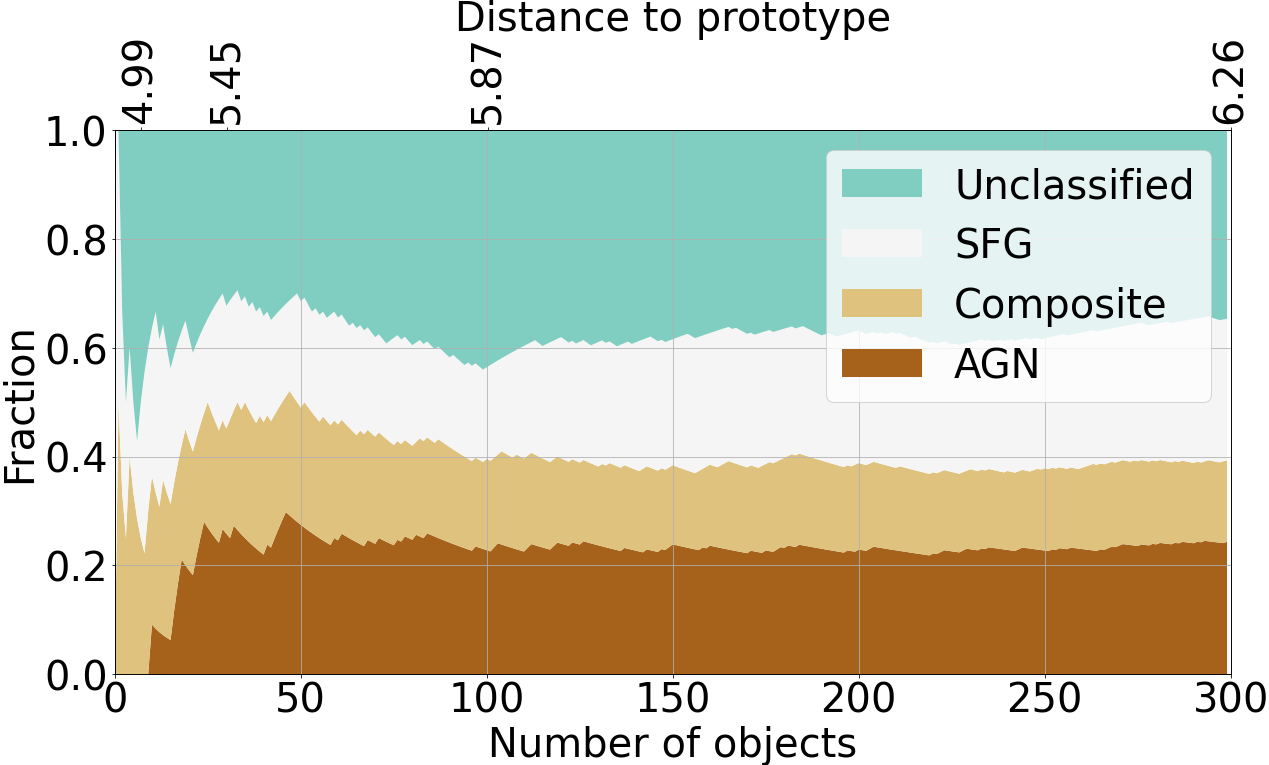}} &
  \parbox{8cm}{\centering\includegraphics[width=1\linewidth]{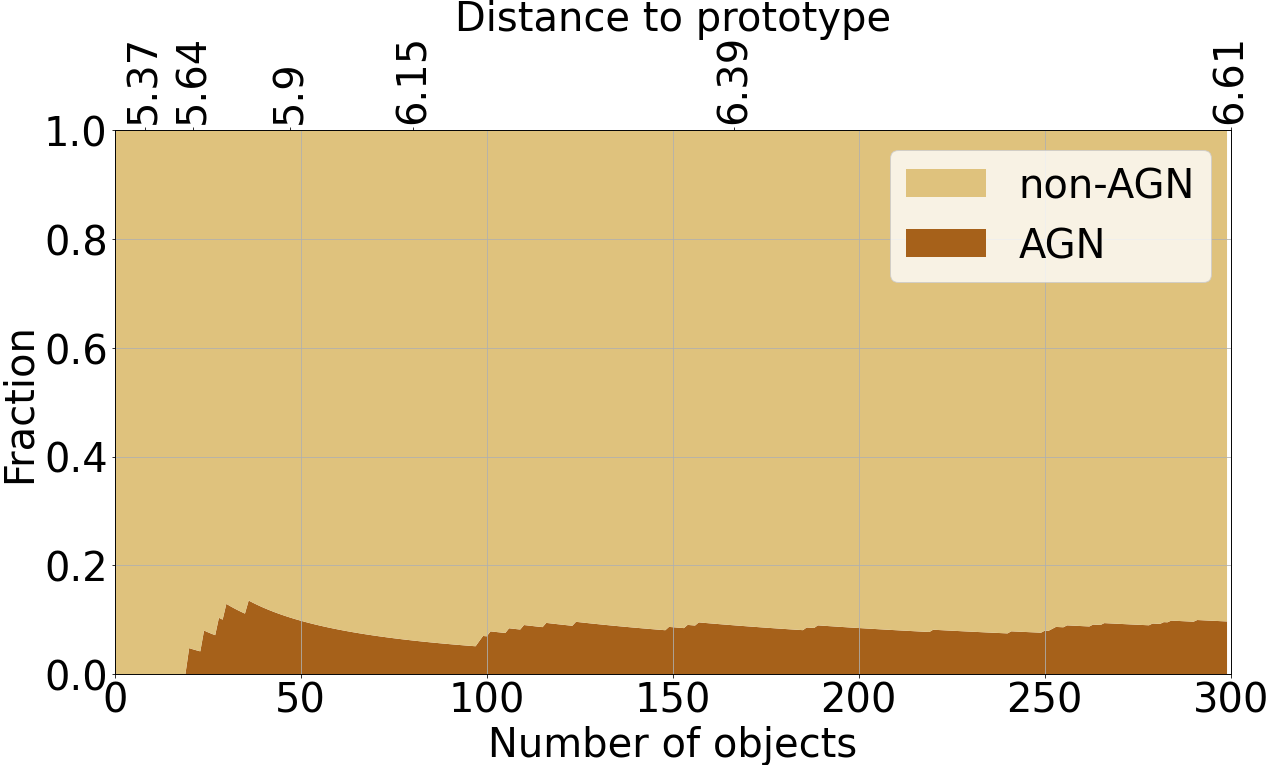}}
  \\
  8 & \parbox{3.5cm}{\includegraphics[width=3.5cm]{images/queries/olenaq9.jpeg}}  &  \parbox{8cm}{\centering\includegraphics[width=1\linewidth]{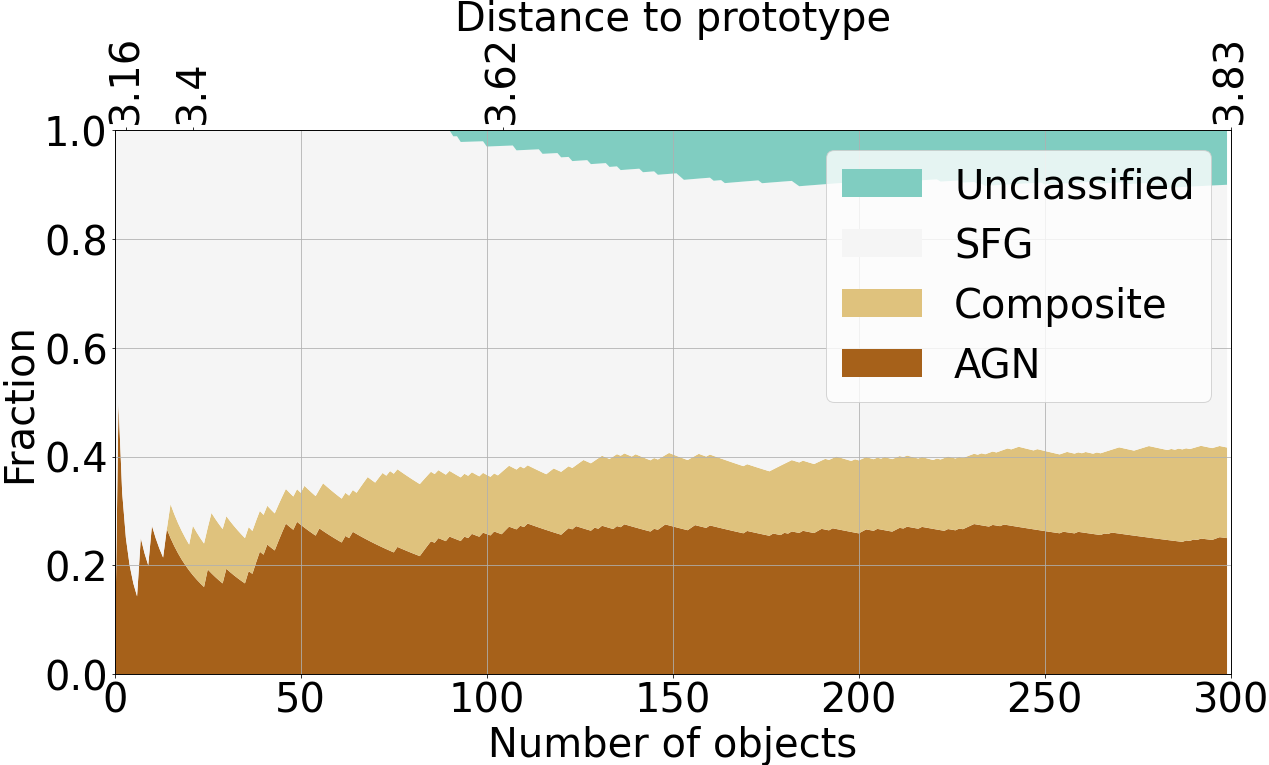}} &
  \parbox{8cm}{\centering\includegraphics[width=1\linewidth]{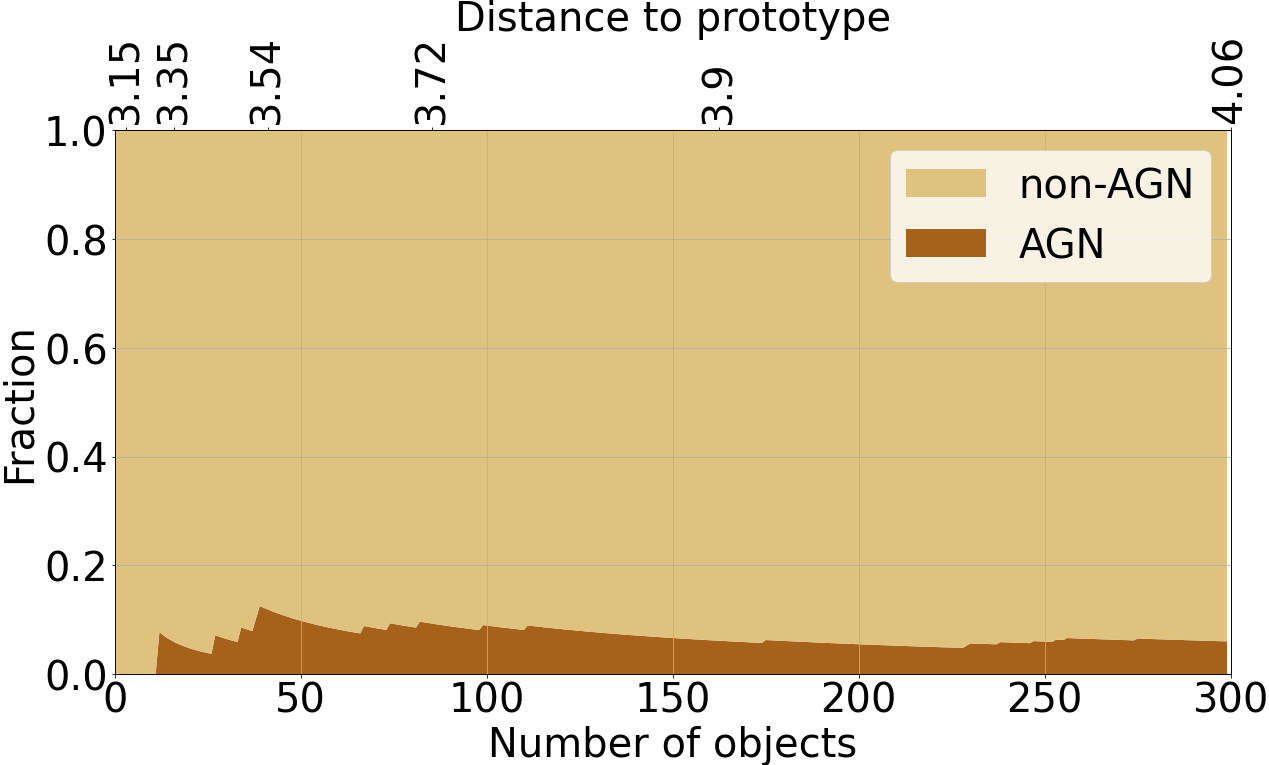}}
  \\\\

\hline \hline
\end{tabular}
}
\end{table*}

\begin{table*}[!ht]
\caption{The obtained results for non-AGN prototypes\,\#9-13 presented in Table~\ref{tab:reference}. }\label{tab:refs3}
\centering
\resizebox{\textwidth}{!}{  
\setlength\tabcolsep{4 pt}

\begin{tabular}{cccc} 
\hline\hline             
\,\# & Thumbnail & Random & X-ray MOC \\
\hline \hline \\
  9 & \parbox{3.5cm}{\includegraphics[width=3.5cm]{images/queries/quiescent1.jpeg}}  &  
     \parbox{8cm}{\centering\includegraphics[width=1\linewidth]{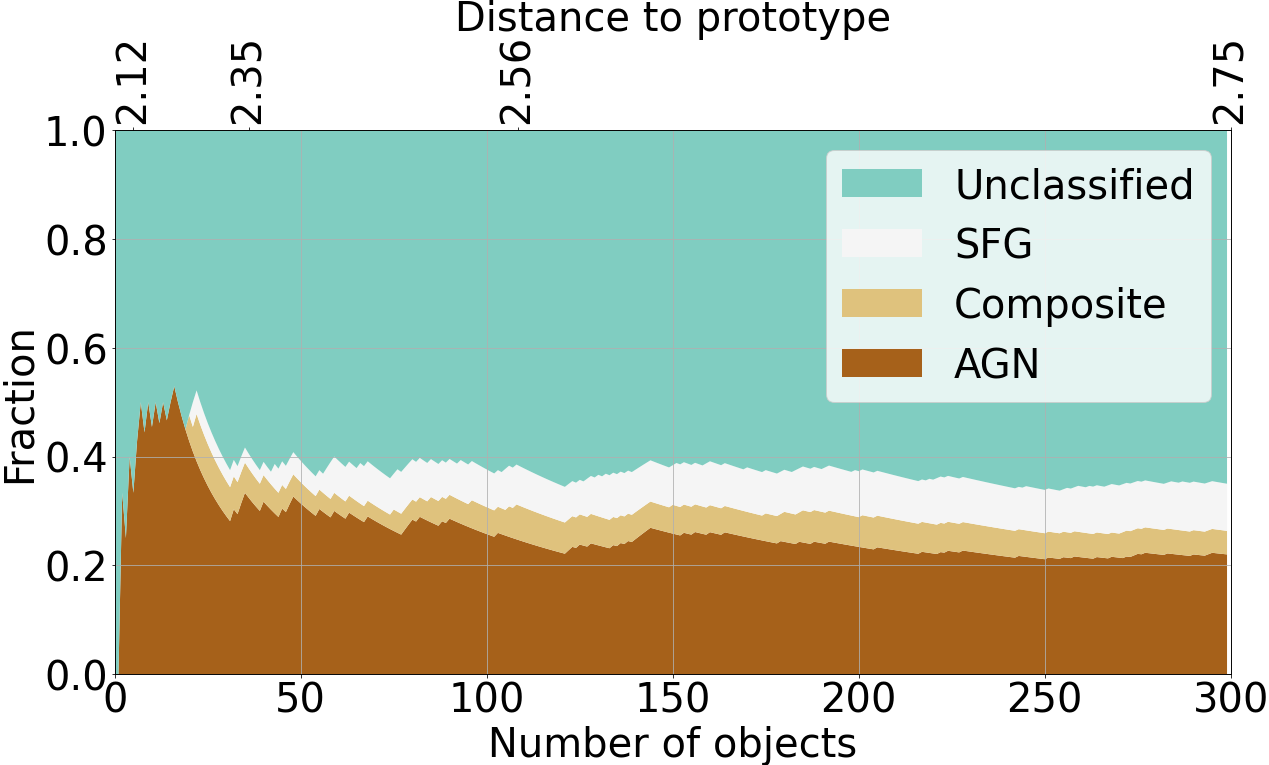}} & \parbox{8cm}{
      \centering\includegraphics[width=1\linewidth]{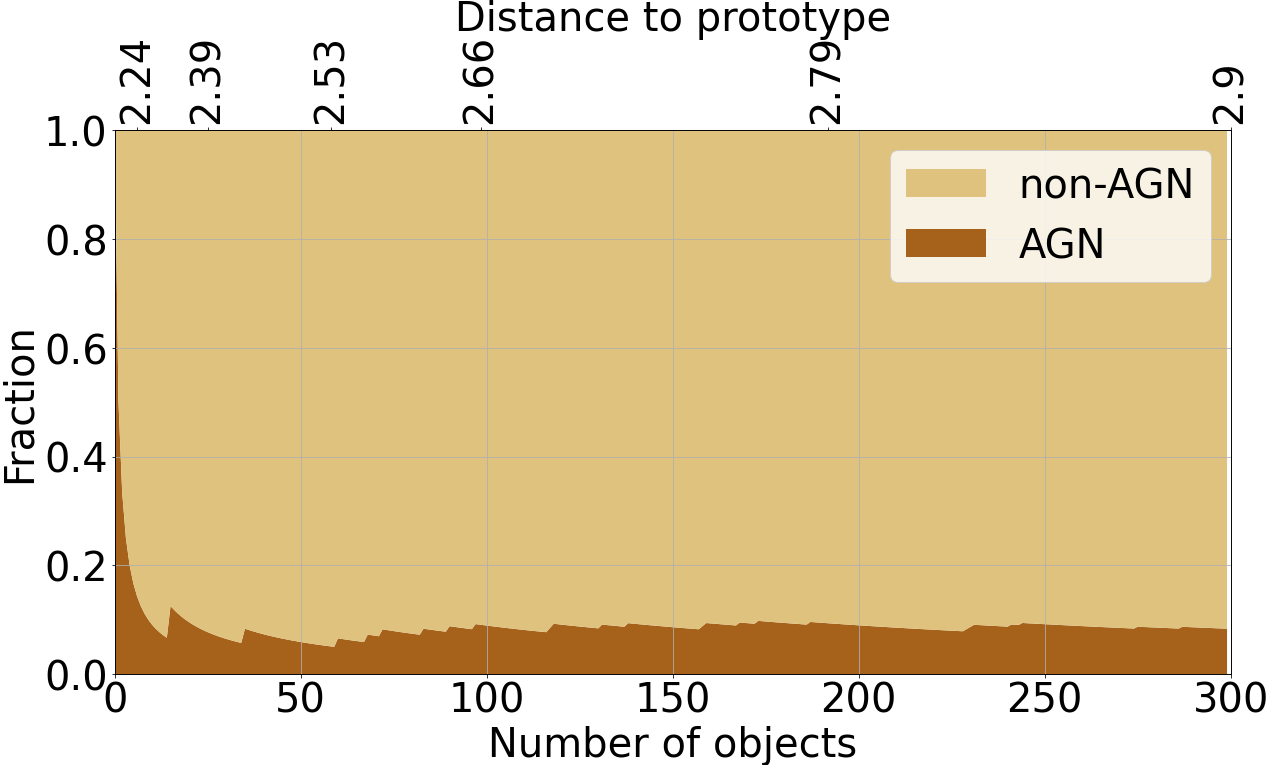}} \\
 10 & \parbox{3.5cm}{\includegraphics[width=3.5cm]{images/queries/quiescent2.jpeg}}  &  
     \parbox{8cm}{\centering\includegraphics[width=1\linewidth]{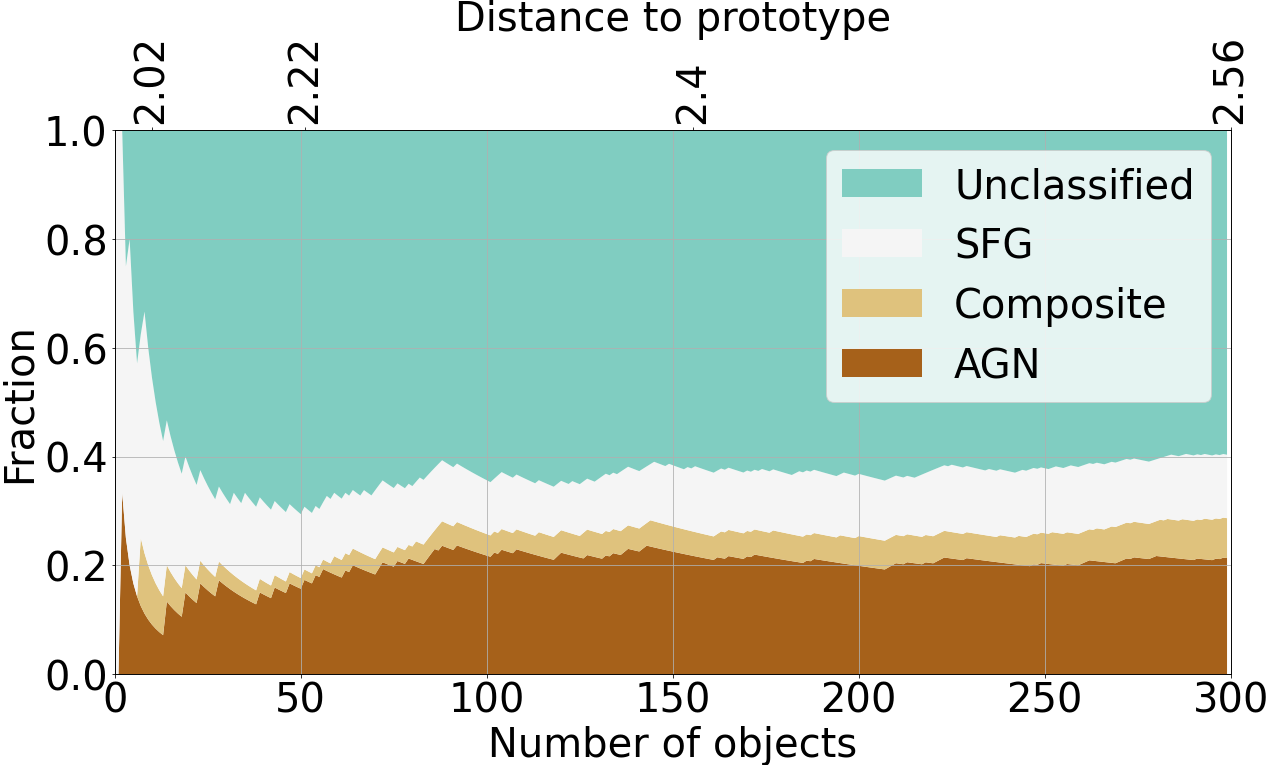}} & \parbox{8cm}{
      \centering\includegraphics[width=1\linewidth]{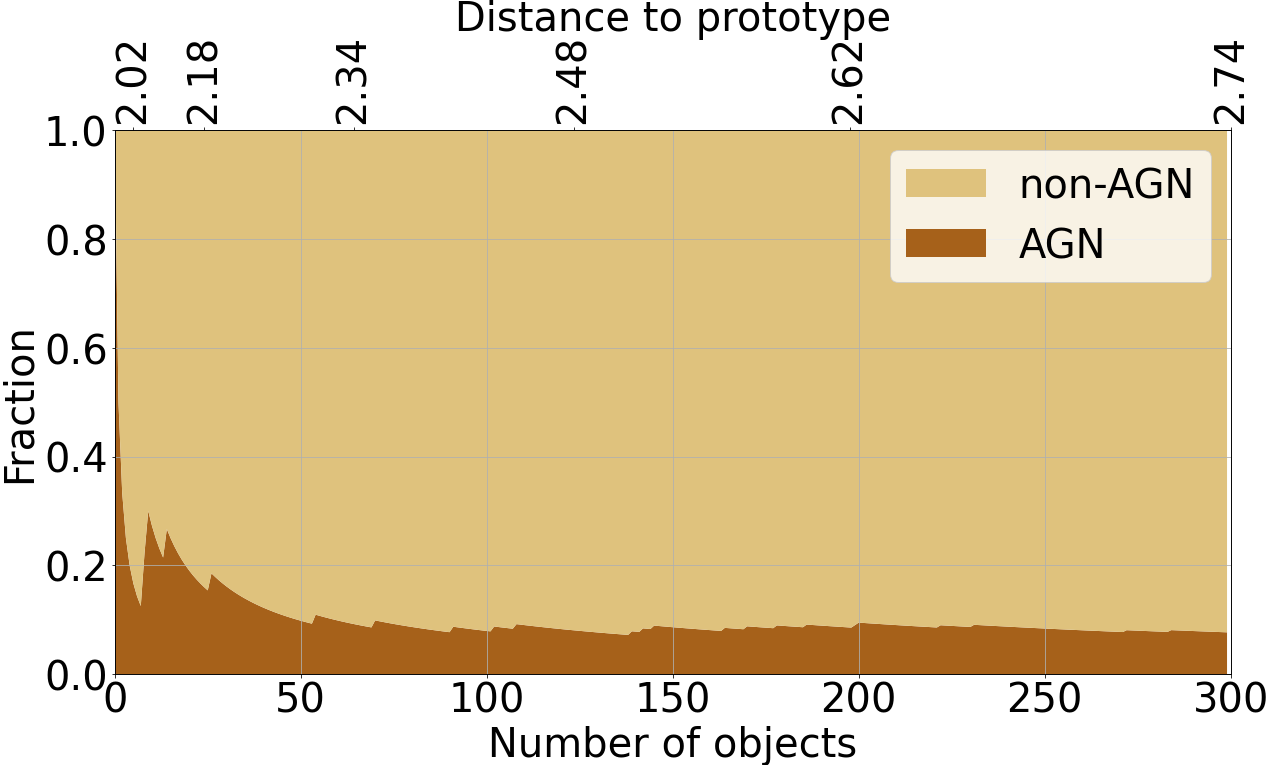}} \\
 11 & \parbox{3.5cm}{\includegraphics[width=3.5cm]{images/queries/quiescent3.jpeg}}  &  
     \parbox{8cm}{\centering\includegraphics[width=1\linewidth]{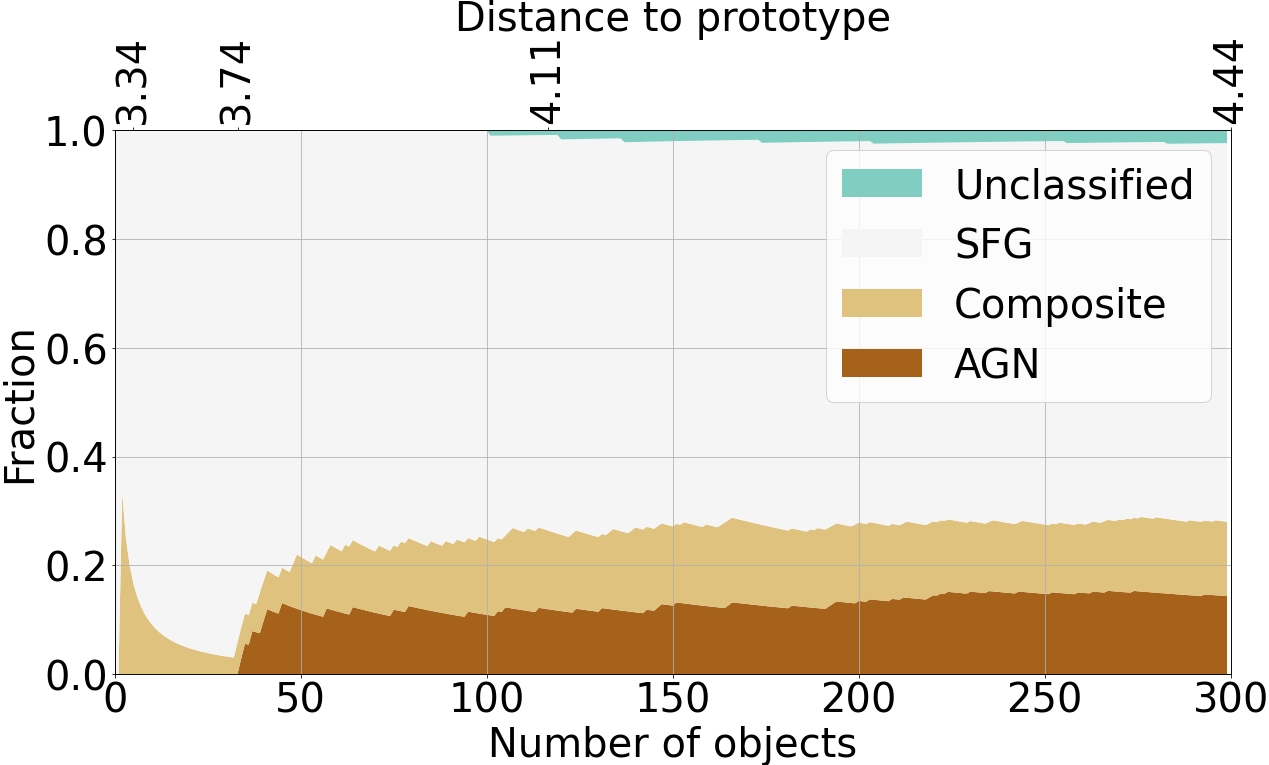}} & \parbox{8cm}{
      \centering\includegraphics[ width=1\linewidth]{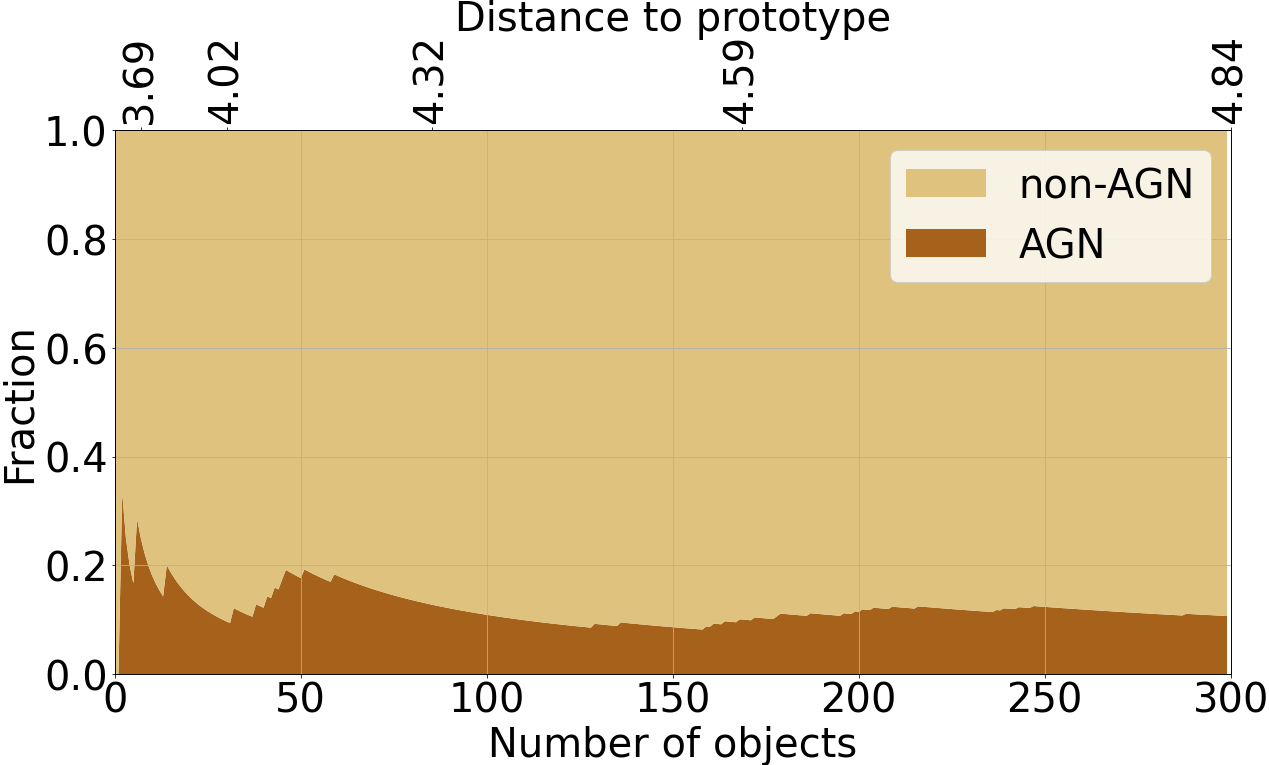}} \\
 12 & \parbox{3.5cm}{\includegraphics[width=3.5cm]{images/queries/quiescent4.jpeg}}  &  
     \parbox{8cm}{\centering\includegraphics[width=1\linewidth]{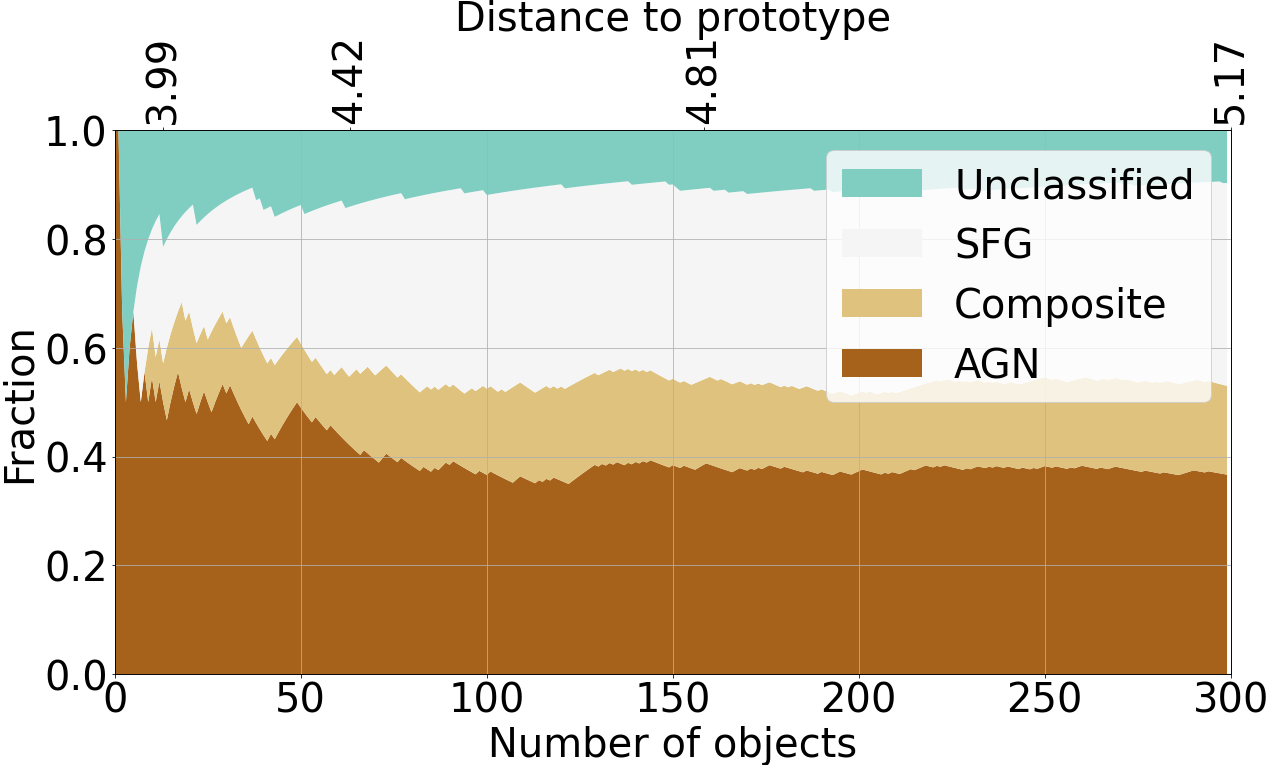}} & \parbox{8cm}{
      \centering\includegraphics[ width=1\linewidth]{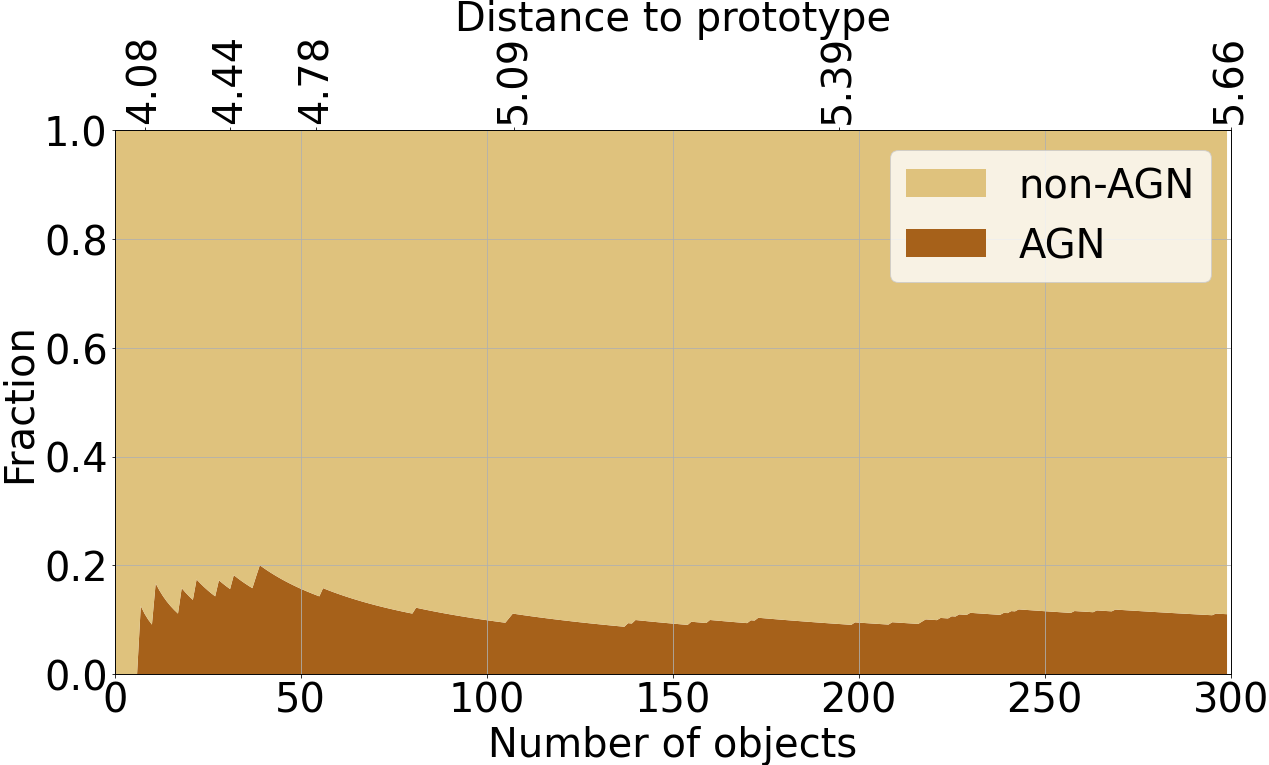}} \\
 13 & \parbox{3.5cm}{\includegraphics[width=3.5cm]{images/queries/quiescent5.jpeg}}  &  
     \parbox{8cm}{\centering\includegraphics[width=1\linewidth]{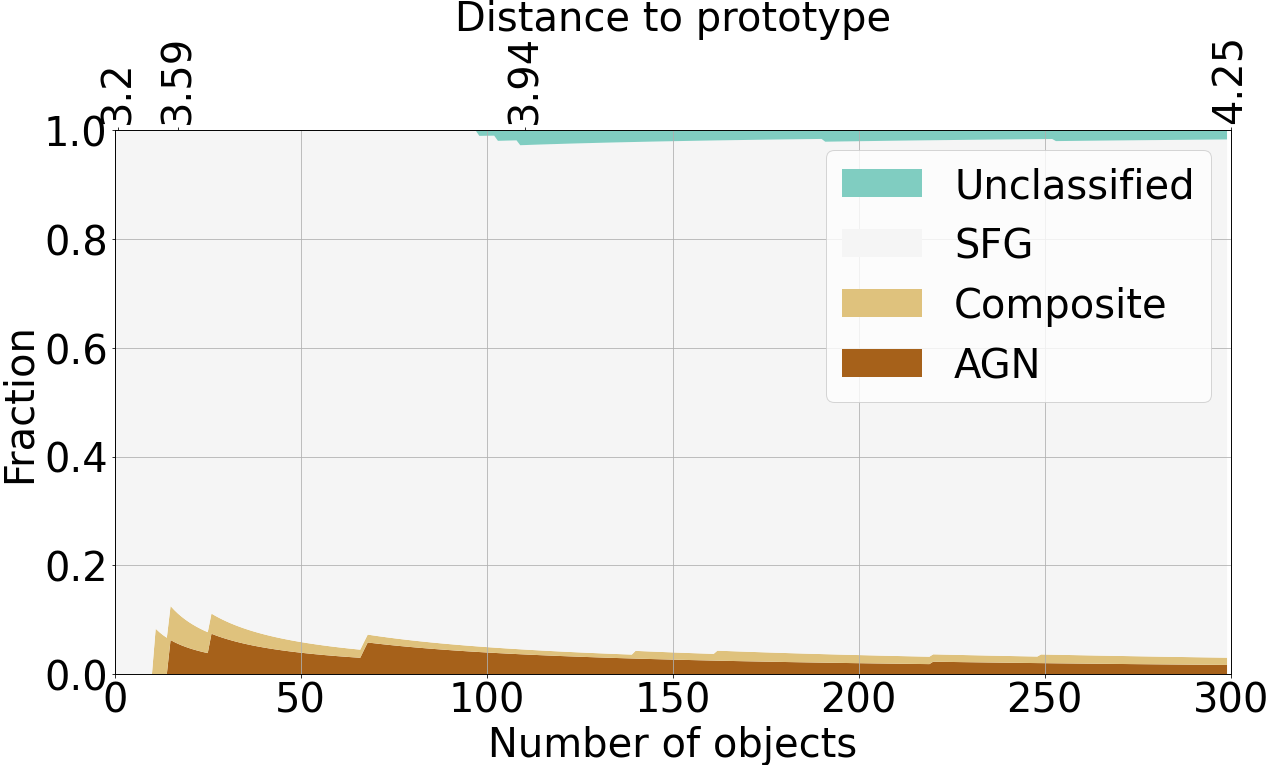}} & \parbox{8cm}{
      \centering\includegraphics[ width=1\linewidth]{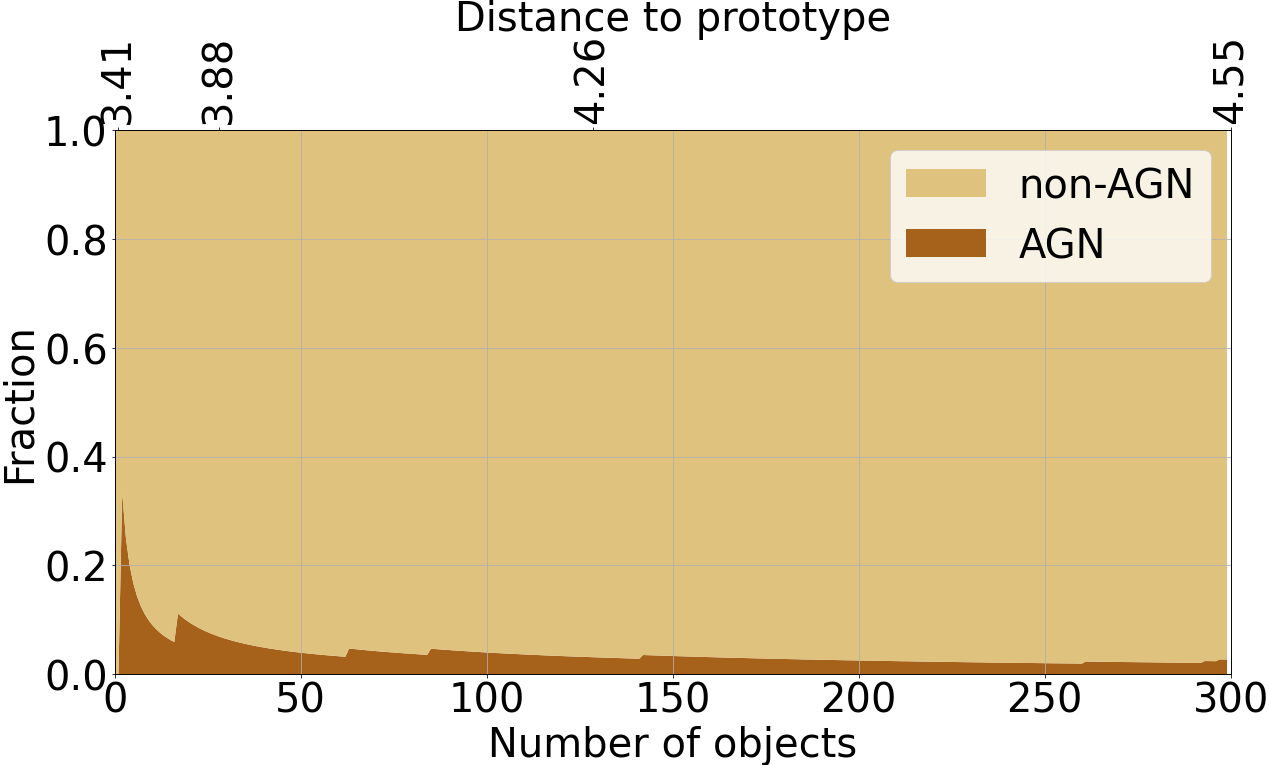}} \\\\
\hline \hline
\end{tabular}
}
\end{table*}

\begin{table*} [!ht] 
\caption{The results for AGN prototype\,\#2,\,3,\,6 obtained by our method based on one single $g$, $r$ or $i$-band in comparison with the results obtained based on three bands.}\label{tab:morphcolor}
\centering        
\renewcommand{\arraystretch}{1.5}
\setlength\tabcolsep{4 pt}
\begin{tabular}{lccc} 
\hline\hline
\# & 2 & 3 & 6\\
\hline\hline\\[-2ex]
\rotatebox[origin=c]{90}{\centering Thumbnail} & \parbox{3.5cm}{\includegraphics[width=3.5cm]{images/queries/olenaq2.jpeg}} & \parbox{3.5cm}{\includegraphics[width=3.5cm]{images/queries/olenaq3.jpeg}} & \parbox{3.5cm}{\includegraphics[width=3.5cm]{images/queries/olenaq7.jpeg}} \\\\[-2ex]
\hline\\[-2ex]
\rotatebox[origin=c]{90}{\centering $g$-band} & \parbox{5.65cm}{\centering\includegraphics[width=1\linewidth]{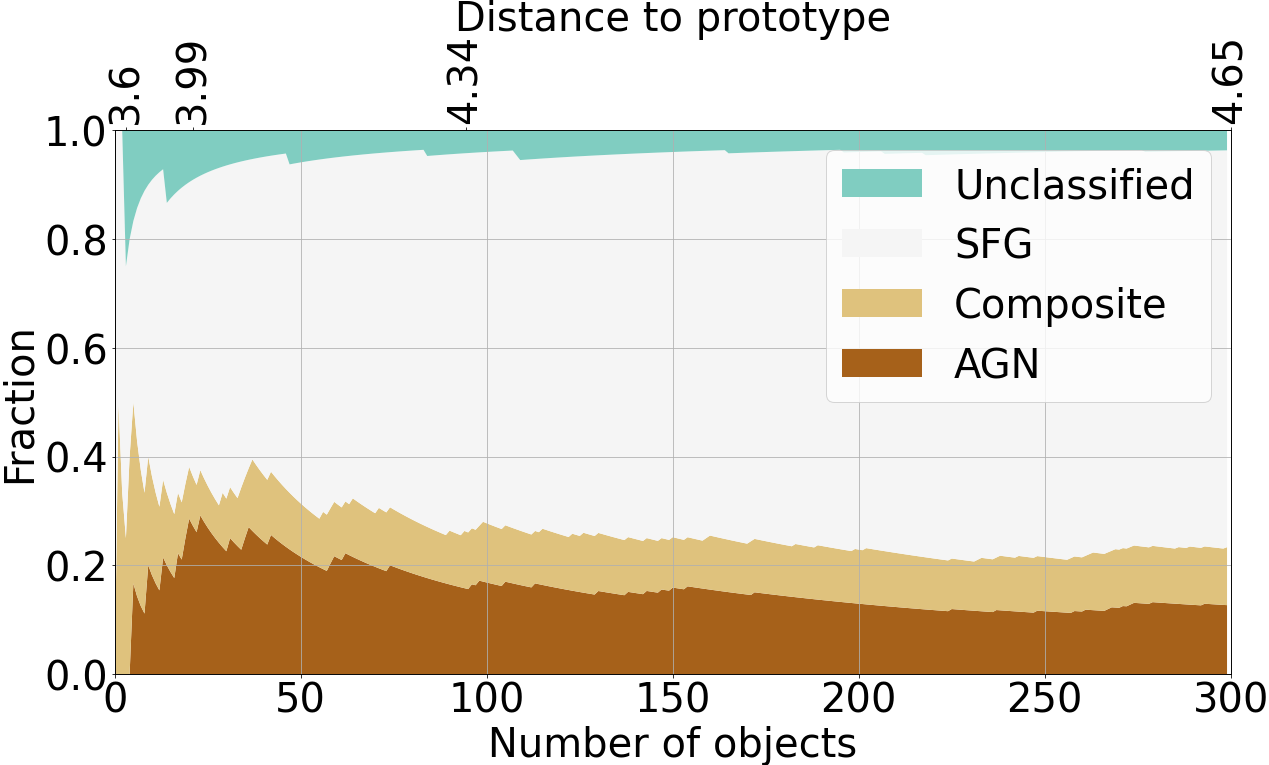}} & \parbox{5.65cm}{\centering\includegraphics[width=1\linewidth]{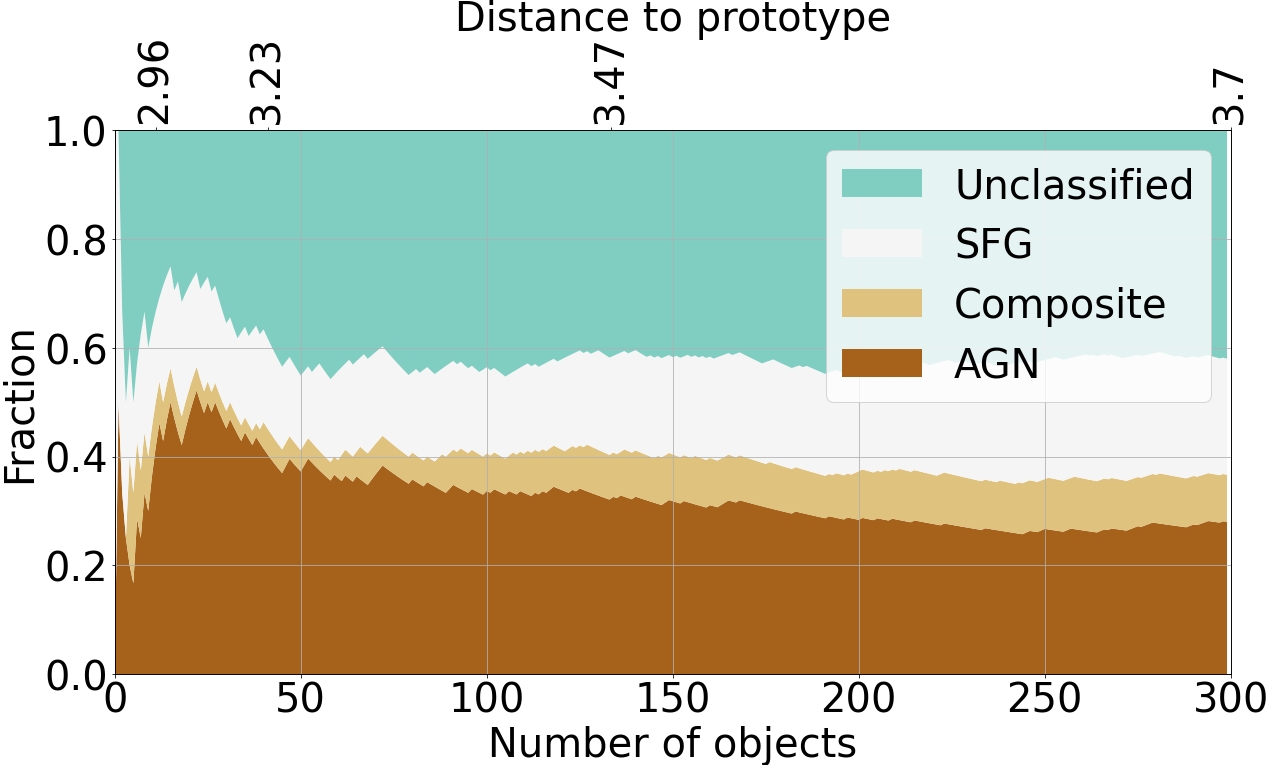}} & \parbox{5.65cm}{\centering\includegraphics[width=1\linewidth]{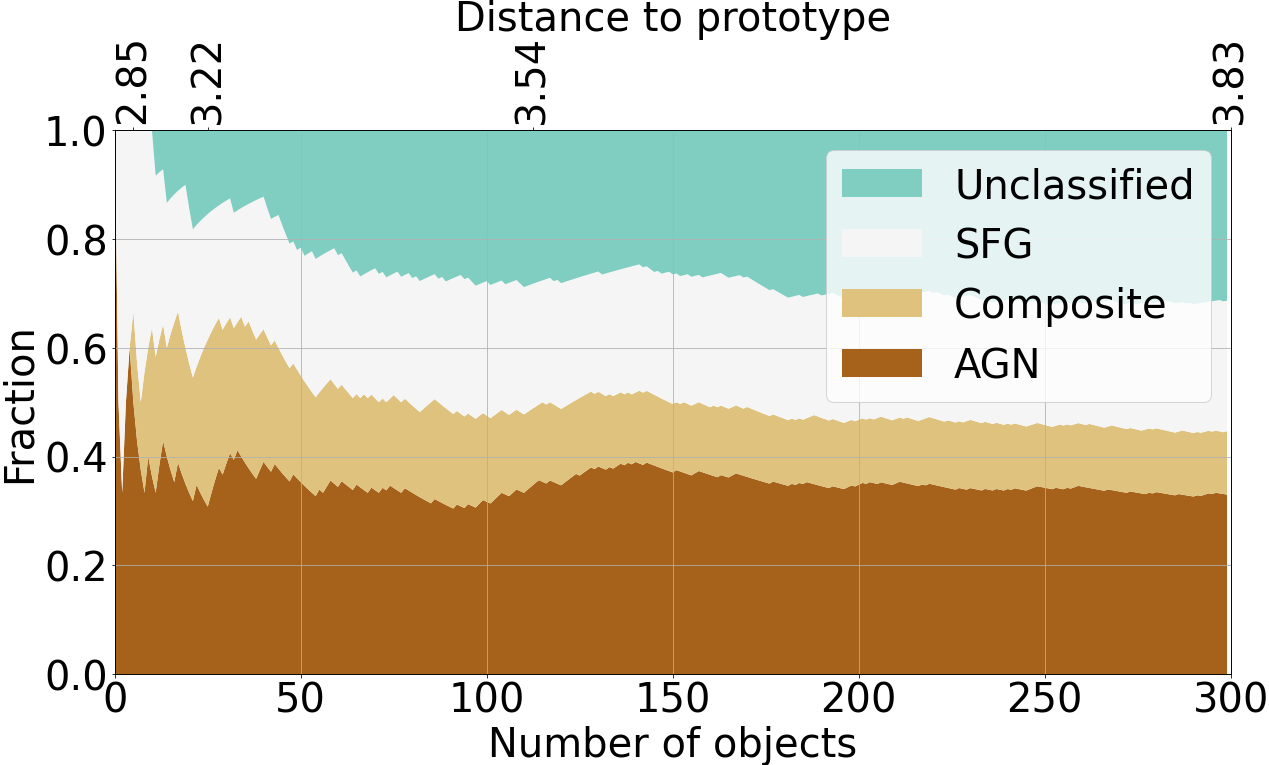}}\\\\[-2ex]
\rotatebox[origin=c]{90}{\centering $r$-band} & \parbox{5.65cm}{\centering\includegraphics[width=1\linewidth]{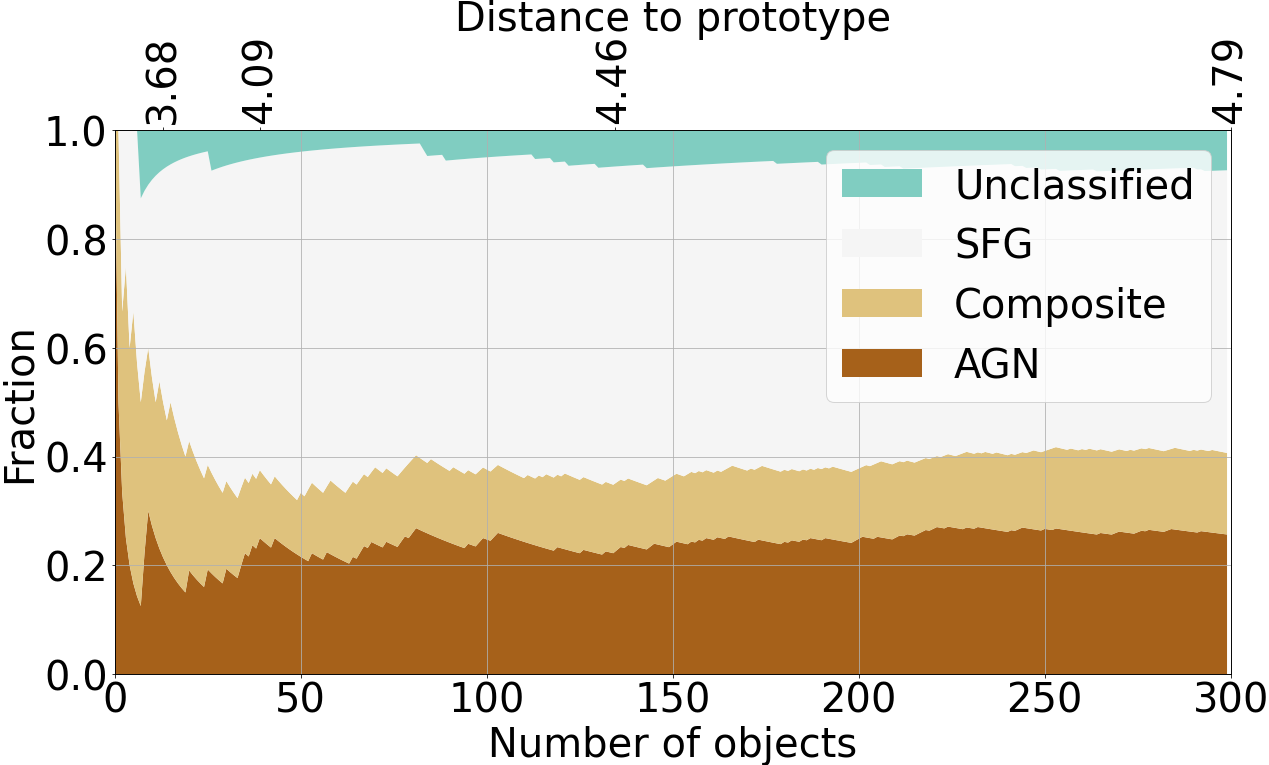}} & \parbox{5.65cm}{\centering\includegraphics[width=1\linewidth]{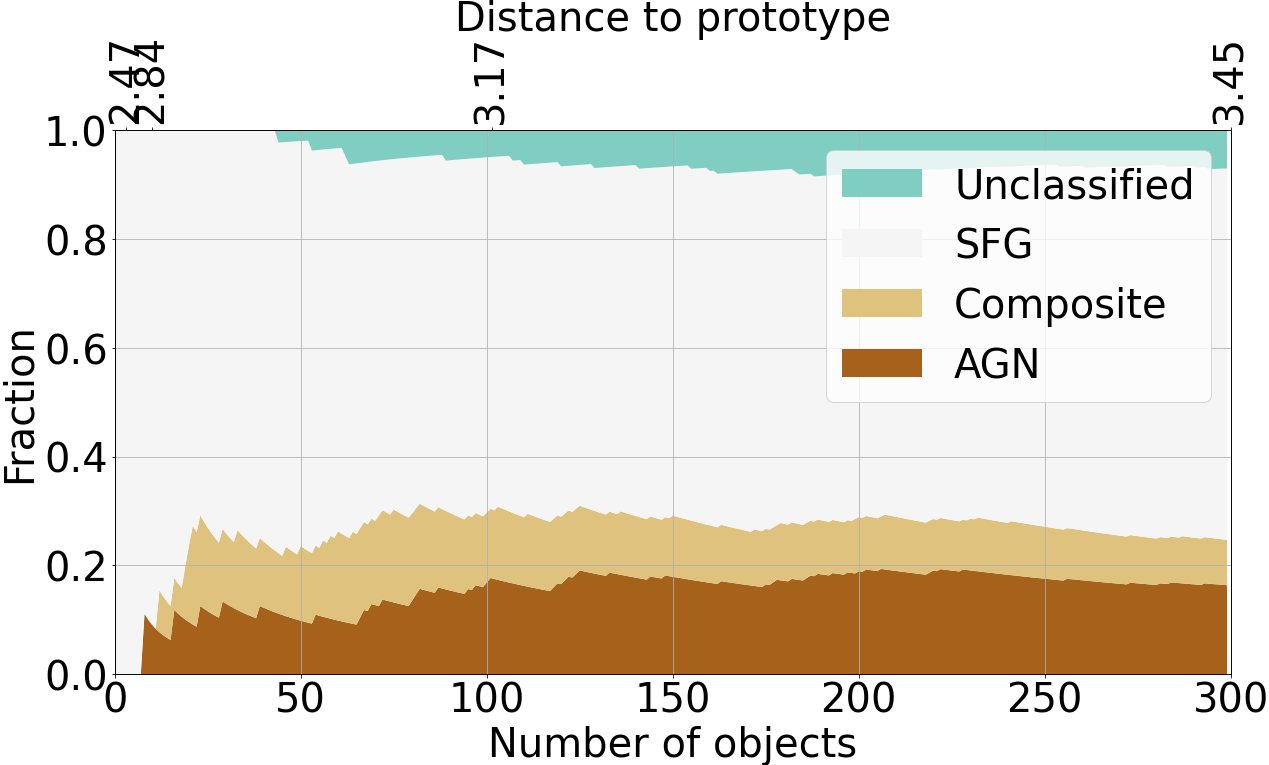}} & \parbox{5.65cm}{\centering\includegraphics[width=1\linewidth]{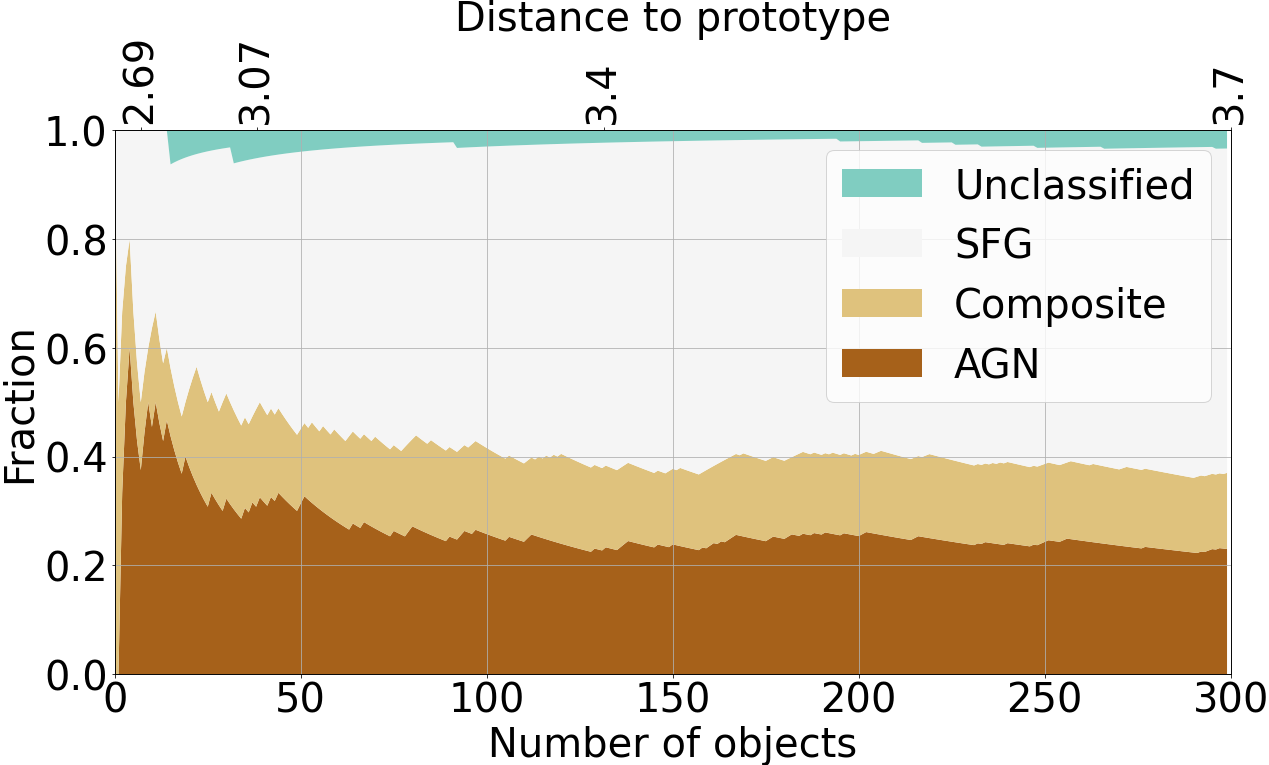}} \\\\[-2ex]
\rotatebox[origin=c]{90}{\centering $i$-band} & \parbox{5.65cm}{\centering\includegraphics[width=1\linewidth]{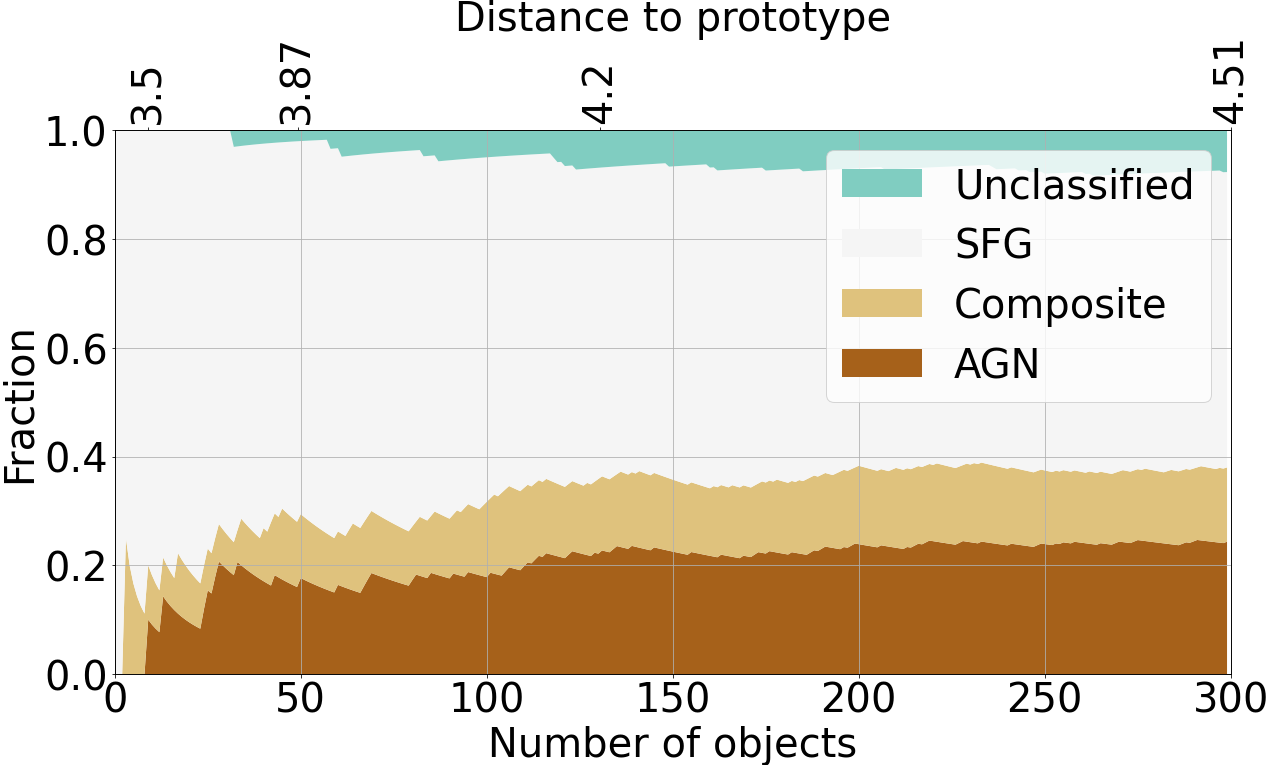}} & \parbox{5.65cm}{\centering\includegraphics[width=1\linewidth]{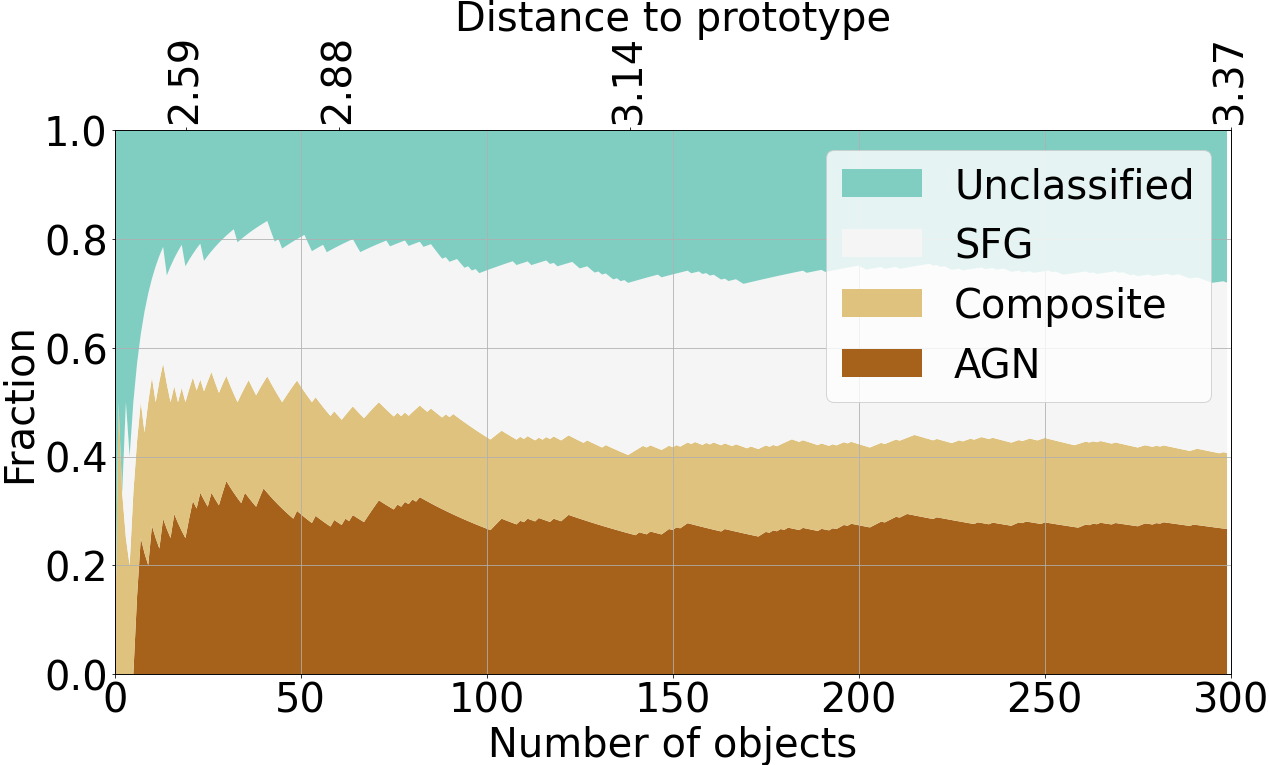}} & \parbox{5.65cm}{\centering\includegraphics[width=1\linewidth]{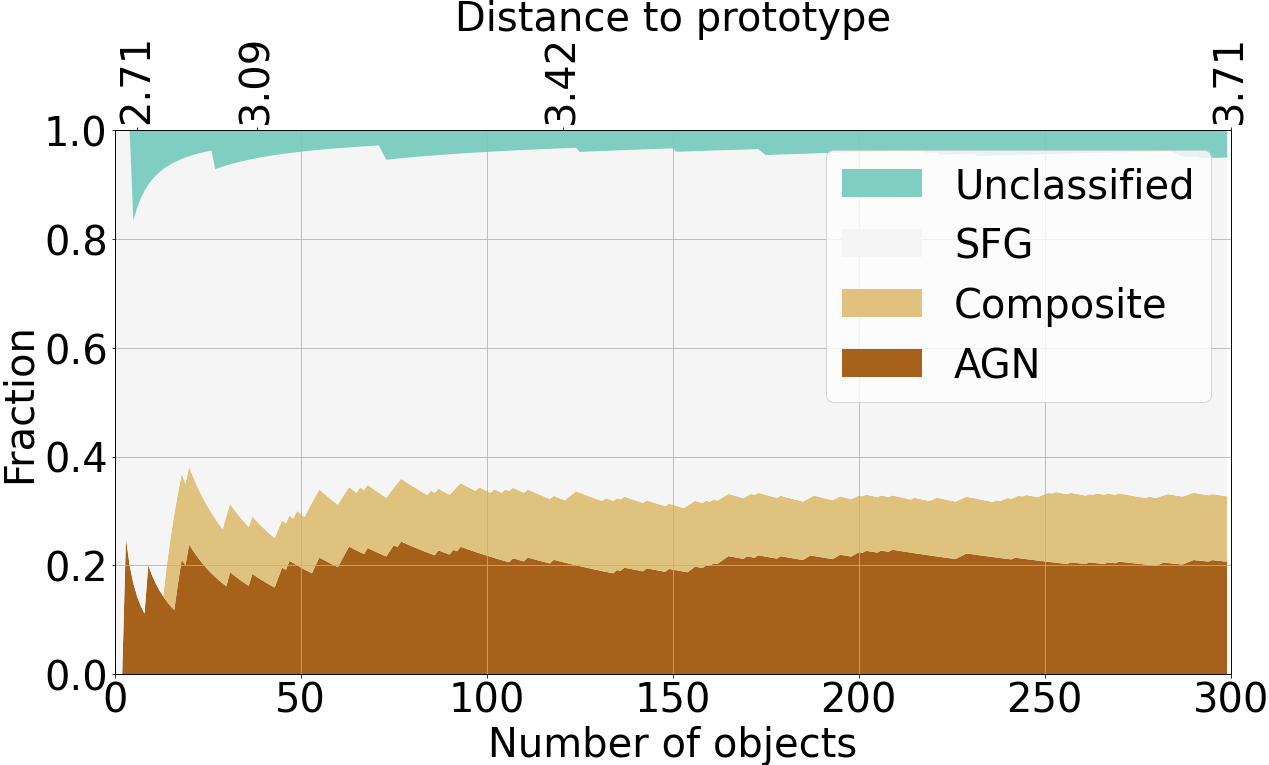}} \\\\[-2ex]
\hline\\[-2ex]
\rotatebox[origin=c]{90}{\centering Three bands} & \parbox{5.65cm}{\centering\includegraphics[width=1\linewidth]{images/x-ray/agn_bpt_fractions_2.png}} & \parbox{5.65cm}{\centering\includegraphics[width=1\linewidth]{images/x-ray/agn_bpt_fractions_3.png}} & \parbox{5.65cm}{\centering\includegraphics[width=1\linewidth]{images/x-ray/agn_bpt_fractions_7.png}}\\\\[-2ex]
\hline \hline
\end{tabular}
\end{table*}

\begin{table*}  
\caption{The application of the recursive technique to our method on the example of prototype\,\#2. The reference object for each next step is selected as the closest object to the reference prototype of the current step (see description in the text). $N_{new}$ is the number of objects among the 25 nearest neighbours not selected in the previous step.}\label{tab:recursiveLowest}
\setlength\tabcolsep{2 pt}
\centering
\begin{tabular}{cccc ccc} 
\hline\hline             
\multirow{2}{*}{Step}  & SDSS & \multirow{2}{*}{Position} & \multirow{2}{*}{Prototype} &    4 Nearest & \multirow{2}{*}{Fractions} & \multirow{2}{*}{$N_{new}$} \\ 
    & ID &      &       & neighbours&  &\\
\hline \hline
  1 & \rotatebox[origin=c]{90}{\parbox{4cm}{\centering SDSS\\J164607.00+422737.4}} 
  & \rotatebox[origin=c]{90}{ \parbox{4cm}{\centering ra: 251.52917 \\ dec: 42.46041}} &
 \parbox{4cm}{\includegraphics[width=4cm]{images/queries/olenaq2.jpeg}} 
 & \parbox{4cm}{\includegraphics[width=4cm]{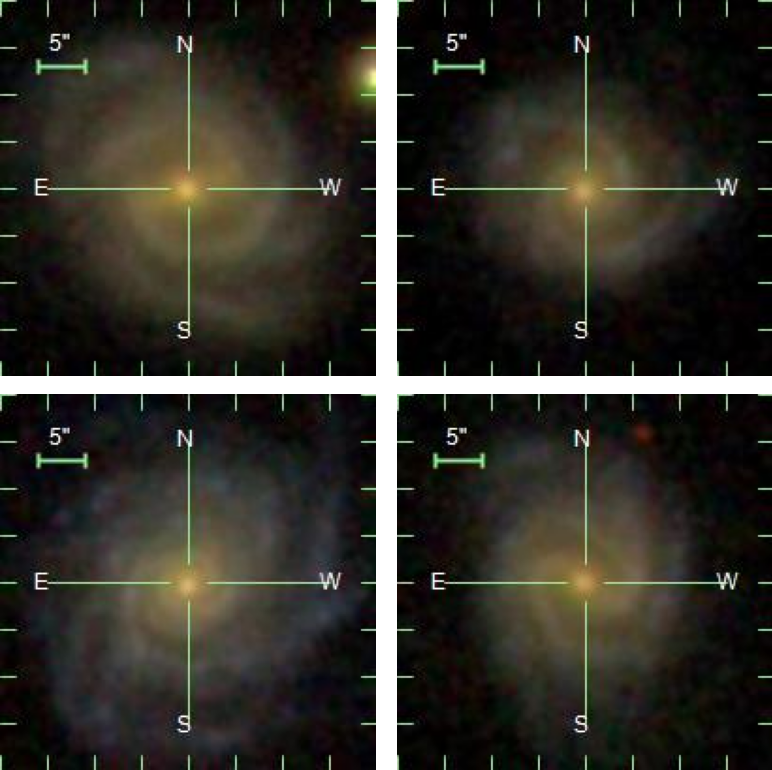}}     & \parbox{6cm}{\includegraphics[width=6cm]{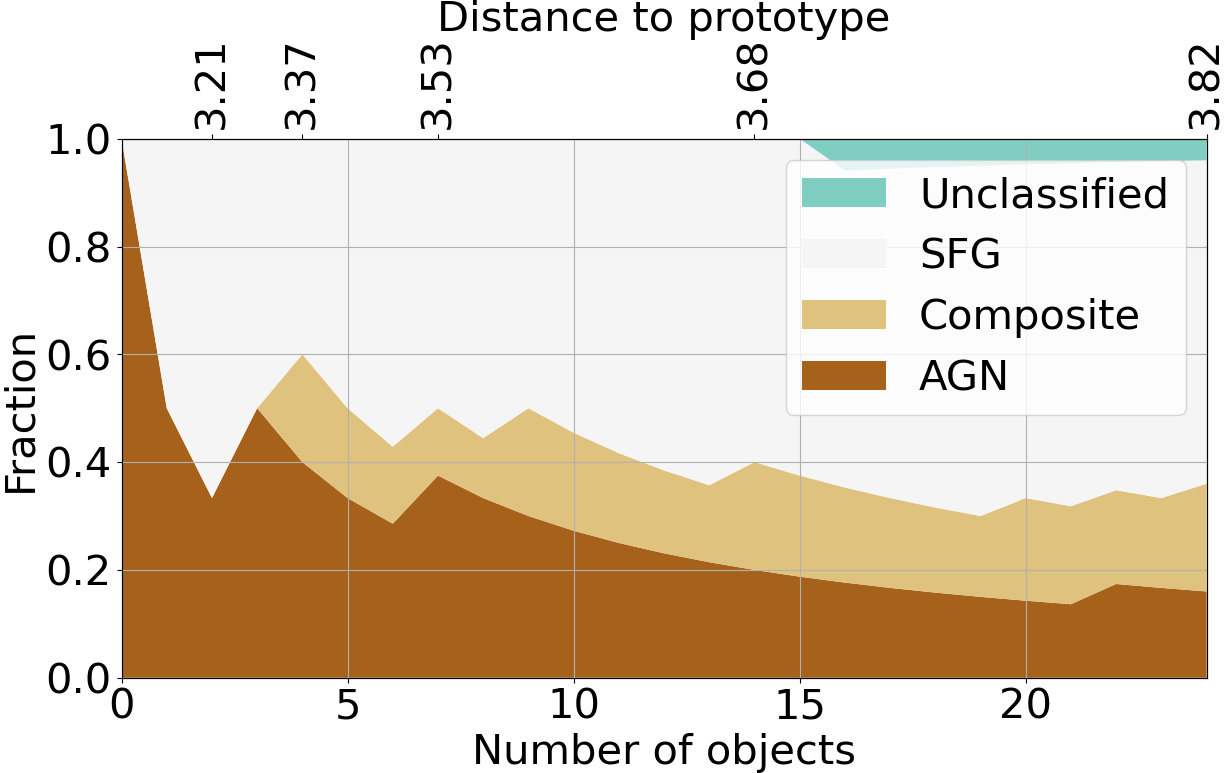}} & 25\\
  2 & \rotatebox[origin=c]{90}{\parbox{4cm}{\centering SDSS\\J111859.64+613538.1}} 
  & \rotatebox[origin=c]{90}{ \parbox{4cm}{\centering ra: 169.748518407 \\ dec: 61.593940720}} &
 \parbox{4cm}{\includegraphics[width=4cm]{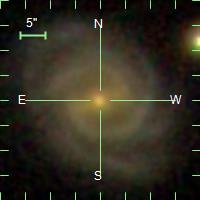}} 
 & \parbox{4cm}{\includegraphics[width=4cm]{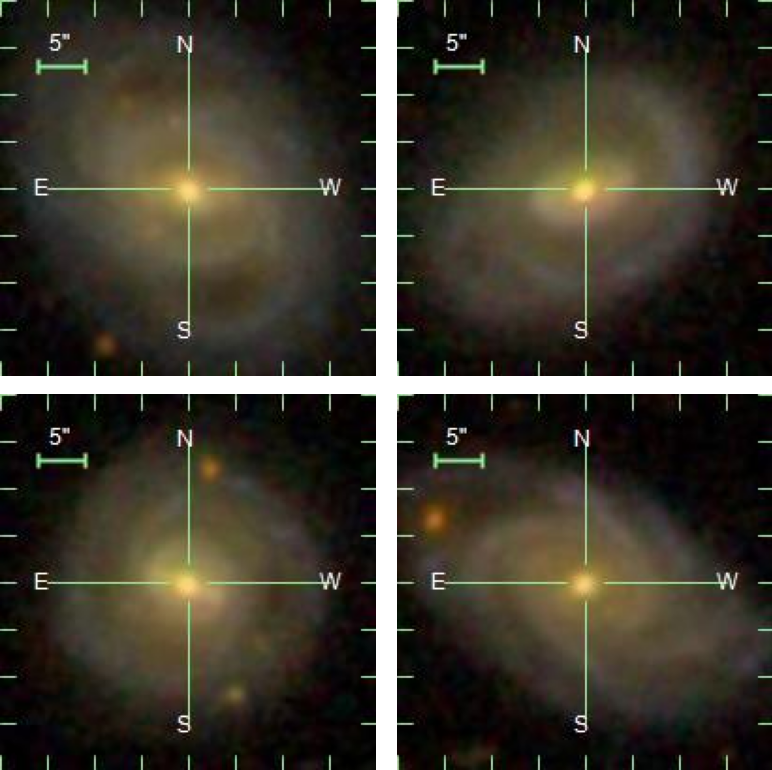}}     & \parbox{6cm}{\includegraphics[width=6cm]{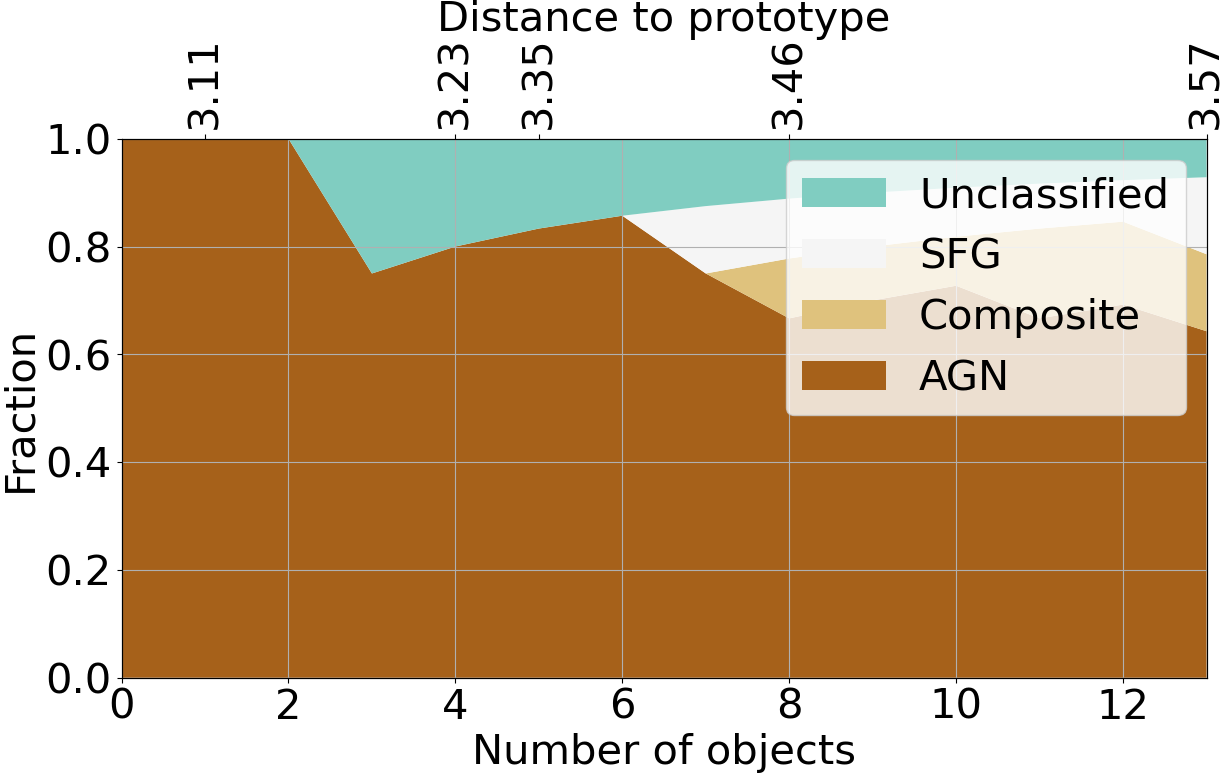}}  & 14 \\ 
  3 & \rotatebox[origin=c]{90}{\parbox{4cm}{\centering SDSS\\J161525.19+260637.2}} 
  & \rotatebox[origin=c]{90}{ \parbox{4cm}{\centering ra: 243.854969001 \\ dec: 26.110337075 }} &
 \parbox{4cm}{\includegraphics[width=4cm]{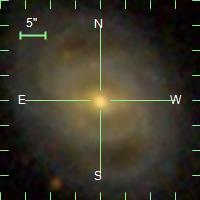}} 
 & \parbox{4cm}{\includegraphics[width=4cm]{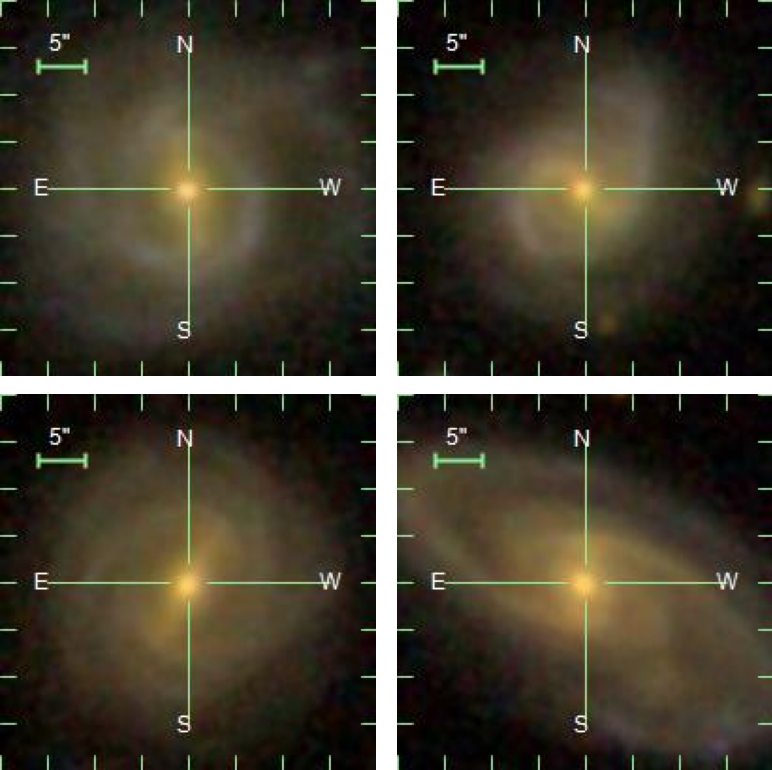}}     & \parbox{6cm}{\includegraphics[width=6cm]{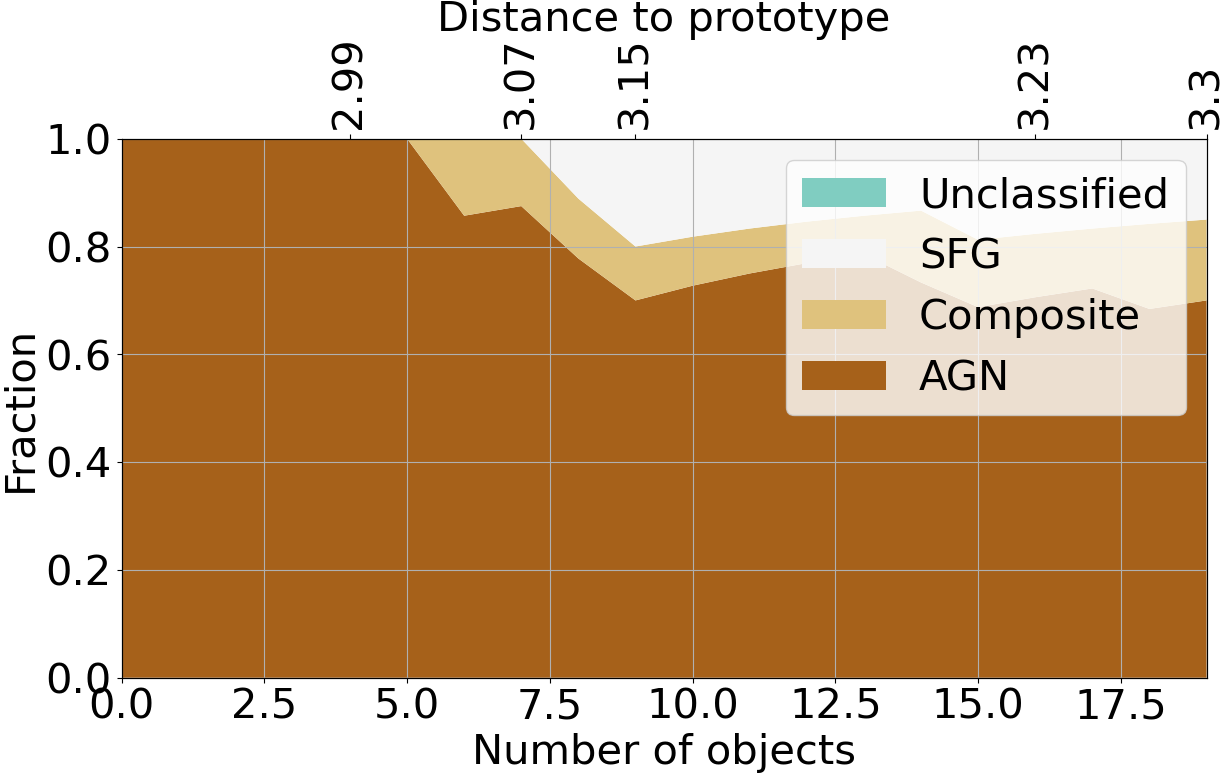}}  & 20 \\ 
  4 & \rotatebox[origin=c]{90}{\parbox{4cm}{\centering SDSS\\J125348.60+293518.1}} 
  & \rotatebox[origin=c]{90}{ \parbox{4cm}{\centering ra: 193.452509978 \\ dec: 29.588383685}} &
 \parbox{4cm}{\includegraphics[width=4cm]{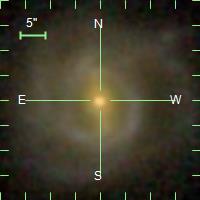}} 
 & \parbox{4cm}{\includegraphics[width=4cm]{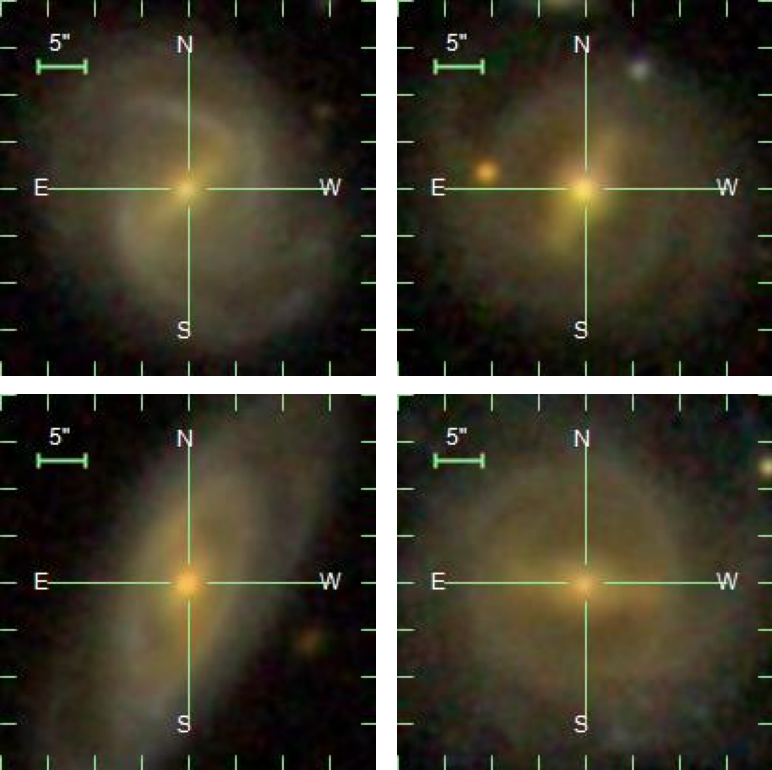}}     & \parbox{6cm}{\includegraphics[width=6cm]{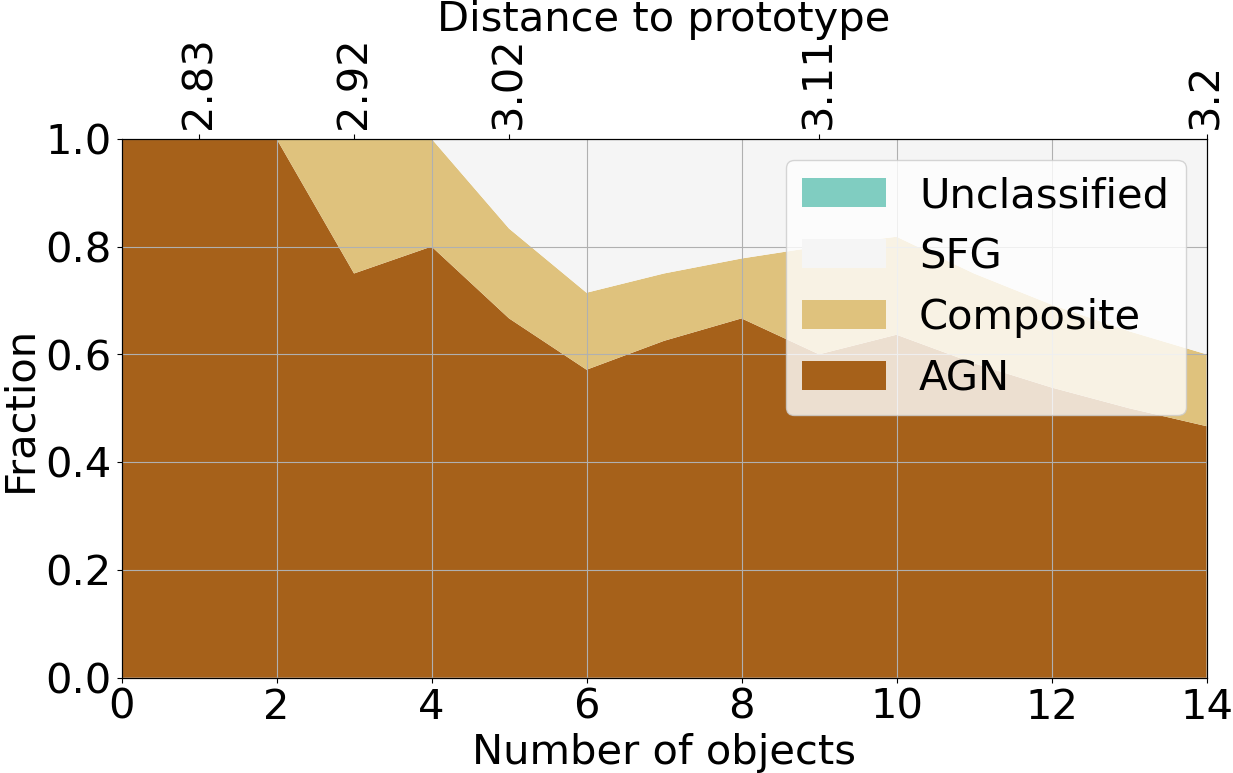}}  & 15 \\ 
  5 & \rotatebox[origin=c]{90}{\parbox{4cm}{\centering SDSS\\J094458.29+385711.3}} 
  & \rotatebox[origin=c]{90}{ \parbox{4cm}{\centering ra: 146.242880914 \\ dec: 38.953138897}} &
 \parbox{4cm}{\includegraphics[width=4cm]{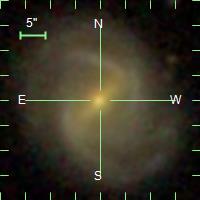}} 
 & \parbox{4cm}{\includegraphics[width=4cm]{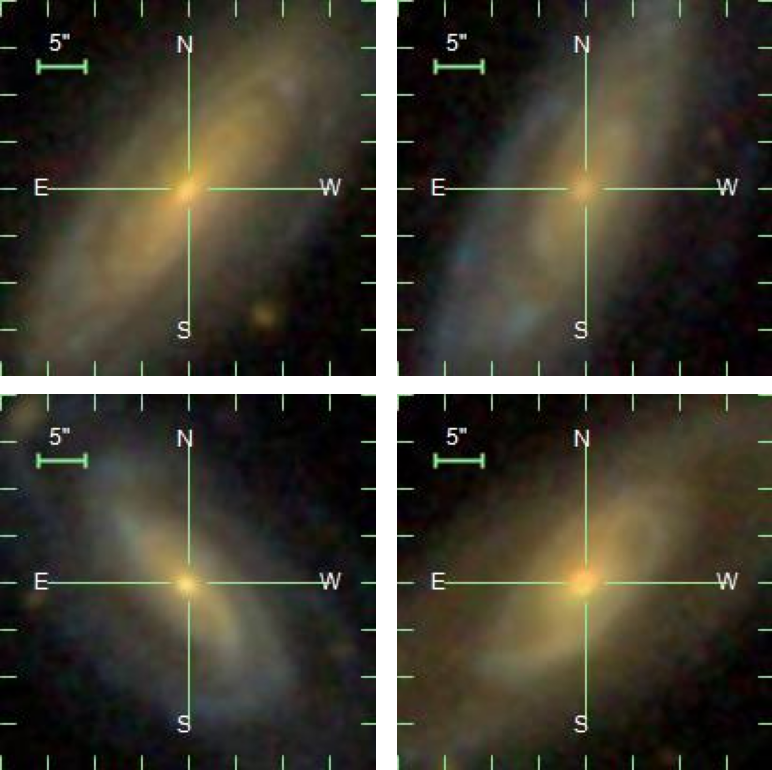}}     & \parbox{6cm}{\includegraphics[width=6cm]{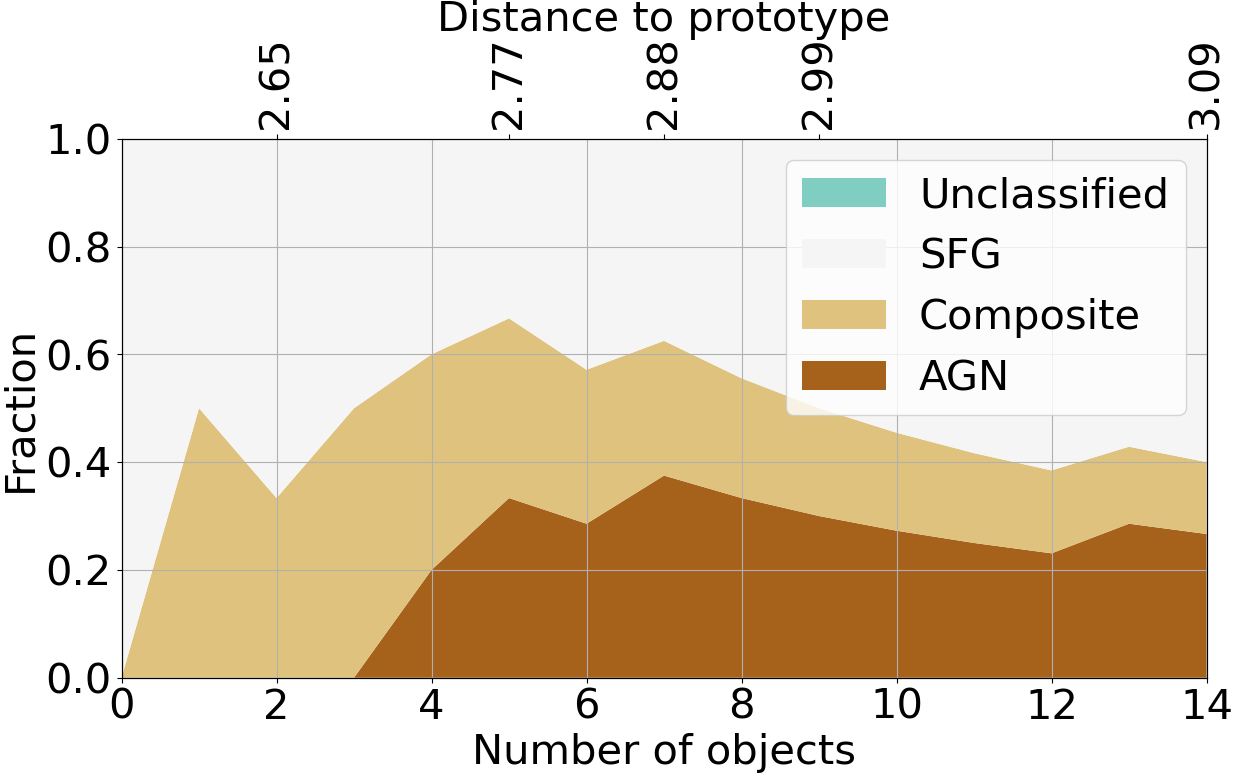}} & 15  \\  
 
\hline \hline
\end{tabular}

\end{table*}
\clearpage

\section{Neighbours}\label{sec:appendix}
In this section we show the 25 nearest neighbours for the 13 prototypes.

\begin{figure*}[!hb]
\centering
      \centering
      \includegraphics[width=4cm]{images/queries/q1.jpeg}
        \includegraphics[width=0.5\linewidth]{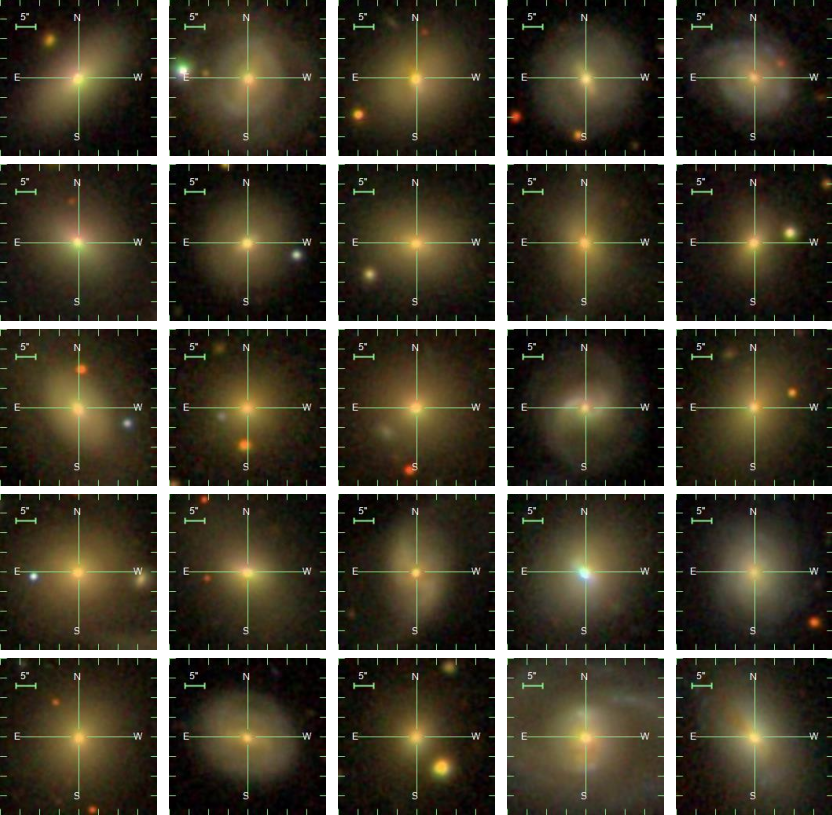}
    \caption{The closest 25 neighbours for AGN prototype\,\#1.}
    \label{fig:nns1}
\end{figure*}

\begin{figure*}
\centering
      \centering
            \includegraphics[width=4cm]{images/queries/olenaq2.jpeg}
        \includegraphics[width=0.5\linewidth]{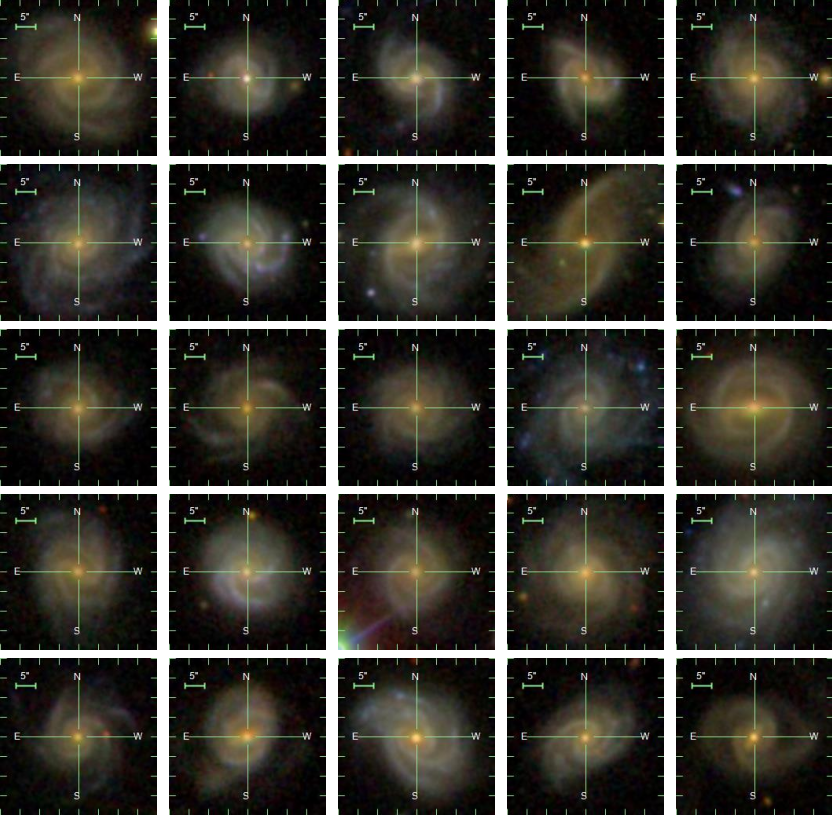}
    \caption{The closest 25 neighbours for AGN prototype\,\#2.}
    \label{fig:nns2}
\end{figure*}

\begin{figure*}
\centering
      \centering
            \includegraphics[width=4cm]{images/queries/olenaq3.jpeg}
        \includegraphics[width=0.5\linewidth]{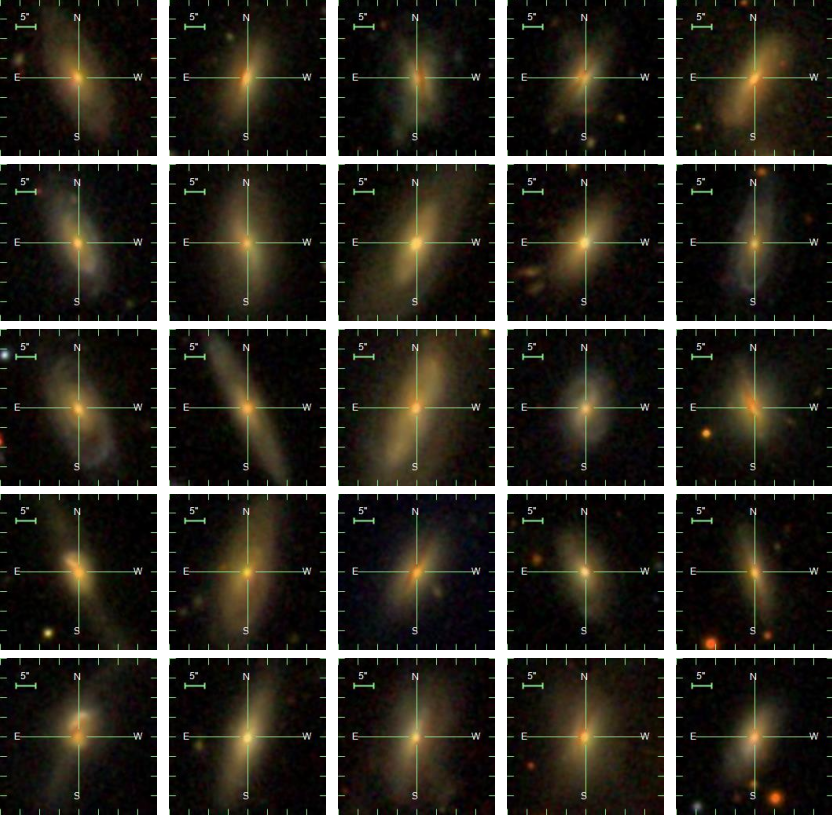}
    \caption{The closest 25 neighbours for AGN prototype\,\#3.}
    \label{fig:nns3}
\end{figure*}

\begin{figure*}
\centering
      \centering
            \includegraphics[width=4cm]{images/queries/olenaq4.jpeg}
        \includegraphics[width=0.5\linewidth]{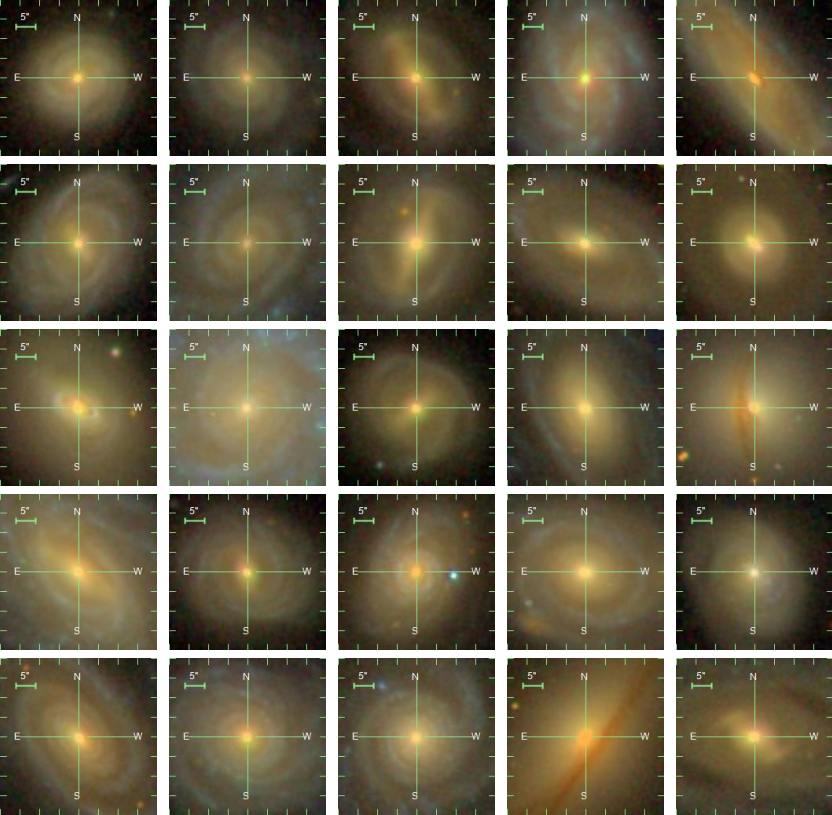}
    \caption{The closest 25 neighbours for AGN prototype\,\#4.}
    \label{fig:nns4}
\end{figure*}

\begin{figure*}
\centering
      \centering
            \includegraphics[width=4cm]{images/queries/olenaq5.jpeg}
        \includegraphics[width=0.5\linewidth]{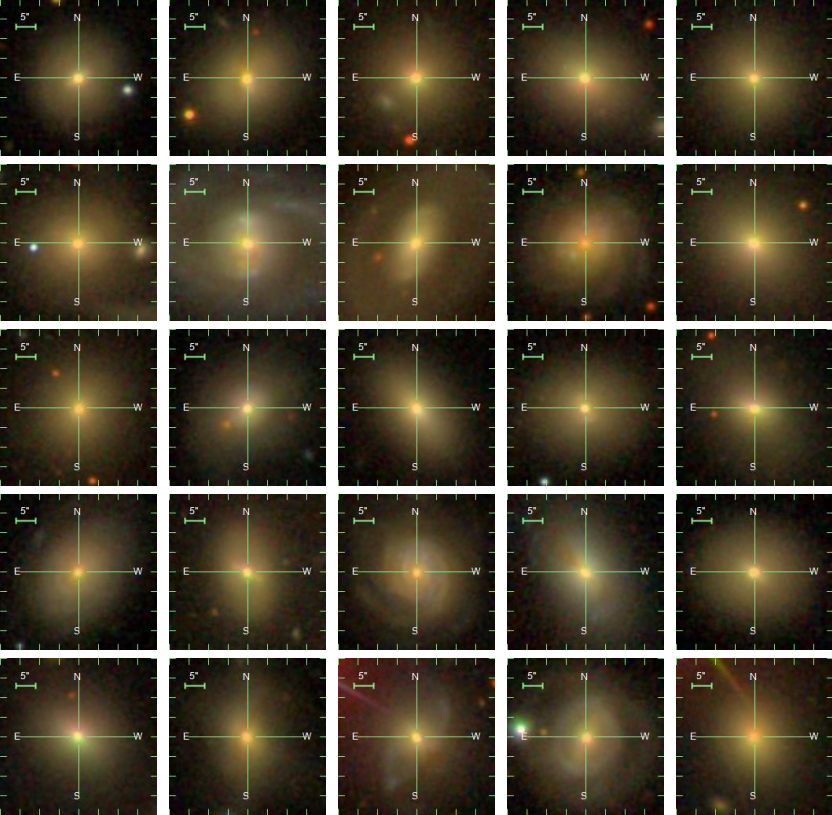}
    \caption{The closest 25 neighbours for AGN prototype\,\#5.}
    \label{fig:nns5}
\end{figure*}

\begin{figure*}
\centering
      \centering
            \includegraphics[width=4cm]{images/queries/olenaq7.jpeg}
        \includegraphics[width=0.5\linewidth]{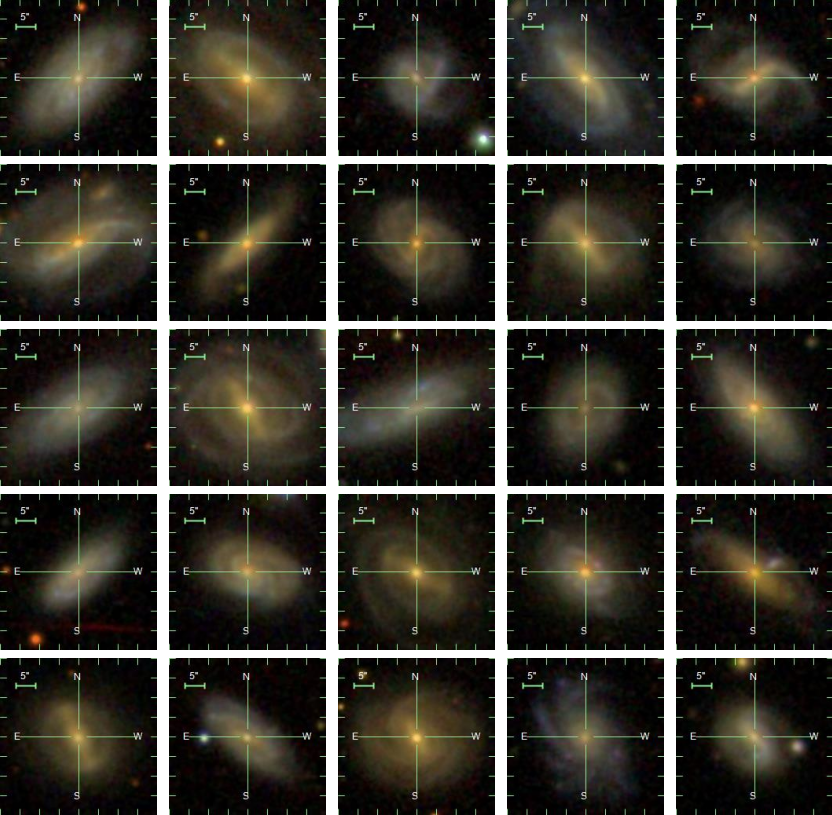}
    \caption{The closest 25 neighbours for AGN prototype\,\#6.}
    \label{fig:nns6}
\end{figure*}

\begin{figure*}
\centering
      \centering
            \includegraphics[width=4cm]{images/queries/olenaq8.jpeg}
        \includegraphics[width=0.5\linewidth]{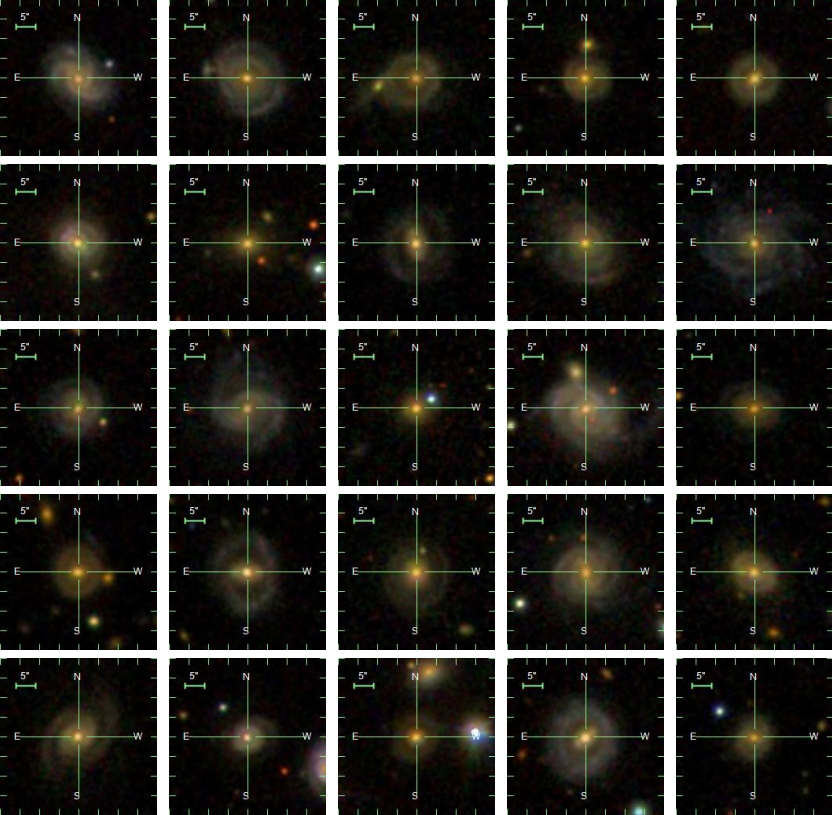}
    \caption{The closest 25 neighbours for AGN prototype\,\#7.}
    \label{fig:nns7}
\end{figure*}

\begin{figure*}
\centering
      \centering
            \includegraphics[width=4cm]{images/queries/olenaq9.jpeg}
        \includegraphics[width=0.5\linewidth]{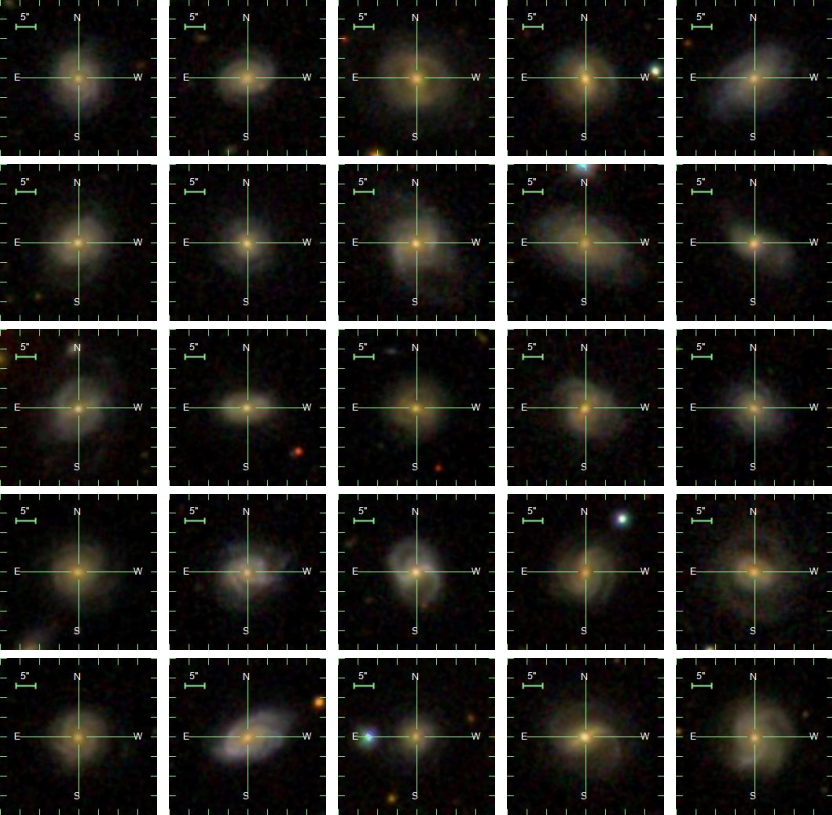}
    \caption{The closest 25 neighbours for AGN prototype\,\#8.}
    \label{fig:nns8}
\end{figure*}

\begin{figure*}
\centering
      \centering
            \includegraphics[width=4cm]{images/queries/quiescent1.jpeg}
        \includegraphics[width=0.5\linewidth]{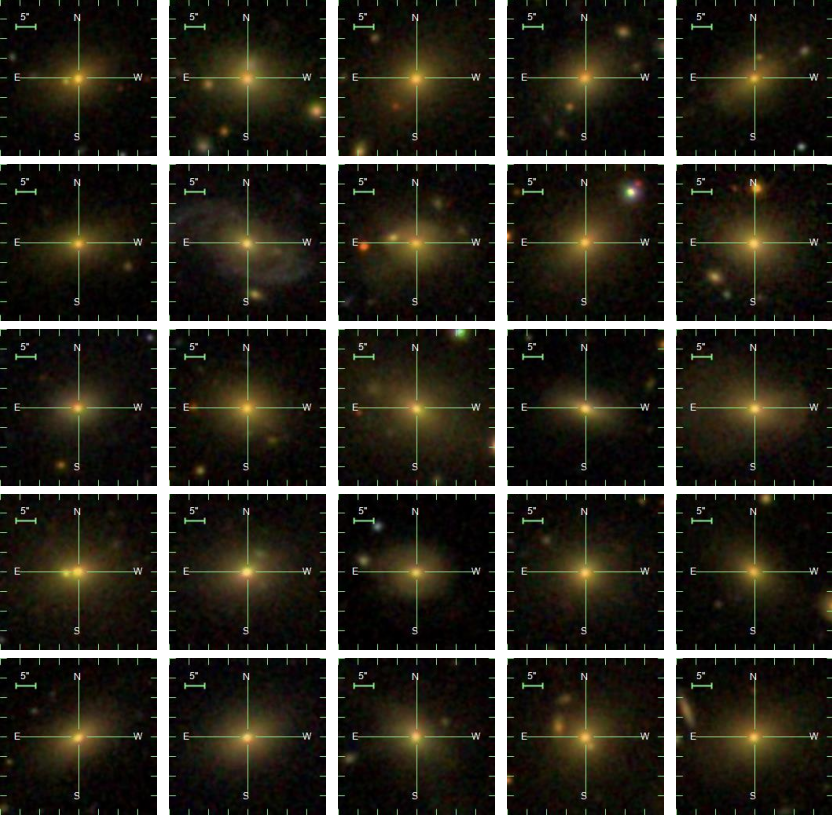}
    \caption{The closest 25 neighbours for non-AGN prototype\,\#9.}
    \label{fig:nns9}
\end{figure*}

\begin{figure*}
\centering
      \centering
            \includegraphics[width=4cm]{images/queries/quiescent2.jpeg}
        \includegraphics[width=0.5\linewidth]{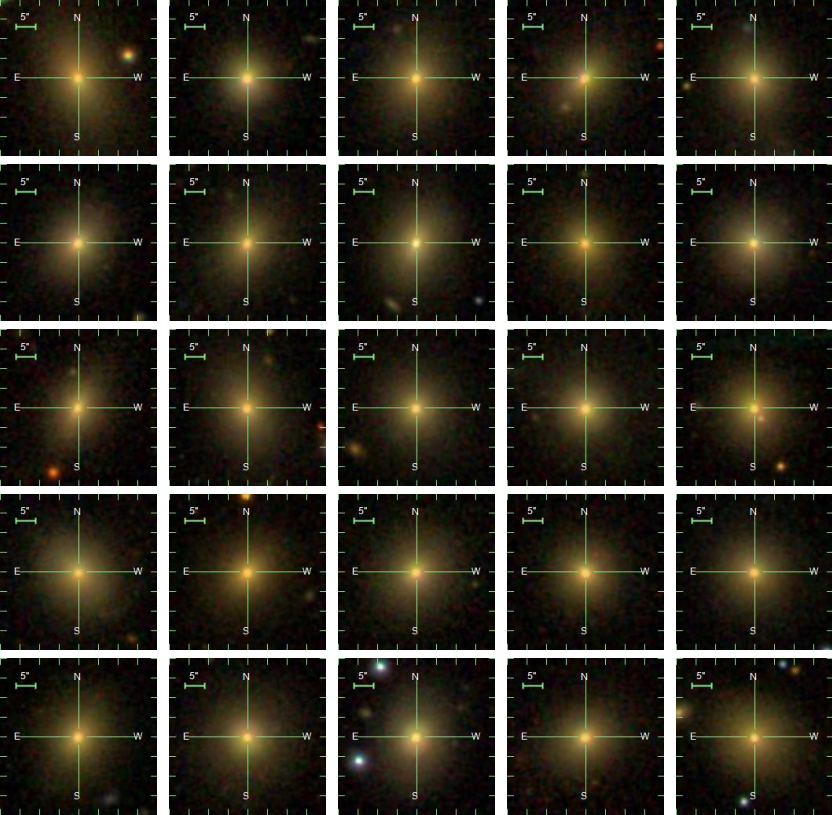}
    \caption{The closest 25 neighbours for non-AGN prototype\,\#10.}
    \label{fig:nns10}
\end{figure*}

\begin{figure*}
\centering
      \centering
            \includegraphics[width=4cm]{images/queries/quiescent3.jpeg}
        \includegraphics[width=0.5\linewidth]{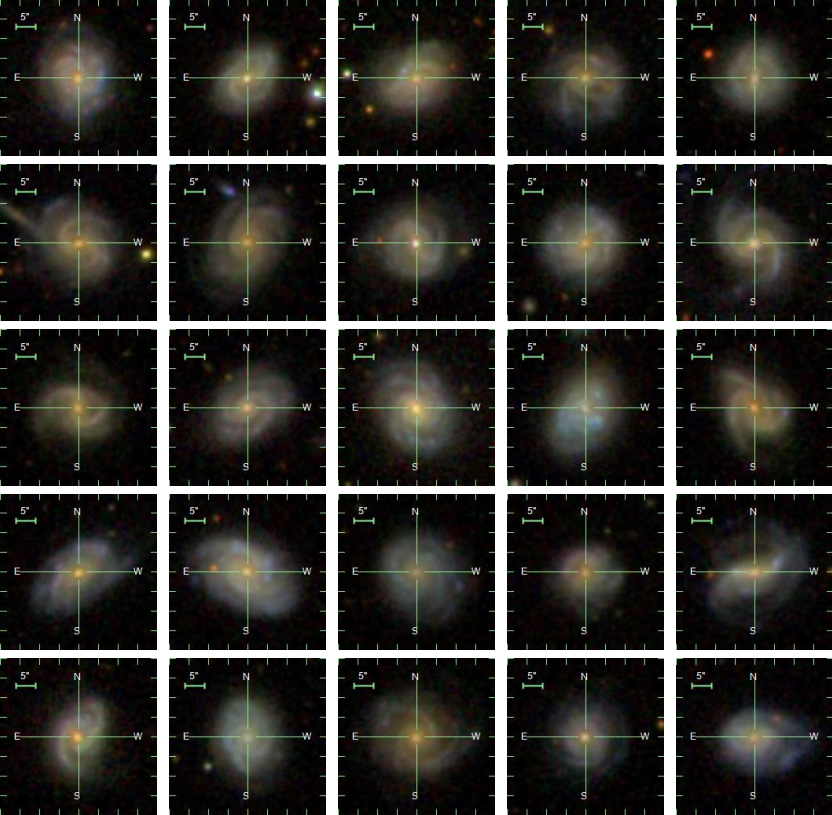}
    \caption{The closest 25 neighbours for non-AGN prototype\,\#11.}
    \label{fig:nns11}
\end{figure*}

\begin{figure*}
\centering
      \centering
            \includegraphics[width=4cm]{images/queries/quiescent4.jpeg}
        \includegraphics[width=0.5\linewidth]{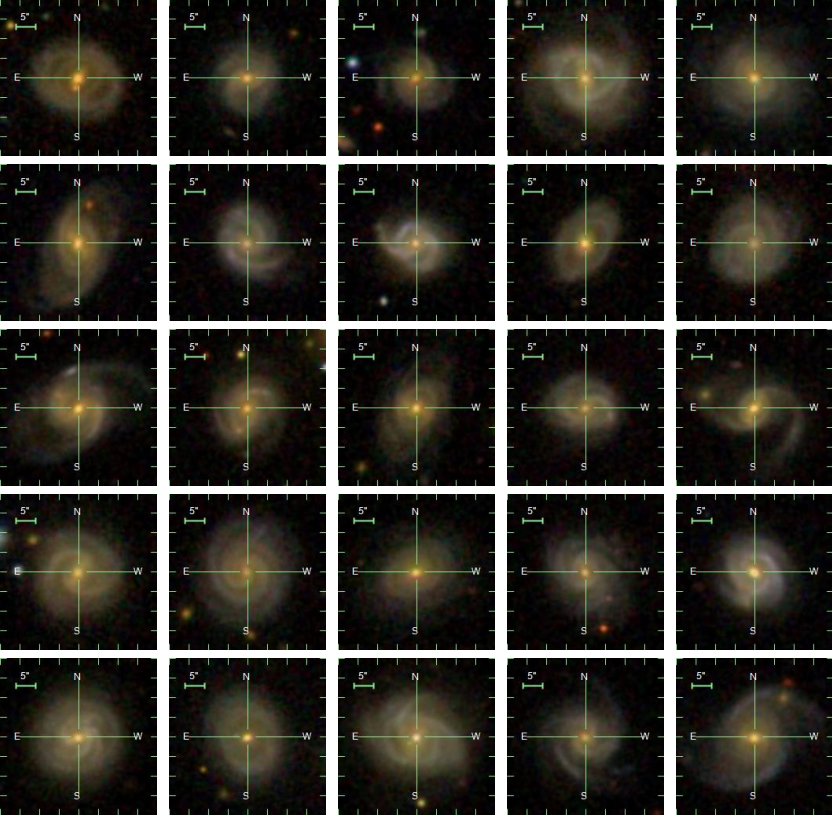}
    \caption{The closest 25 neighbours for non-AGN prototype\,\#12.}
    \label{fig:nns12}
\end{figure*}

\begin{figure*}
\centering
      \centering
            \includegraphics[width=4cm]{images/queries/quiescent5.jpeg}
        \includegraphics[width=0.5\linewidth]{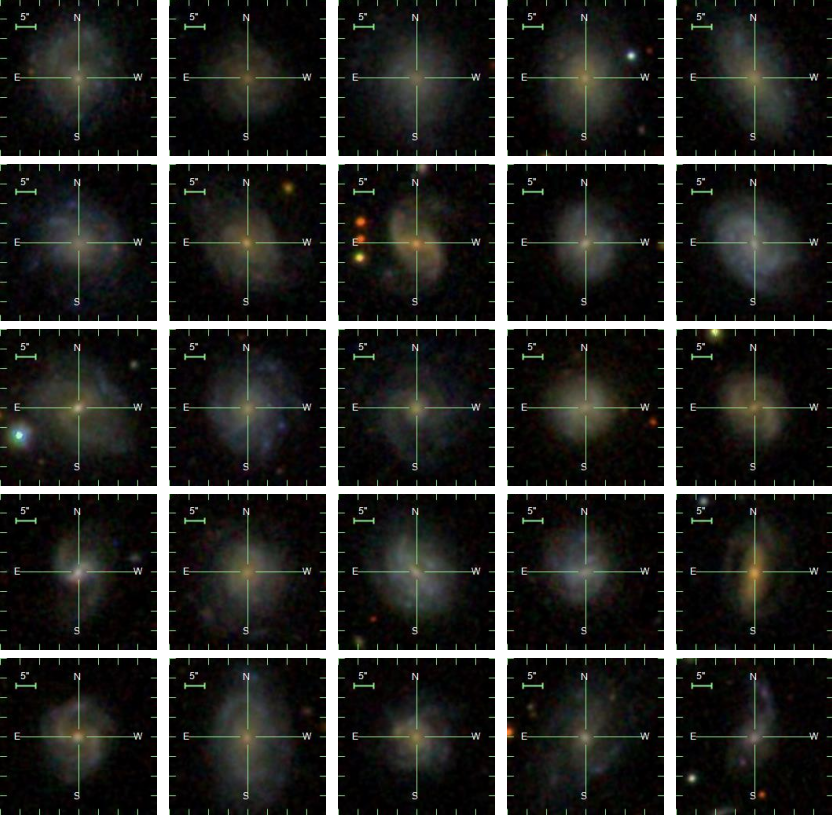}
    \caption{The closest 25 neighbours for non-AGN prototype\,\#13.}
    \label{fig:nns13}
\end{figure*}

%% file: main.bbl
\begin{thebibliography}{107}
\expandafter\ifx\csname natexlab\endcsname\relax\def\natexlab#1{#1}\fi

\bibitem[{{Agostino} \& {Salim}(2019)}]{Agostino2019}
{Agostino}, C.~J. \& {Salim}, S. 2019, \apj, 876, 12

\bibitem[{{Ahumada} {et~al.}(2020){Ahumada}, {Prieto}, {Almeida}, {Anders},
  {Anderson}, {Andrews}, {Anguiano}, {Arcodia}, {Armengaud}, {Aubert}, {Avila},
  {Avila-Reese}, {Badenes}, {Balland}, {Barger}, {Barrera-Ballesteros}, {Basu},
  {Bautista}, {Beaton}, {Beers}, {Benavides}, {Bender}, {Bernardi}, {Bershady},
  {Beutler}, {Bidin}, {Bird}, {Bizyaev}, {Blanc}, {Blanton}, {Boquien},
  {Borissova}, {Bovy}, {Brandt}, {Brinkmann}, {Brownstein}, {Bundy}, {Bureau},
  {Burgasser}, {Burtin}, {Cano-D{\'\i}az}, {Capasso}, {Cappellari}, {Carrera},
  {Chabanier}, {Chaplin}, {Chapman}, {Cherinka}, {Chiappini}, {Doohyun Choi},
  {Chojnowski}, {Chung}, {Clerc}, {Coffey}, {Comerford}, {Comparat}, {da
  Costa}, {Cousinou}, {Covey}, {Crane}, {Cunha}, {Ilha}, {Dai}, {Damsted},
  {Darling}, {Davidson}, {Davies}, {Dawson}, {De}, {de la Macorra}, {De Lee},
  {Queiroz}, {Deconto Machado}, {de la Torre}, {Dell'Agli}, {du Mas des
  Bourboux}, {Diamond-Stanic}, {Dillon}, {Donor}, {Drory}, {Duckworth},
  {Dwelly}, {Ebelke}, {Eftekharzadeh}, {Davis Eigenbrot}, {Elsworth},
  {Eracleous}, {Erfanianfar}, {Escoffier}, {Fan}, {Farr},
  {Fern{\'a}ndez-Trincado}, {Feuillet}, {Finoguenov}, {Fofie},
  {Fraser-McKelvie}, {Frinchaboy}, {Fromenteau}, {Fu}, {Galbany}, {Garcia},
  {Garc{\'\i}a-Hern{\'a}ndez}, {Oehmichen}, {Ge}, {Maia}, {Geisler}, {Gelfand},
  {Goddy}, {Gonzalez-Perez}, {Grabowski}, {Green}, {Grier}, {Guo}, {Guy},
  {Harding}, {Hasselquist}, {Hawken}, {Hayes}, {Hearty}, {Hekker}, {Hogg},
  {Holtzman}, {Horta}, {Hou}, {Hsieh}, {Huber}, {Hunt}, {Chitham}, {Imig},
  {Jaber}, {Angel}, {Johnson}, {Jones}, {J{\"o}nsson}, {Jullo}, {Kim},
  {Kinemuchi}, {Kirkpatrick}, {Kite}, {Klaene}, {Kneib}, {Kollmeier}, {Kong},
  {Kounkel}, {Krishnarao}, {Lacerna}, {Lan}, {Lane}, {Law}, {Le Goff}, {Leung},
  {Lewis}, {Li}, {Lian}, {Lin}, {Long}, {Longa-Pe{\~n}a}, {Lundgren}, {Lyke},
  {Ted Mackereth}, {MacLeod}, {Majewski}, {Manchado}, {Maraston}, {Martini},
  {Masseron}, {Masters}, {Mathur}, {McDermid}, {Merloni}, {Merrifield},
  {M{\'e}sz{\'a}ros}, {Miglio}, {Minniti}, {Minsley}, {Miyaji}, {Mohammad},
  {Mosser}, {Mueller}, {Muna}, {Mu{\~n}oz-Guti{\'e}rrez}, {Myers}, {Nadathur},
  {Nair}, {Nandra}, {do Nascimento}, {Nevin}, {Newman}, {Nidever}, {Nitschelm},
  {Noterdaeme}, {O'Connell}, {Olmstead}, {Oravetz}, {Oravetz}, {Osorio},
  {Pace}, {Padilla}, {Palanque-Delabrouille}, {Palicio}, {Pan}, {Pan},
  {Parker}, {Paviot}, {Peirani}, {Ram{\'r}ez}, {Penny}, {Percival},
  {Perez-Fournon}, {P{\'e}rez-R{\`a}fols}, {Petitjean}, {Pieri},
  {Pinsonneault}, {Poovelil}, {Povick}, {Prakash}, {Price-Whelan}, {Raddick},
  {Raichoor}, {Ray}, {Rembold}, {Rezaie}, {Riffel}, {Riffel}, {Rix}, {Robin},
  {Roman-Lopes}, {Rom{\'a}n-Z{\'u}{\~n}iga}, {Rose}, {Ross}, {Rossi},
  {Rowlands}, {Rubin}, {Salvato}, {S{\'a}nchez}, {S{\'a}nchez-Menguiano},
  {S{\'a}nchez-Gallego}, {Sayres}, {Schaefer}, {Schiavon}, {Schimoia},
  {Schlafly}, {Schlegel}, {Schneider}, {Schultheis}, {Schwope}, {Seo},
  {Serenelli}, {Shafieloo}, {Shamsi}, {Shao}, {Shen}, {Shetrone}, {Shirley},
  {Aguirre}, {Simon}, {Skrutskie}, {Slosar}, {Smethurst}, {Sobeck}, {Sodi},
  {Souto}, {Stark}, {Stassun}, {Steinmetz}, {Stello}, {Stermer},
  {Storchi-Bergmann}, {Streblyanska}, {Stringfellow}, {Stutz}, {Su{\'a}rez},
  {Sun}, {Taghizadeh-Popp}, {Talbot}, {Tayar}, {Thakar}, {Theriault}, {Thomas},
  {Thomas}, {Tinker}, {Tojeiro}, {Toledo}, {Tremonti}, {Troup}, {Tuttle},
  {Unda-Sanzana}, {Valentini}, {Vargas-Gonz{\'a}lez}, {Vargas-Maga{\~n}a},
  {V{\'a}zquez-Mata}, {Vivek}, {Wake}, {Wang}, {Weaver}, {Weijmans}, {Wild},
  {Wilson}, {Wilson}, {Wolthuis}, {Wood-Vasey}, {Yan}, {Yang}, {Y{\`e}che},
  {Zamora}, {Zarrouk}, {Zasowski}, {Zhang}, {Zhao}, {Zhao}, {Zheng}, {Zheng},
  {Zhu}, \& {Zou}}]{Ahumada2020}
{Ahumada}, R., {Prieto}, C.~A., {Almeida}, A., {et~al.} 2020, \apjs, 249, 3

\bibitem[{{Aihara} {et~al.}(2019){Aihara}, {AlSayyad}, {Ando}, {Armstrong},
  {Bosch}, {Egami}, {Furusawa}, {Furusawa}, {Goulding}, {Harikane}, {Hikage},
  {Ho}, {Hsieh}, {Huang}, {Ikeda}, {Imanishi}, {Ito}, {Iwata}, {Jaelani},
  {Kakuma}, {Kawana}, {Kikuta}, {Kobayashi}, {Koike}, {Komiyama}, {Li},
  {Liang}, {Lin}, {Luo}, {Lupton}, {Lust}, {MacArthur}, {Matsuoka}, {Mineo},
  {Miyatake}, {Miyazaki}, {More}, {Murata}, {Namiki}, {Nishizawa}, {Oguri},
  {Okabe}, {Okamoto}, {Okura}, {Ono}, {Onodera}, {Onoue}, {Osato}, {Ouchi},
  {Shibuya}, {Strauss}, {Sugiyama}, {Suto}, {Takada}, {Takagi}, {Takata},
  {Takita}, {Tanaka}, {Terai}, {Toba}, {Uchiyama}, {Utsumi}, {Wang}, {Wang}, \&
  {Yamada}}]{Aihara2019}
{Aihara}, H., {AlSayyad}, Y., {Ando}, M., {et~al.} 2019, \pasj, 71, 114

\bibitem[{{{\'A}lvarez-M{\'a}rquez} {et~al.}(2019){{\'A}lvarez-M{\'a}rquez},
  {Colina}, {Marques-Chaves}, {Ceverino}, {Alonso-Herrero}, {Caputi},
  {Garc{\'\i}a-Mar{\'\i}n}, {Labiano}, {Le F{\`e}vre}, {Norgaard-Nielsen},
  {{\"O}stlin}, {P{\'e}rez-Gonz{\'a}lez}, {Pye}, {Tikkanen}, {van der Werf},
  {Walter}, \& {Wright}}]{AlvarezMarquez2019}
{{\'A}lvarez-M{\'a}rquez}, J., {Colina}, L., {Marques-Chaves}, R., {et~al.}
  2019, \aap, 629, A9

\bibitem[{{Awang Iskandar} {et~al.}(2020){Awang Iskandar}, {Zijlstra},
  {McDonald}, {Abdullah}, {Fuller}, {Fauzi}, \& {Abdullah}}]{Awang2020}
{Awang Iskandar}, D. N.~F., {Zijlstra}, A.~A., {McDonald}, I., {et~al.} 2020,
  Galaxies, 8, 88

\bibitem[{Baldwin {et~al.}(1981)Baldwin, Phillips, \& Terlevich}]{Baldwin1981}
Baldwin, J.~A., Phillips, M.~M., \& Terlevich, R. 1981, Publications of the
  Astronomical Society of the Pacific, 93, 5

\bibitem[{{Baron}(2019)}]{Baron2019}
{Baron}, D. 2019, arXiv e-prints, arXiv:1904.07248

\bibitem[{{Baron} \& {Poznanski}(2017)}]{Baron2017}
{Baron}, D. \& {Poznanski}, D. 2017, \mnras, 465, 4530

\bibitem[{{Birchall} {et~al.}(2020){Birchall}, {Watson}, \&
  {Aird}}]{Birchall2020}
{Birchall}, K.~L., {Watson}, M.~G., \& {Aird}, J. 2020, \mnras, 492, 2268

\bibitem[{Bishop(2006)}]{Bishop2006}
Bishop, C.~M. 2006, Pattern Recognition and Machine Learning (Springer)

\bibitem[{{Blanton} {et~al.}(2017){Blanton}, {Bershady}, {Abolfathi},
  {Albareti}, {Allende Prieto}, {Almeida}, {Alonso-Garc{\'{\i}}a}, {Anders},
  {Anderson}, {Andrews}, \& et~al.}]{Blanton2017}
{Blanton}, M.~R., {Bershady}, M.~A., {Abolfathi}, B., {et~al.} 2017, \aj, 154,
  28

\bibitem[{Brandt \& Hasinger(2005)}]{Brandt2005}
Brandt, W. \& Hasinger, G. 2005, Annual Review of Astronomy and Astrophysics,
  43, 827

\bibitem[{{Brescia} {et~al.}(2013){Brescia}, {Cavuoti}, {D'Abrusco}, {Longo},
  \& {Mercurio}}]{Brescia2013}
{Brescia}, M., {Cavuoti}, S., {D'Abrusco}, R., {Longo}, G., \& {Mercurio}, A.
  2013, \apj, 772, 140

\bibitem[{{Brescia} {et~al.}(2019){Brescia}, {Salvato}, {Cavuoti}, {Ananna},
  {Riccio}, {LaMassa}, {Urry}, \& {Longo}}]{Brescia2019}
{Brescia}, M., {Salvato}, M., {Cavuoti}, S., {et~al.} 2019, \mnras, 489, 663

\bibitem[{Brinchmann {et~al.}(2004)Brinchmann, Charlot, White, Tremonti,
  Kauffmann, Heckman, \& Brinkmann}]{Brinchmann2004}
Brinchmann, J., Charlot, S., White, S., {et~al.} 2004, Monthly Notices of the
  Royal Astronomical Society, 351, 1151

\bibitem[{{Byrne} {et~al.}(2019){Byrne}, {Christensen}, {Tsekitsidis},
  {Brooks}, \& {Quinn}}]{Byrne2019}
{Byrne}, L., {Christensen}, C., {Tsekitsidis}, M., {Brooks}, A., \& {Quinn}, T.
  2019, \apj, 871, 213

\bibitem[{{Castro-Ginard, A.} {et~al.}(2018){Castro-Ginard, A.}, {Jordi, C.},
  {Luri, X.}, {Julbe, F.}, {Morvan, M.}, {Balaguer-N\'u\~nez, L.}, \&
  {Cantat-Gaudin, T.}}]{CastroGinard2018}
{Castro-Ginard, A.}, {Jordi, C.}, {Luri, X.}, {et~al.} 2018, A\&A, 618, A59

\bibitem[{Cavuoti {et~al.}(2013)Cavuoti, Brescia, D'Abrusco, Longo, \&
  Paolillo}]{Cavuoti2013}
Cavuoti, S., Brescia, M., D'Abrusco, R., Longo, G., \& Paolillo, M. 2013,
  Monthly Notices of the Royal Astronomical Society, 437, 968

\bibitem[{{Chang} {et~al.}(2021){Chang}, {Hsieh}, {Wang}, {Lin}, {Lim}, {Toba},
  {Zhong}, \& {Chang}}]{Chang2021}
{Chang}, Y.-Y., {Hsieh}, B.-C., {Wang}, W.-H., {et~al.} 2021, \apj, 920, 68

\bibitem[{{Chen}(2021)}]{Chen2021}
{Chen}, Y.~C. 2021, \apjs, 256, 34

\bibitem[{Chung {et~al.}(2014)Chung, Kochanek, Assef, Brown, Stern, Jannuzi,
  Gonzalez, Hickox, \& Moustakas}]{Chung2014}
Chung, S.~M., Kochanek, C.~S., Assef, R., {et~al.} 2014, The Astrophysical
  Journal, 790, 54

\bibitem[{{Dark Energy Survey Collaboration} {et~al.}(2016){Dark Energy Survey
  Collaboration}, {Abbott}, {Abdalla}, {Aleksi{\'c}}, {Allam}, {Amara},
  {Bacon}, {Balbinot}, {Banerji}, {Bechtol}, {Benoit-L{\'e}vy}, {Bernstein},
  {Bertin}, {Blazek}, {Bonnett}, {Bridle}, {Brooks}, {Brunner}, {Buckley-Geer},
  {Burke}, {Caminha}, {Capozzi}, {Carlsen}, {Carnero-Rosell}, {Carollo},
  {Carrasco-Kind}, {Carretero}, {Castander}, {Clerkin}, {Collett}, {Conselice},
  {Crocce}, {Cunha}, {D'Andrea}, {da Costa}, {Davis}, {Desai}, {Diehl},
  {Dietrich}, {Dodelson}, {Doel}, {Drlica-Wagner}, {Estrada}, {Etherington},
  {Evrard}, {Fabbri}, {Finley}, {Flaugher}, {Foley}, {Fosalba}, {Frieman},
  {Garc{\'\i}a-Bellido}, {Gaztanaga}, {Gerdes}, {Giannantonio}, {Goldstein},
  {Gruen}, {Gruendl}, {Guarnieri}, {Gutierrez}, {Hartley}, {Honscheid}, {Jain},
  {James}, {Jeltema}, {Jouvel}, {Kessler}, {King}, {Kirk}, {Kron}, {Kuehn},
  {Kuropatkin}, {Lahav}, {Li}, {Lima}, {Lin}, {Maia}, {Makler}, {Manera},
  {Maraston}, {Marshall}, {Martini}, {McMahon}, {Melchior}, {Merson}, {Miller},
  {Miquel}, {Mohr}, {Morice-Atkinson}, {Naidoo}, {Neilsen}, {Nichol}, {Nord},
  {Ogando}, {Ostrovski}, {Palmese}, {Papadopoulos}, {Peiris}, {Peoples},
  {Percival}, {Plazas}, {Reed}, {Refregier}, {Romer}, {Roodman}, {Ross},
  {Rozo}, {Rykoff}, {Sadeh}, {Sako}, {S{\'a}nchez}, {Sanchez}, {Santiago},
  {Scarpine}, {Schubnell}, {Sevilla-Noarbe}, {Sheldon}, {Smith}, {Smith},
  {Soares-Santos}, {Sobreira}, {Soumagnac}, {Suchyta}, {Sullivan}, {Swanson},
  {Tarle}, {Thaler}, {Thomas}, {Thomas}, {Tucker}, {Vieira}, {Vikram},
  {Walker}, {Wechsler}, {Weller}, {Wester}, {Whiteway}, {Wilcox}, {Yanny},
  {Zhang}, \& {Zuntz}}]{Abbott2016}
{Dark Energy Survey Collaboration}, {Abbott}, T., {Abdalla}, F.~B., {et~al.}
  2016, \mnras, 460, 1270

\bibitem[{De~Cicco {et~al.}(2021)De~Cicco, Bauer, Paolillo, Cavuoti,
  Sánchez-Sáez, Brandt, Pignata, Vaccari, \& Radovich}]{DeCicco2021}
De~Cicco, D., Bauer, F., Paolillo, M., {et~al.} 2021, Astronomy and
  Astrophysics, 645

\bibitem[{{de Jong} {et~al.}(2015){de Jong}, {Verdoes Kleijn}, {Boxhoorn},
  {Buddelmeijer}, {Capaccioli}, {Getman}, {Grado}, {Helmich}, {Huang},
  {Irisarri}, {Kuijken}, {La Barbera}, {McFarland}, {Napolitano}, {Radovich},
  {Sikkema}, {Valentijn}, {Begeman}, {Brescia}, {Cavuoti}, {Choi}, {Cordes},
  {Covone}, {Dall'Ora}, {Hildebrandt}, {Longo}, {Nakajima}, {Paolillo},
  {Puddu}, {Rifatto}, {Tortora}, {van Uitert}, {Buddendiek},
  {Harnois-D{\'e}raps}, {Erben}, {Eriksen}, {Heymans}, {Hoekstra}, {Joachimi},
  {Kitching}, {Klaes}, {Koopmans}, {K{\"o}hlinger}, {Roy}, {Sif{\'o}n},
  {Schneider}, {Sutherland}, {Viola}, \& {Vriend}}]{deJong2015}
{de Jong}, J. T.~A., {Verdoes Kleijn}, G.~A., {Boxhoorn}, D.~R., {et~al.} 2015,
  \aap, 582, A62

\bibitem[{{Delli Veneri} {et~al.}(2019){Delli Veneri}, {Cavuoti}, {Brescia},
  {Longo}, \& {Riccio}}]{DelliVeneri2019}
{Delli Veneri}, M., {Cavuoti}, S., {Brescia}, M., {Longo}, G., \& {Riccio}, G.
  2019, \mnras, 486, 1377

\bibitem[{Deng {et~al.}(2009)Deng, Dong, Socher, Li, Li, \&
  Fei-Fei}]{deng2009imagenet}
Deng, J., Dong, W., Socher, R., {et~al.} 2009, in 2009 IEEE conference on
  computer vision and pattern recognition, Ieee, 248--255

\bibitem[{Ding {et~al.}(2019)Ding, Sohn, Kawczynski, Trivedi, Harnish, Jenkins,
  Lituiev, Copeland, Aboian, Mari~Aparici, {et~al.}}]{ding2019deep}
Ding, Y., Sohn, J.~H., Kawczynski, M.~G., {et~al.} 2019, Radiology, 290, 456

\bibitem[{{D'Isanto} {et~al.}(2016){D'Isanto}, {Cavuoti}, {Brescia}, {Donalek},
  {Longo}, {Riccio}, \& {Djorgovski}}]{Disanto2016}
{D'Isanto}, A., {Cavuoti}, S., {Brescia}, M., {et~al.} 2016, \mnras, 457, 3119

\bibitem[{{D'Isanto} {et~al.}(2018){D'Isanto}, {Cavuoti}, {Gieseke}, \&
  {Polsterer}}]{Disanto2018b}
{D'Isanto}, A., {Cavuoti}, S., {Gieseke}, F., \& {Polsterer}, K.~L. 2018, \aap,
  616, A97

\bibitem[{{D'Isanto} \& {Polsterer}(2018)}]{Disanto2018}
{D'Isanto}, A. \& {Polsterer}, K.~L. 2018, \aap, 609, A111

\bibitem[{{Eisenstein} {et~al.}(2011){Eisenstein}, {Weinberg}, {Agol},
  {Aihara}, {Allende Prieto}, {Anderson}, {Arns}, {Aubourg}, {Bailey},
  {Balbinot}, \& et~al.}]{Eisenstein2011}
{Eisenstein}, D.~J., {Weinberg}, D.~H., {Agol}, E., {et~al.} 2011, \aj, 142, 72

\bibitem[{Esteva {et~al.}(2017)Esteva, Kuprel, Novoa, Ko, Swetter, Blau, \&
  Thrun}]{esteva2017dermatologist}
Esteva, A., Kuprel, B., Novoa, R.~A., {et~al.} 2017, nature, 542, 115

\bibitem[{{Euclid Collaboration} {et~al.}(2020){Euclid Collaboration},
  {Desprez}, {Paltani}, {Coupon}, {Almosallam}, {Alvarez-Ayllon}, {Amaro},
  {Brescia}, {Brodwin}, {Cavuoti}, {De Vicente-Albendea}, {Fotopoulou},
  {Hatfield}, {Hartley}, {Ilbert}, {Jarvis}, {Longo}, {Rau}, {Saha}, {Speagle},
  {Tramacere}, {Castellano}, {Dubath}, {Galametz}, {Kuemmel}, {Laigle},
  {Merlin}, {Mohr}, {Pilo}, {Salvato}, {Andreon}, {Auricchio}, {Baccigalupi},
  {Balaguera-Antol{\'\i}nez}, {Baldi}, {Bardelli}, {Bender}, {Biviano},
  {Bodendorf}, {Bonino}, {Bozzo}, {Branchini}, {Brinchmann}, {Burigana},
  {Cabanac}, {Camera}, {Capobianco}, {Cappi}, {Carbone}, {Carretero},
  {Carvalho}, {Casas}, {Casas}, {Castander}, {Castignani}, {Cimatti},
  {Cledassou}, {Colodro-Conde}, {Congedo}, {Conselice}, {Conversi}, {Copin},
  {Corcione}, {Courtois}, {Cuby}, {Da Silva}, {de la Torre}, {Degaudenzi}, {Di
  Ferdinando}, {Douspis}, {Duncan}, {Dupac}, {Ealet}, {Fabbian}, {Fabricius},
  {Farrens}, {Ferreira}, {Finelli}, {Fosalba}, {Fourmanoit}, {Frailis},
  {Franceschi}, {Fumana}, {Galeotta}, {Garilli}, {Gillard}, {Gillis},
  {Giocoli}, {Gozaliasl}, {Graci{\'a}-Carpio}, {Grupp}, {Guzzo}, {Hailey},
  {Haugan}, {Holmes}, {Hormuth}, {Humphrey}, {Jahnke}, {Keihanen}, {Kermiche},
  {Kilbinger}, {Kirkpatrick}, {Kitching}, {Kohley}, {Kubik}, {Kunz},
  {Kurki-Suonio}, {Ligori}, {Lilje}, {Lloro}, {Maino}, {Maiorano}, {Marggraf},
  {Markovic}, {Martinet}, {Marulli}, {Massey}, {Maturi}, {Mauri},
  {Maurogordato}, {Medinaceli}, {Mei}, {Meneghetti}, {Metcalf}, {Meylan},
  {Moresco}, {Moscardini}, {Munari}, {Niemi}, {Padilla}, {Pasian}, {Patrizii},
  {Pettorino}, {Pires}, {Polenta}, {Poncet}, {Popa}, {Potter}, {Pozzetti},
  {Raison}, {Renzi}, {Rhodes}, {Riccio}, {Rossetti}, {Saglia}, {Sapone},
  {Schneider}, {Scottez}, {Secroun}, {Serrano}, {Sirignano}, {Sirri}, {Stanco},
  {Stern}, {Sureau}, {Tallada Cresp{\'\i}}, {Tavagnacco}, {Taylor}, {Tenti},
  {Tereno}, {Toledo-Moreo}, {Torradeflot}, {Valenziano}, {Valiviita},
  {Vassallo}, {Viel}, {Wang}, {Welikala}, {Whittaker}, {Zacchei}, {Zamorani},
  {Zoubian}, \& {Zucca}}]{Desprez2020}
{Euclid Collaboration}, {Desprez}, G., {Paltani}, S., {et~al.} 2020, \aap, 644,
  A31

\bibitem[{{Fabian}(2012)}]{Fabian2012}
{Fabian}, A.~C. 2012, \araa, 50, 455

\bibitem[{Faisst {et~al.}(2019)Faisst, Prakash, Capak, \& Lee}]{Faisst2019}
Faisst, A., Prakash, A., Capak, P., \& Lee, B. 2019, Astrophysical Journal
  Letters, 881

\bibitem[{{Falocco} {et~al.}(2022){Falocco}, {Carrera}, \&
  {Larsson}}]{Falocco2022}
{Falocco}, S., {Carrera}, F.~J., \& {Larsson}, J. 2022, \mnras, 510, 161

\bibitem[{{Fluke} \& {Jacobs}(2020)}]{Fluke2020}
{Fluke}, C.~J. \& {Jacobs}, C. 2020, WIREs Data Mining and Knowledge Discovery,
  10, e1349

\bibitem[{{Fotopoulou} \& {Paltani}(2018)}]{Fotopoulou2018}
{Fotopoulou}, S. \& {Paltani}, S. 2018, \aap, 619, A14

\bibitem[{{Frontera-Pons, J.} {et~al.}(2017){Frontera-Pons, J.}, {Sureau, F.},
  {Bobin, J.}, \& {Le Floc\'{}h, E.}}]{FronteraPons2017}
{Frontera-Pons, J.}, {Sureau, F.}, {Bobin, J.}, \& {Le Floc\'{}h, E.} 2017,
  A\&A, 603, A60

\bibitem[{Goebel {et~al.}(2018)Goebel, Chander, Holzinger, Lecue, Akata,
  Stumpf, Kieseberg, \& Holzinger}]{Goebel2018}
Goebel, R., Chander, A., Holzinger, K., {et~al.} 2018, in Machine Learning and
  Knowledge Extraction, ed. A.~Holzinger, P.~Kieseberg, A.~M. Tjoa, \&
  E.~Weippl (Cham: Springer International Publishing), 295--303

\bibitem[{Goodfellow {et~al.}(2016)Goodfellow, Bengio, \&
  Courville}]{Goodfellow-et-al-2016}
Goodfellow, I., Bengio, Y., \& Courville, A. 2016, Deep Learning (MIT Press),
  \url{http://www.deeplearningbook.org}

\bibitem[{{Goudfrooij}(1995)}]{Goudfrooij1995}
{Goudfrooij}, P. 1995, \pasp, 107, 502

\bibitem[{{Green} {et~al.}(2012){Green}, {Schechter}, {Baltay}, {Bean},
  {Bennett}, {Brown}, {Conselice}, {Donahue}, {Fan}, {Gaudi}, {Hirata},
  {Kalirai}, {Lauer}, {Nichol}, {Padmanabhan}, {Perlmutter}, {Rauscher},
  {Rhodes}, {Roellig}, {Stern}, {Sumi}, {Tanner}, {Wang}, {Weinberg}, {Wright},
  {Gehrels}, {Sambruna}, {Traub}, {Anderson}, {Cook}, {Garnavich},
  {Hillenbrand}, {Ivezic}, {Kerins}, {Lunine}, {McDonald}, {Penny}, {Phillips},
  {Rieke}, {Riess}, {van der Marel}, {Barry}, {Cheng}, {Content}, {Cutri},
  {Goullioud}, {Grady}, {Helou}, {Jackson}, {Kruk}, {Melton}, {Peddie},
  {Rioux}, \& {Seiffert}}]{Green2012}
{Green}, J., {Schechter}, P., {Baltay}, C., {et~al.} 2012, arXiv e-prints,
  arXiv:1208.4012

\bibitem[{{H{\"a}ring} \& {Rix}(2004)}]{Haring2004}
{H{\"a}ring}, N. \& {Rix}, H.-W. 2004, \apjl, 604, L89

\bibitem[{Hastie {et~al.}(2009)Hastie, Tibshirani, Friedman, \&
  Friedman}]{hastie2009elements}
Hastie, T., Tibshirani, R., Friedman, J.~H., \& Friedman, J.~H. 2009, The
  elements of statistical learning: data mining, inference, and prediction,
  Vol.~2 (Springer)

\bibitem[{{Heckman}(1980)}]{Heckman1980}
{Heckman}, T.~M. 1980, \aap, 500, 187

\bibitem[{{Heckman} \& {Best}(2014)}]{Heckman2014}
{Heckman}, T.~M. \& {Best}, P.~N. 2014, \araa, 52, 589

\bibitem[{{Heinis} {et~al.}(2016){Heinis}, {Gezari}, {Kumar}, {Burgett},
  {Flewelling}, {Huber}, {Kaiser}, {Wainscoat}, \& {Waters}}]{Heinis2016}
{Heinis}, S., {Gezari}, S., {Kumar}, S., {et~al.} 2016, \apj, 826, 62

\bibitem[{{Hickox} \& {Alexander}(2018)}]{Hickox2018}
{Hickox}, R.~C. \& {Alexander}, D.~M. 2018, \araa, 56, 625

\bibitem[{{Hirashita} {et~al.}(2015){Hirashita}, {Nozawa}, {Villaume}, \&
  {Srinivasan}}]{Hirashita2015}
{Hirashita}, H., {Nozawa}, T., {Villaume}, A., \& {Srinivasan}, S. 2015,
  \mnras, 454, 1620

\bibitem[{{Ivezi{\'c}} {et~al.}(2019){Ivezi{\'c}}, {Kahn}, {Tyson}, {Abel},
  {Acosta}, {Allsman}, {Alonso}, {AlSayyad}, {Anderson}, {Andrew}, \&
  et~al.}]{Ivezic2019}
{Ivezi{\'c}}, {\v Z}., {Kahn}, S.~M., {Tyson}, J.~A., {et~al.} 2019, \apj, 873,
  111

\bibitem[{{Ji} {et~al.}(2022){Ji}, {Giavalisco}, {Kirkpatrick}, {Kocevski},
  {Daddi}, {Delvecchio}, \& {Hatcher}}]{Ji2022}
{Ji}, Z., {Giavalisco}, M., {Kirkpatrick}, A., {et~al.} 2022, \apj, 925, 74

\bibitem[{{Kauffmann} {et~al.}(2003){Kauffmann}, {Heckman}, {Tremonti},
  {Brinchmann}, {Charlot}, {White}, {Ridgway}, {Brinkmann}, {Fukugita}, {Hall},
  {Ivezi{\'c}}, {Richards}, \& {Schneider}}]{Kauffmann:03c}
{Kauffmann}, G., {Heckman}, T.~M., {Tremonti}, C., {et~al.} 2003, \mnras, 346,
  1055

\bibitem[{{Kewley} {et~al.}(2006){Kewley}, {Groves}, {Kauffmann}, \&
  {Heckman}}]{Kewley:06}
{Kewley}, L.~J., {Groves}, B., {Kauffmann}, G., \& {Heckman}, T. 2006, \mnras,
  372, 961

\bibitem[{{Kim} {et~al.}(2011){Kim}, {Protopapas}, {Byun}, {Alcock}, {Khardon},
  \& {Trichas}}]{Kim2011}
{Kim}, D.-W., {Protopapas}, P., {Byun}, Y.-I., {et~al.} 2011, \apj, 735, 68

\bibitem[{{Kim} {et~al.}(2021){Kim}, {Bae}, {Chung}, {Park}, {Kim}, \&
  {Kim}}]{Kim2021tl}
{Kim}, Y.~J., {Bae}, J.~P., {Chung}, J.-W., {et~al.} 2021, Scientific Reports,
  11, 3605

\bibitem[{{Kinson} {et~al.}(2021){Kinson}, {Oliveira}, \& {van
  Loon}}]{Kinson2021}
{Kinson}, D.~A., {Oliveira}, J.~M., \& {van Loon}, J.~T. 2021, \mnras, 507,
  5106

\bibitem[{{Kormendy} \& {Ho}(2013)}]{Kormendy2013}
{Kormendy}, J. \& {Ho}, L.~C. 2013, \araa, 51, 511

\bibitem[{Lecun {et~al.}(1998)Lecun, Bottou, Bengio, \& Haffner}]{Lecun1998}
Lecun, Y., Bottou, L., Bengio, Y., \& Haffner, P. 1998, Proceedings of the
  IEEE, 86, 2278

\bibitem[{{Lianou} {et~al.}(2019){Lianou}, {Barmby}, {Mosenkov}, {Lehnert}, \&
  {Karczewski}}]{Lianou2019}
{Lianou}, S., {Barmby}, P., {Mosenkov}, A.~A., {Lehnert}, M., \& {Karczewski},
  O. 2019, \aap, 631, A38

\bibitem[{{Longo} {et~al.}(2019){Longo}, {Mer{\'e}nyi}, \&
  {Ti{\v{n}}o}}]{Longo2019}
{Longo}, G., {Mer{\'e}nyi}, E., \& {Ti{\v{n}}o}, P. 2019, \pasp, 131, 100101

\bibitem[{{Lutz} {et~al.}(2010){Lutz}, {Mainieri}, {Rafferty}, {Shao},
  {Hasinger}, {Wei{\ss}}, {Walter}, {Smail}, {Alexander}, {Brandt}, {Chapman},
  {Coppin}, {F{\"o}rster Schreiber}, {Gawiser}, {Genzel}, {Greve}, {Ivison},
  {Koekemoer}, {Kurczynski}, {Menten}, {Nordon}, {Popesso}, {Schinnerer},
  {Silverman}, {Wardlow}, \& {Xue}}]{Lutz2010}
{Lutz}, D., {Mainieri}, V., {Rafferty}, D., {et~al.} 2010, \apj, 712, 1287

\bibitem[{{Magnier} {et~al.}(2020){Magnier}, {Schlafly}, {Finkbeiner}, {Tonry},
  {Goldman}, {R{\"o}ser}, {Schilbach}, {Casertano}, {Chambers}, {Flewelling},
  {Huber}, {Price}, {Sweeney}, {Waters}, {Denneau}, {Draper}, {Hodapp},
  {Jedicke}, {Kaiser}, {Kudritzki}, {Metcalfe}, {Stubbs}, \&
  {Wainscoat}}]{Magnier2020}
{Magnier}, E.~A., {Schlafly}, E.~F., {Finkbeiner}, D.~P., {et~al.} 2020, \apjs,
  251, 6

\bibitem[{{Martinazzo} {et~al.}(2020){Martinazzo}, {Espadoto}, \&
  {Hirata}}]{Martinazzo2020}
{Martinazzo}, A., {Espadoto}, M., \& {Hirata}, N. S.~T. 2020, arXiv e-prints,
  arXiv:2004.11336

\bibitem[{{Masters} {et~al.}(2015){Masters}, {Capak}, {Stern}, {Ilbert},
  {Salvato}, {Schmidt}, {Longo}, {Rhodes}, {Paltani}, {Mobasher}, {Hoekstra},
  {Hildebrandt}, {Coupon}, {Steinhardt}, {Speagle}, {Faisst}, {Kalinich},
  {Brodwin}, {Brescia}, \& {Cavuoti}}]{Masters2015}
{Masters}, D., {Capak}, P., {Stern}, D., {et~al.} 2015, \apj, 813, 53

\bibitem[{{Mateos} {et~al.}(2012){Mateos}, {Alonso-Herrero}, {Carrera},
  {Blain}, {Watson}, {Barcons}, {Braito}, {Severgnini}, {Donley}, \&
  {Stern}}]{Mateos2012}
{Mateos}, S., {Alonso-Herrero}, A., {Carrera}, F.~J., {et~al.} 2012, \mnras,
  426, 3271

\bibitem[{{McConnell} \& {Ma}(2013)}]{McConnell2013}
{McConnell}, N.~J. \& {Ma}, C.-P. 2013, \apj, 764, 184

\bibitem[{McCulloch \& Pitts(1943)}]{McCulloch1943}
McCulloch, W.~S. \& Pitts, W. 1943, Bulletin of Mathematical Biophysics, 5, 115

\bibitem[{{Mendez} {et~al.}(2013){Mendez}, {Coil}, {Aird}, {Diamond-Stanic},
  {Moustakas}, {Blanton}, {Cool}, {Eisenstein}, {Wong}, \& {Zhu}}]{Mendez2013}
{Mendez}, A.~J., {Coil}, A.~L., {Aird}, J., {et~al.} 2013, \apj, 770, 40

\bibitem[{Menegola {et~al.}(2017)Menegola, Fornaciali, Pires, Bittencourt,
  Avila, \& Valle}]{menegola2017knowledge}
Menegola, A., Fornaciali, M., Pires, R., {et~al.} 2017, in 2017 IEEE 14th
  international symposium on biomedical imaging (ISBI 2017), IEEE, 297--300

\bibitem[{Merloni(2016)}]{Merloni2016}
Merloni, A. 2016, in Astrophysical Black Holes, ed. F.~Haardt, V.~Gorini,
  U.~Moschella, A.~Treves, \& M.~Colpi (Cham: Springer International
  Publishing), 101--143

\bibitem[{Mislis {et~al.}(2018)Mislis, Pyrzas, \& Alsubai}]{Mislis2018}
Mislis, D., Pyrzas, S., \& Alsubai, K.~A. 2018, Monthly Notices of the Royal
  Astronomical Society, 481, 1624

\bibitem[{{Mullaney} {et~al.}(2012){Mullaney}, {Pannella}, {Daddi}, {Alexand
  er}, {Elbaz}, {Hickox}, {Bournaud}, {Altieri}, {Aussel}, {Coia},
  {Dannerbauer}, {Dasyra}, {Dickinson}, {Hwang}, {Kartaltepe}, {Leiton},
  {Magdis}, {Magnelli}, {Popesso}, {Valtchanov}, {Bauer}, {Brand t}, {Del
  Moro}, {Hanish}, {Ivison}, {Juneau}, {Luo}, {Lutz}, {Sargent}, {Scott}, \&
  {Xue}}]{Mullaney2012a}
{Mullaney}, J.~R., {Pannella}, M., {Daddi}, E., {et~al.} 2012, \mnras, 419, 95

\bibitem[{{Ofman} {et~al.}(2022){Ofman}, {Averbuch}, {Shliselberg}, {Benaun},
  {Segev}, \& {Rissman}}]{Ofman2022}
{Ofman}, L., {Averbuch}, A., {Shliselberg}, A., {et~al.} 2022, \na, 91, 101693

\bibitem[{Palaversa {et~al.}(2013)Palaversa, Ivezi{\'{c}}, Eyer, Ru{\v{z}}djak,
  Sudar, Galin, Kroflin, Mesari{\'{c}}, Munk, Vrbanec, Bo{\v{z}}i{\'{c}},
  Loebman, Sesar, Rimoldini, Hunt-Walker, VanderPlas, Westman, Stuart, Becker,
  Srdo{\v{c}}, Wozniak, \& Oluseyi}]{Palaversa2013}
Palaversa, L., Ivezi{\'{c}}, {\v{Z}}., Eyer, L., {et~al.} 2013, The
  Astronomical Journal, 146, 101

\bibitem[{Pan \& Yang(2009)}]{pan2009survey}
Pan, S.~J. \& Yang, Q. 2009, IEEE Transactions on knowledge and data
  engineering, 22, 1345

\bibitem[{{Prima} \& {Bouhorma}(2020)}]{Prima2020}
{Prima}, B. \& {Bouhorma}, M. 2020, ISPRS - International Archives of the
  Photogrammetry, Remote Sensing and Spatial Information Sciences, 4443, 343

\bibitem[{{Razim} {et~al.}(2021){Razim}, {Cavuoti}, {Brescia}, {Riccio},
  {Salvato}, \& {Longo}}]{Razim2021}
{Razim}, O., {Cavuoti}, S., {Brescia}, M., {et~al.} 2021, \mnras, 507, 5034

\bibitem[{{Richards} {et~al.}(2005){Richards}, {Croom}, {Anderson},
  {Bland-Hawthorn}, {Boyle}, {De Propris}, {Drinkwater}, {Fan}, {Gunn},
  {Ivezi{\'c}}, {Jester}, {Loveday}, {Meiksin}, {Miller}, {Myers}, {Nichol},
  {Outram}, {Pimbblet}, {Roseboom}, {Ross}, {Schneider}, {Shanks}, {Sharp},
  {Stoughton}, {Strauss}, {Szalay}, {Vanden Berk}, \& {York}}]{Richards2005}
{Richards}, G.~T., {Croom}, S.~M., {Anderson}, S.~F., {et~al.} 2005, \mnras,
  360, 839

\bibitem[{{Richards} {et~al.}(2002){Richards}, {Fan}, {Newberg}, {Strauss},
  {Vanden Berk}, {Schneider}, {Yanny}, {Boucher}, {Burles}, {Frieman}, {Gunn},
  {Hall}, {Ivezi{\'c}}, {Kent}, {Loveday}, {Lupton}, {Rockosi}, {Schlegel},
  {Stoughton}, {SubbaRao}, \& {York}}]{Richards2002}
{Richards}, G.~T., {Fan}, X., {Newberg}, H.~J., {et~al.} 2002, \aj, 123, 2945

\bibitem[{{Rosario} {et~al.}(2013){Rosario}, {Santini}, {Lutz}, {Netzer},
  {Bauer}, {Berta}, {Magnelli}, {Popesso}, {Alexander}, {Brandt}, {Genzel},
  {Maiolino}, {Mullaney}, {Nordon}, {Saintonge}, {Tacconi}, \&
  {Wuyts}}]{Rosario2013}
{Rosario}, D.~J., {Santini}, P., {Lutz}, D., {et~al.} 2013, \apj, 771, 63

\bibitem[{{Rosen} {et~al.}(2016){Rosen}, {Webb}, {Watson}, {Ballet}, {Barret},
  {Braito}, {Carrera}, {Ceballos}, {Coriat}, {Della Ceca}, {Denkinson},
  {Esquej}, {Farrell}, {Freyberg}, {Gris{\'e}}, {Guillout}, {Heil},
  {Koliopanos}, {Law-Green}, {Lamer}, {Lin}, {Martino}, {Michel}, {Motch},
  {Nebot Gomez-Moran}, {Page}, {Page}, {Page}, {Pakull}, {Pye}, {Read},
  {Rodriguez}, {Sakano}, {Saxton}, {Schwope}, {Scott}, {Sturm}, {Traulsen},
  {Yershov}, \& {Zolotukhin}}]{Rosen2016}
{Rosen}, S.~R., {Webb}, N.~A., {Watson}, M.~G., {et~al.} 2016, \aap, 590, A1

\bibitem[{Rosenblatt(1958)}]{Rosenblatt1958}
Rosenblatt, F. 1958, Psychological Review, 65

\bibitem[{{Scaramella} {et~al.}(2021){Scaramella}, {Amiaux}, {Mellier},
  {Burigana}, {Carvalho}, {Cuillandre}, {Da Silva}, {Derosa}, {Dinis},
  {Maiorano}, {Maris}, {Tereno}, {Laureijs}, {Boenke}, {Buenadicha}, {Dupac},
  {Gaspar Venancio}, {G{\'o}mez-{\'A}lvarez}, {Hoar}, {Alvarez}, {Racca},
  {Saavedra-Criado}, {Schwartz}, {Vavrek}, {Schirmer}, {Aussel}, {Azzollini},
  {Cardone}, {Cropper}, {Ealet}, {Garilli}, {Gillard}, {Granett}, {Guzzo},
  {Hoekstra}, {Jahnke}, {Kitching}, {Meneghetti}, {Miller}, {Nakajima},
  {Niemi}, {Pasian}, {Percival}, {Sauvage}, {Scodeggio}, {Wachter}, {Zacchei},
  {Aghanim}, {Amara}, {Auphan}, {Auricchio}, {Awan}, {Balestra}, {Bender},
  {Bodendorf}, {Bonino}, {Branchini}, {Brau-Nogue}, {Brescia}, {Candini},
  {Capobianco}, {Carbone}, {Carlberg}, {Carretero}, {Casas}, {Castander},
  {Castellano}, {Cavuoti}, {Cimatti}, {Cledassou}, {Congedo}, {Conselice},
  {Conversi}, {Copin}, {Corcione}, {Costille}, {Courbin}, {Degaudenzi},
  {Douspis}, {Dubath}, {Duncan}, {Dusini}, {Farrens}, {Ferriol}, {Fosalba},
  {Fourmanoit}, {Frailis}, {Franceschi}, {Franzetti}, {Fumana}, {Gillis},
  {Giocoli}, {Grazian}, {Grupp}, {Haugan}, {Holmes}, {Hormuth}, {Hudelot},
  {Kermiche}, {Kiessling}, {Kilbinger}, {Kohley}, {Kubik}, {K{\"u}mmel},
  {Kunz}, {Kurki-Suonio}, {Ligori}, {Lilje}, {Lloro}, {Mansutti}, {Marggraf},
  {Markovic}, {Marulli}, {Massey}, {Maurogordato}, {Melchior}, {Merlin},
  {Meylan}, {Mohr}, {Moresco}, {Morin}, {Moscardini}, {Munari}, {Nichol},
  {Padilla}, {Paltani}, {Peacock}, {Pedersen}, {Pettorino}, {Pires}, {Poncet},
  {Popa}, {Pozzetti}, {Raison}, {Rebolo}, {Rhodes}, {Rix}, {Roncarelli},
  {Rossetti}, {Saglia}, {Schneider}, {Schrabback}, {Secroun}, {Seidel},
  {Serrano}, {Sirignano}, {Sirri}, {Skottfelt}, {Stanco}, {Starck},
  {Tallada-Cresp{\'\i}}, {Tavagnacco}, {Taylor}, {Teplitz}, {Toledo-Moreo},
  {Torradeflot}, {Trifoglio}, {Valentijn}, {Valenziano}, {Verdoes Kleijn},
  {Wang}, {Welikala}, {Weller}, {Wetzstein}, {Zamorani}, {Zoubian}, {Andreon},
  {Baldi}, {Bardelli}, {Boucaud}, {Camera}, {Fabbian}, {Farinelli},
  {Graci{\'a}-Carpio}, {Maino}, {Medinaceli}, {Mei}, {Neissner}, {Polenta},
  {Renzi}, {Romelli}, {Rosset}, {Sureau}, {Tenti}, {Vassallo}, {Zucca},
  {Baccigalupi}, {Balaguera-Antol{\'\i}nez}, {Battaglia}, {Biviano}, {Borgani},
  {Bozzo}, {Cabanac}, {Cappi}, {Casas}, {Castignani}, {Colodro-Conde},
  {Coupon}, {Courtois}, {Cuby}, {de la Torre}, {Desai}, {Di Ferdinando},
  {Dole}, {Fabricius}, {Farina}, {Ferreira}, {Finelli}, {Flose-Reimberg},
  {Fotopoulou}, {Galeotta}, {Ganga}, {Gozaliasl}, {Hook}, {Keihanen},
  {Kirkpatrick}, {Liebing}, {Lindholm}, {Mainetti}, {Martinelli}, {Martinet},
  {Maturi}, {McCracken}, {Metcalf}, {Morgante}, {Nightingale}, {Nucita},
  {Patrizii}, {Potter}, {Riccio}, {S{\'a}nchez}, {Sapone}, {Schewtschenko},
  {Schultheis}, {Scottez}, {Teyssier}, {Tutusaus}, {Valiviita}, {Viel},
  {Vriend}, \& {Whittaker}}]{Scaramella2021}
{Scaramella}, R., {Amiaux}, J., {Mellier}, Y., {et~al.} 2021, arXiv e-prints,
  arXiv:2108.01201

\bibitem[{{Schaefer, C.} {et~al.}(2018){Schaefer, C.}, {Geiger, M.}, {Kuntzer,
  T.}, \& {Kneib, J.-P.}}]{Schaefer2018}
{Schaefer, C.}, {Geiger, M.}, {Kuntzer, T.}, \& {Kneib, J.-P.} 2018, A\&A, 611,
  A2

\bibitem[{Schmidhuber(2015)}]{Schmidhuber2015}
Schmidhuber, J. 2015, Neural Networks, 61, 85

\bibitem[{{Schmidt} {et~al.}(2020){Schmidt}, {Malz}, {Soo}, {Almosallam},
  {Brescia}, {Cavuoti}, {Cohen-Tanugi}, {Connolly}, {DeRose}, {Freeman},
  {Graham}, {Iyer}, {Jarvis}, {Kalmbach}, {Kovacs}, {Lee}, {Longo}, {Morrison},
  {Newman}, {Nourbakhsh}, {Nuss}, {Pospisil}, {Tranin}, {Wechsler}, {Zhou},
  {Izbicki}, \& {LSST Dark Energy Science Collaboration}}]{Schmidt2020}
{Schmidt}, S.~J., {Malz}, A.~I., {Soo}, J.~Y.~H., {et~al.} 2020, \mnras, 499,
  1587

\bibitem[{{Schneider} {et~al.}(2007){Schneider}, {Hall}, {Richards}, {Strauss},
  {Vanden Berk}, {Anderson}, {Brandt}, {Fan}, {Jester}, {Gray}, {Gunn},
  {SubbaRao}, {Thakar}, {Stoughton}, {Szalay}, {Yanny}, {York}, {Bahcall},
  {Barentine}, {Blanton}, {Brewington}, {Brinkmann}, {Brunner}, {Castander},
  {Csabai}, {Frieman}, {Fukugita}, {Harvanek}, {Hogg}, {Ivezi{\'c}}, {Kent},
  {Kleinman}, {Knapp}, {Kron}, {Krzesi{\'n}ski}, {Long}, {Lupton}, {Nitta},
  {Pier}, {Saxe}, {Shen}, {Snedden}, {Weinberg}, \& {Wu}}]{Schneider2007}
{Schneider}, D.~P., {Hall}, P.~B., {Richards}, G.~T., {et~al.} 2007, \aj, 134,
  102

\bibitem[{{Schneider} {et~al.}(2010){Schneider}, {Richards}, {Hall}, {Strauss},
  {Anderson}, {Boroson}, {Ross}, {Shen}, {Brandt}, {Fan}, {Inada}, {Jester},
  {Knapp}, {Krawczyk}, {Thakar}, {Vanden Berk}, {Voges}, {Yanny}, {York},
  {Bahcall}, {Bizyaev}, {Blanton}, {Brewington}, {Brinkmann}, {Eisenstein},
  {Frieman}, {Fukugita}, {Gray}, {Gunn}, {Hibon}, {Ivezi{\'c}}, {Kent}, {Kron},
  {Lee}, {Lupton}, {Malanushenko}, {Malanushenko}, {Oravetz}, {Pan}, {Pier},
  {Price}, {Saxe}, {Schlegel}, {Simmons}, {Snedden}, {SubbaRao}, {Szalay}, \&
  {Weinberg}}]{Schneider2010}
{Schneider}, D.~P., {Richards}, G.~T., {Hall}, P.~B., {et~al.} 2010, \aj, 139,
  2360

\bibitem[{{Shakura} \& {Sunyaev}(1973)}]{Shakura1973}
{Shakura}, N.~I. \& {Sunyaev}, R.~A. 1973, \aap, 24, 337

\bibitem[{{Shimizu} {et~al.}(2015){Shimizu}, {Mushotzky}, {Mel{\'e}ndez},
  {Koss}, \& {Rosario}}]{Shimizu2015}
{Shimizu}, T.~T., {Mushotzky}, R.~F., {Mel{\'e}ndez}, M., {Koss}, M., \&
  {Rosario}, D.~J. 2015, \mnras, 452, 1841

\bibitem[{Stein {et~al.}(2021)Stein, Blaum, Harrington, Medan, \&
  Lukic}]{stein2021mining}
Stein, G., Blaum, J., Harrington, P., Medan, T., \& Lukic, Z. 2021, arXiv
  preprint arXiv:2110.00023

\bibitem[{{Stemo} {et~al.}(2020){Stemo}, {Comerford}, {Barrows}, {Stern},
  {Assef}, \& {Griffith}}]{Stemo2020}
{Stemo}, A., {Comerford}, J.~M., {Barrows}, R.~S., {et~al.} 2020, \apj, 888, 78

\bibitem[{Sánchez-Sáez {et~al.}(2019)Sánchez-Sáez, Lira, Cartier, Miranda,
  Ho, Arévalo, Bauer, Coppi, \& Yovaniniz}]{Sanchez-Saez2019}
Sánchez-Sáez, P., Lira, P., Cartier, R., {et~al.} 2019, Astrophysical
  Journal, Supplement Series, 242

\bibitem[{Tan \& Le(2019)}]{tan2019efficientnet}
Tan, M. \& Le, Q. 2019, in International Conference on Machine Learning, PMLR,
  6105--6114

\bibitem[{{Thacker} {et~al.}(2014){Thacker}, {MacMackin}, {Wurster}, \&
  {Hobbs}}]{Tracker2014}
{Thacker}, R.~J., {MacMackin}, C., {Wurster}, J., \& {Hobbs}, A. 2014, \mnras,
  443, 1125

\bibitem[{{Thom} {et~al.}(2012){Thom}, {Tumlinson}, {Werk}, {Prochaska},
  {Oppenheimer}, {Peeples}, {Tripp}, {Katz}, {O'Meara}, {Ford}, {Dav{\'e}},
  {Sembach}, \& {Weinberg}}]{Thom2012}
{Thom}, C., {Tumlinson}, J., {Werk}, J.~K., {et~al.} 2012, \apjl, 758, L41

\bibitem[{{Thomas} {et~al.}(2002){Thomas}, {Maraston}, \&
  {Bender}}]{Thomas2002}
{Thomas}, D., {Maraston}, C., \& {Bender}, R. 2002, \apss, 281, 371

\bibitem[{Torbaniuk {et~al.}(2021)Torbaniuk, Paolillo, Carrera, Cavuoti,
  Vignali, Longo, \& Aird}]{Torbaniuk2021}
Torbaniuk, O., Paolillo, M., Carrera, F., {et~al.} 2021, Monthly Notices of the
  Royal Astronomical Society, 506, 2619

\bibitem[{{Trump} {et~al.}(2013){Trump}, {Konidaris}, {Barro}, {Koo},
  {Kocevski}, {Juneau}, {Weiner}, {Faber}, {McLean}, {Yan},
  {P{\'e}rez-Gonz{\'a}lez}, \& {Villar}}]{Trump2013}
{Trump}, J.~R., {Konidaris}, N.~P., {Barro}, G., {et~al.} 2013, \apjl, 763, L6

\bibitem[{Wang {et~al.}(2020)Wang, Yao, Kwok, \& Ni}]{wang2020generalizing}
Wang, Y., Yao, Q., Kwok, J.~T., \& Ni, L.~M. 2020, ACM Computing Surveys
  (CSUR), 53, 1

\bibitem[{{Weir} {et~al.}(1995){Weir}, {Fayyad}, \& {Djorgovski}}]{Weir1995}
{Weir}, N., {Fayyad}, U.~M., \& {Djorgovski}, S. 1995, \aj, 109, 2401

\bibitem[{{Wenzl} {et~al.}(2021){Wenzl}, {Schindler}, {Fan}, {Andika},
  {Ba{\~n}ados}, {Decarli}, {Jahnke}, {Mazzucchelli}, {Onoue}, {Venemans},
  {Walter}, \& {Yang}}]{Wenzl2021}
{Wenzl}, L., {Schindler}, J.-T., {Fan}, X., {et~al.} 2021, \aj, 162, 72

\bibitem[{{York} {et~al.}(2000){York}, {Adelman}, {Anderson}, {Anderson},
  {Annis}, {Bahcall}, {Bakken}, {Barkhouser}, {Bastian}, {Berman}, {Boroski},
  {Bracker}, {Briegel}, {Briggs}, {Brinkmann}, {Brunner}, {Burles}, {Carey},
  {Carr}, {Castander}, {Chen}, {Colestock}, {Connolly}, {Crocker}, {Csabai},
  {Czarapata}, {Davis}, {Doi}, {Dombeck}, {Eisenstein}, {Ellman}, {Elms},
  {Evans}, {Fan}, {Federwitz}, {Fiscelli}, {Friedman}, {Frieman}, {Fukugita},
  {Gillespie}, {Gunn}, {Gurbani}, {de Haas}, {Haldeman}, {Harris}, {Hayes},
  {Heckman}, {Hennessy}, {Hindsley}, {Holm}, {Holmgren}, {Huang}, {Hull},
  {Husby}, {Ichikawa}, {Ichikawa}, {Ivezi{\'c}}, {Kent}, {Kim}, {Kinney},
  {Klaene}, {Kleinman}, {Kleinman}, {Knapp}, {Korienek}, {Kron}, {Kunszt},
  {Lamb}, {Lee}, {Leger}, {Limmongkol}, {Lindenmeyer}, {Long}, {Loomis},
  {Loveday}, {Lucinio}, {Lupton}, {MacKinnon}, {Mannery}, {Mantsch}, {Margon},
  {McGehee}, {McKay}, {Meiksin}, {Merelli}, {Monet}, {Munn}, {Narayanan},
  {Nash}, {Neilsen}, {Neswold}, {Newberg}, {Nichol}, {Nicinski}, {Nonino},
  {Okada}, {Okamura}, {Ostriker}, {Owen}, {Pauls}, {Peoples}, {Peterson},
  {Petravick}, {Pier}, {Pope}, {Pordes}, {Prosapio}, {Rechenmacher}, {Quinn},
  {Richards}, {Richmond}, {Rivetta}, {Rockosi}, {Ruthmansdorfer}, {Sandford},
  {Schlegel}, {Schneider}, {Sekiguchi}, {Sergey}, {Shimasaku}, {Siegmund},
  {Smee}, {Smith}, {Snedden}, {Stone}, {Stoughton}, {Strauss}, {Stubbs},
  {SubbaRao}, {Szalay}, {Szapudi}, {Szokoly}, {Thakar}, {Tremonti}, {Tucker},
  {Uomoto}, {Vanden Berk}, {Vogeley}, {Waddell}, {Wang}, {Watanabe},
  {Weinberg}, {Yanny}, {Yasuda}, \& {SDSS Collaboration}}]{York2000}
{York}, D.~G., {Adelman}, J., {Anderson}, Jr., J.~E., {et~al.} 2000, \aj, 120,
  1579

\bibitem[{{Yu} \& {Richards}(2021)}]{Yu2021Challenge}
{Yu}, W. \& {Richards}, G. 2021, LSSTC AGN Data Challenge,
  \url{https://github.com/RichardsGroup/AGN_DataChallenge}

\bibitem[{{Zhang} \& {Hao}(2018)}]{Zhang2018}
{Zhang}, K. \& {Hao}, L. 2018, \apj, 856, 171

\bibitem[{{Zou} \& {Zhong}(2018)}]{Zou2018}
{Zou}, M. \& {Zhong}, Y. 2018, Sensing and Imaging, 19, 6

\end{thebibliography}
